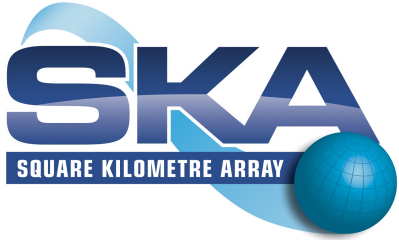

# SKA-Athena Synergy White Paper

## SKA-Athena Synergy Team

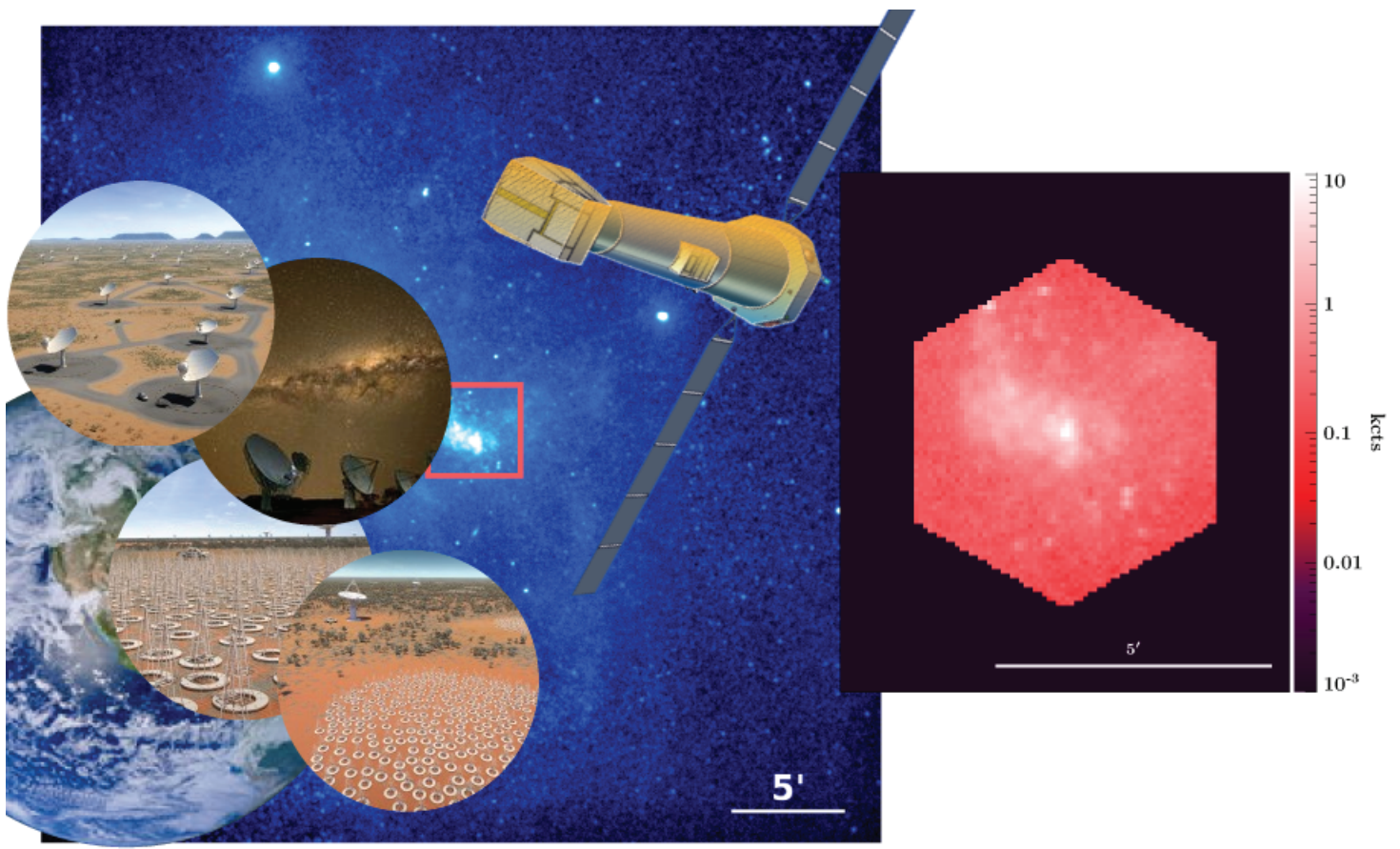

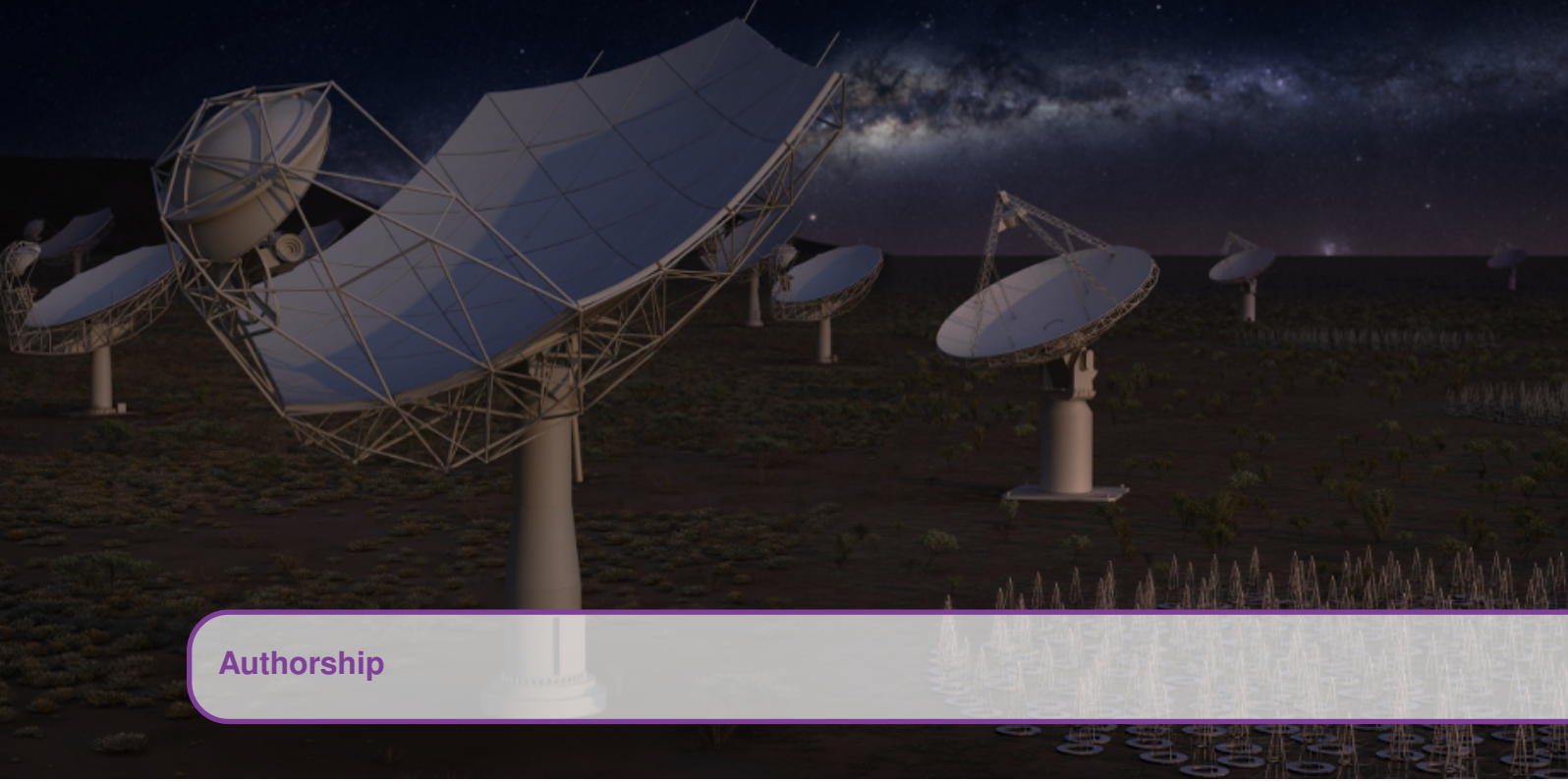

## Authors

- Rossella Cassano (INAF-Istituto di Radioastronomia, Italy).
- Rob Fender (University of Oxford, United Kingdom).
- Chiara Ferrari (Observatoire de la Côte d'Azur, France).
- Andrea Merloni (Max-Planck Institute for Extraterrestrial Physics, Germany).

## Contributors

- Takuya Akahori (Kagoshima University, Japan).
- Hiroki Akamatsu (SRON Netherlands Institute for Space Research, The Netherlands).
- Yago Ascasibar (Universidad Autónoma de Madrid, Spain).
- David Ballantyne (Georgia Institute of Technology, United States).
- Gianfranco Brunetti (INAF-Istituto di Radioastronomia, Italy) and Maxim Markevitch (NASA-Goddard Space Flight Center, United States).
- Judith Croston (The Open University, United Kingdom).
- Imma Donnarumma (Agenzia Spaziale Italiana, Italy) and E. M. Rossi (Leiden Observatory, The Netherlands).
- Robert Ferdman (University of East Anglia, United Kingdom) on behalf of the SKA Pulsar Science Working Group.
- Luigina Feretti (INAF-Istituto di Radioastronomia, Italy) and Federica Govoni (INAF Osservatorio Astronomico,Italy).
- Jan Forbrich (University of Hertfordshire, United Kingdom).
- Giancarlo Ghirlanda (INAF-Osservatorio Astronomico di Brera and University Milano Bicocca, Italy).
- Adriano Ingallinera (INAF-Osservatorio Astrofisico di Catania, Italy).
- Andrei Mesinger (Scuola Normale Superiore, Italy).
- Vanessa Moss and Elaine Sadler (Sydney Institute for Astronomy/CAASTRO and University of Sydney, Australia).
- Fabrizio Nicastro (Osservatorio Astronomico di Roma,Italy), Edvige Corbelli (INAF-Osservatorio Astrofisico di Arcetri, Italy) and Luigi Piro (INAF, Istituto di Astrofisica e Planetologia Spaziali, Italy).
- Paolo Padovani (European Southern Observatory, Germany).
- Francesca Panessa (INAF/Istituto di Astrofisica e Planetologia Spaziali, Italy).
- Gabriele Ponti (Max-Planck Institute for Extraterrestrial Physics, Germany).
- Gabriel Pratt (IRFU and Universitè Paris Diderot, France) and Melanie Johnston-Hollitt (School of Chemical and Physical Sciences, New Zealand).
- Reinout van Weeren (Leiden Observatory, Leiden University, the Netherlands).

- Manami Sasaki (Dr. Karl Remeis Observatory, Erlangen Centre for Astroparticle Physics, Friedrich-Alexander-University Erlangen-Nürnberg, Germany).
- Roberto Soria (ICRAR-Curtin University, Australia and NAOC, Chinese Academy of Sciences, China).
- Ian Stevens (University of Birmingham, United Kingdom).
- Franco Vazza (Università di Bologna, Italy and Universität Hamburg, Germany), Stefano Ettori (INAF Osservatorio Astronomico di Bologna, Italy) and C. Gheller (CSCS-ETHZ, Switzerland).
- Natalie Webb (Institut de Recherche en Astrophysique et Planétologie, France).

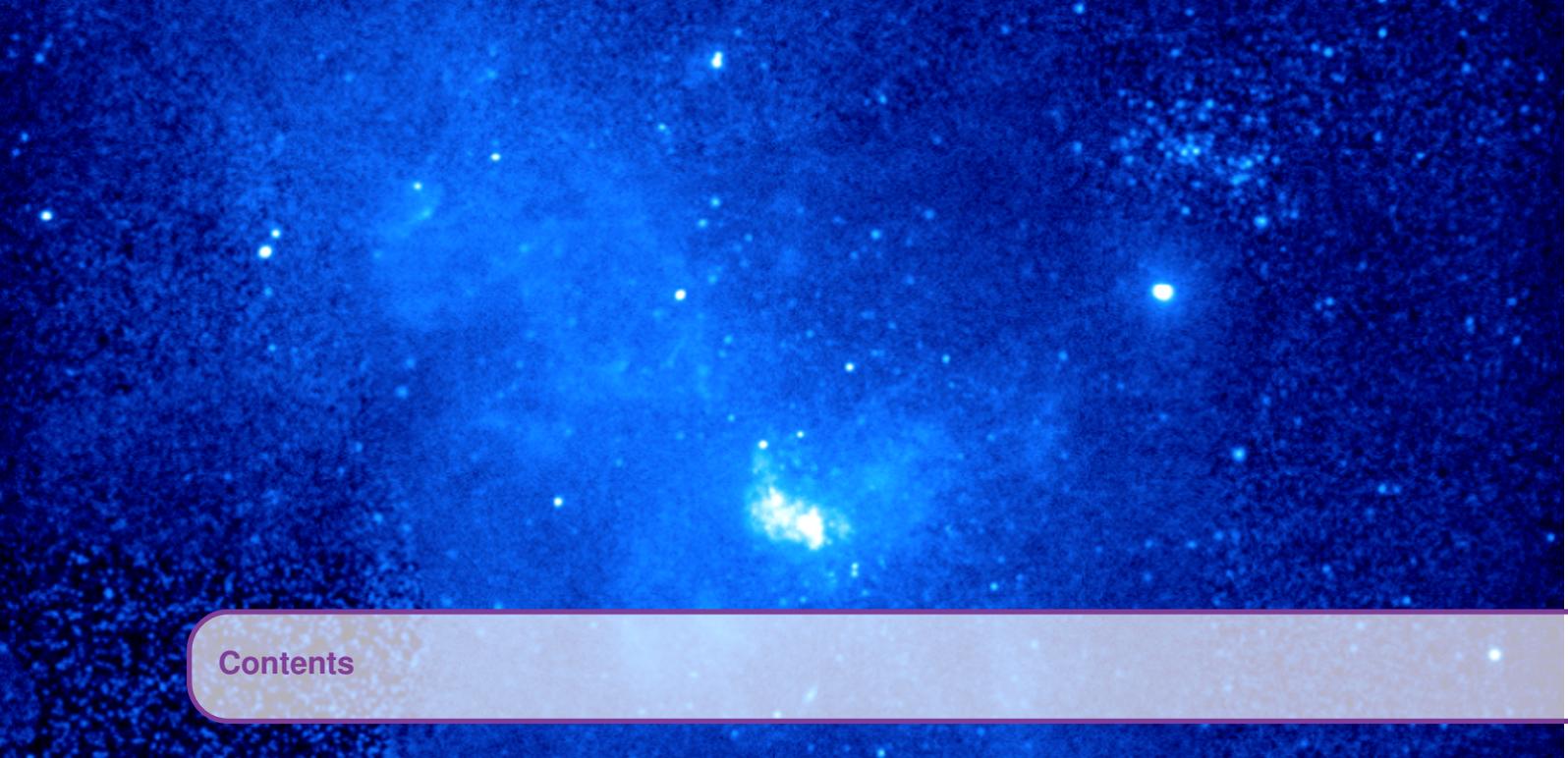

**Contents**

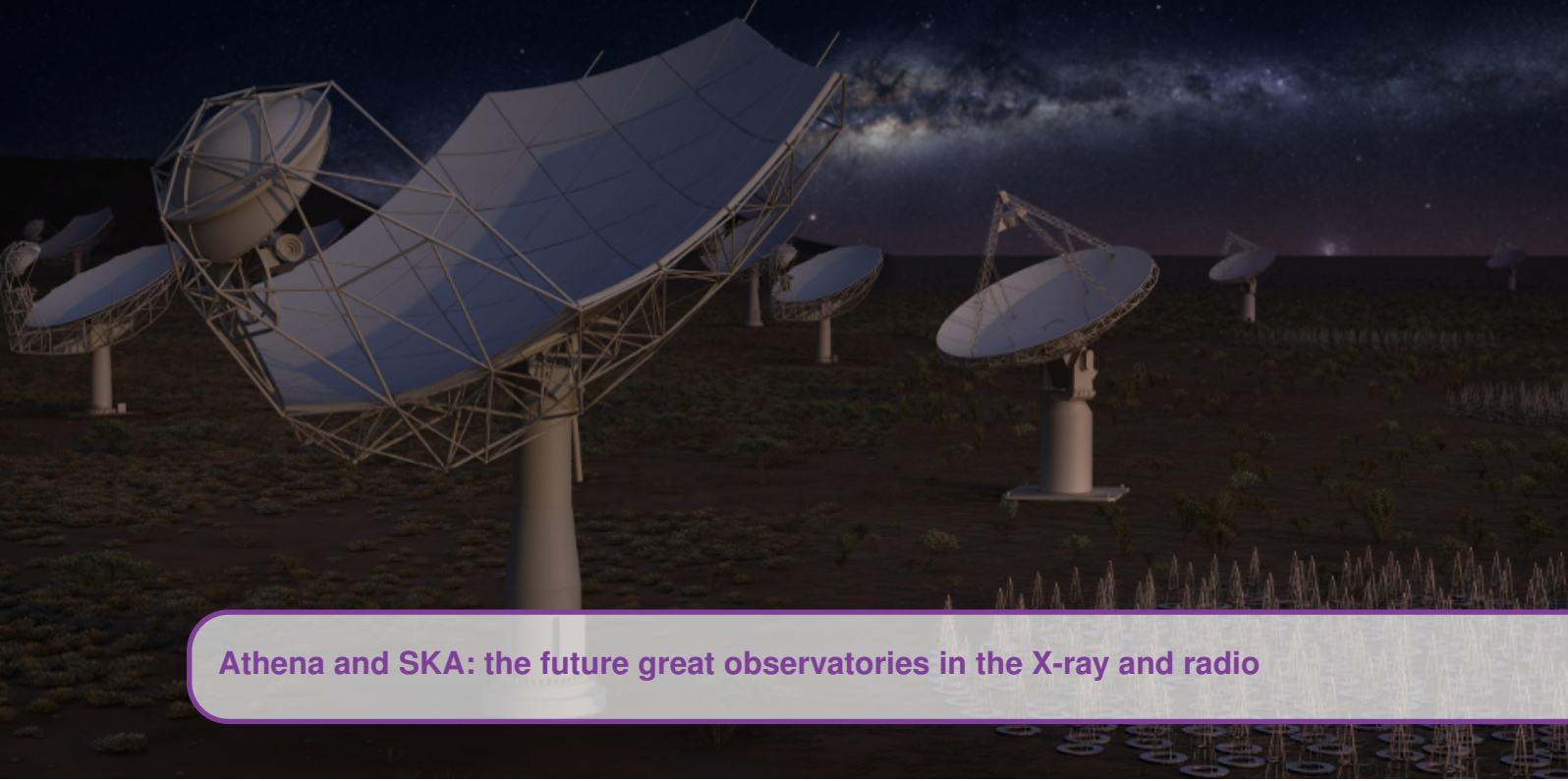

**Athena and SKA: the future great observatories in the X-ray and radio**

*Athena* (Advanced Telescope for High ENergy Astrophysics) is the X-ray observatory large mission selected by the European Space Agency (ESA), within its Cosmic Vision 2015-2025 programme, to address the Hot and Energetic Universe scientific theme (Nandra et al. 2013), and it is provisionally due for launch in the early 2030s. *Athena* will have three key elements to its scientific payload: an X-ray telescope with a focal length of 12 m and two instruments: a Wide Field Imager (*Athena*/WFI) for high count rate, moderate resolution spectroscopy over a large field of view (FoV) and an X-ray Integral Field Unit (*Athena*/X-IFU) for high-spectral resolution imaging. In Tab. 1 we report the main requirements of the *Athena* observatory, as constrained by the main science drivers of the mission[1]. For a detailed description of the instrument capabilities, we refer the reader to the Athena Science Requirements document (Lumb and den Herder 2017)[1].

The Square Kilometer Array (SKA) is the next generation radio observatory and consists of two telescopes, one comprised of dishes operating at mid frequencies (SKA1-MID) and located in South Africa, and the other comprised of Log-Periodic antennas operating at low radio frequencies (SKA1-LOW), which will be located in Australia. The scientific commissioning of the radio telescope is planned to begin in 2021-2022, after the first stations are expected to be available, and will proceed during the roll-out of the complete array. We report in Tab. 2 examples of SKA1-LOW and SKA1-MID capabilities, while for a full description of SKA1 specifications we refer the reader to the Anticipated SKA1 Science Performance document (Braun et al. 2017).

The *Athena* Science Study Team (ASST) and the SKA Organisation (SKAO) agreed to undertake an exercise to identify and develop potential synergies between these two large observatories. To this end an SKA-*Athena* Synergy Team (SAST) was established, composed of four scientists with expertise in different science topics (R. Cassano, R. Fender, C. Ferrari and A. Merloni).

The astrophysical community was involved in this exercise primarily through an SKA-*Athena* Synergy workshop, which took place on April 24-25, 2017 at SKAO, Jodrell Bank, Manchester. Thanks to the many insights gained in preparation for and during the workshop, a number of synergy topics naturally arose within scientific areas where joint radio — X-ray studies should be decisive. These include:

1. Understanding the early Universe, including the sources responsible for the reionization of the Universe at $z > 7$ and the formation of the first generation of stars.
2. Unveiling the growth of supermassive black holes over cosmic time, and determining its relationship to star formation and the evolution of galaxies.
3. Investigating the role of black-hole feedback in shaping galaxy clusters, via the determination of the physical properties of the gas in cluster cores from X-ray observations, and radio studies of the non-thermal cavity contents.
4. Determining the nature of non-thermal phenomena in galaxy clusters, including the relationship between cluster radio halos and turbulence, and the connections between X-ray shock structures and radio relics.

---

[1] All of the expected *Athena* results reported here refer to the mission as described in Nandra et al. 2013.

5. Revealing and illuminating the Cosmic Web of baryons, with the exciting possibility of detecting both thermal and non-thermal emission of cosmic filaments to constrain the plasma conditions at strong accretion shocks, in a hitherto poorly known environment.

6. Combining multiple probes of accretion and outflow physics in X-ray Binaries (XRBs), transients, active galactic nuclei (AGN) and Tidal Disruption Events (TDEs).

7. Pushing forward our understanding of the life cycles of stars in our Galaxy, including young stellar objects (YSOs) and ultra-cool dwarfs, star-planet magnetic interaction, massive stars, pulsars and supernova remnants (SNR).

The final result of the synergy exercise, this White Paper, describes in detail a number of scientific opportunities that will be opened up by the combination of *Athena*[1] and SKA. Some of these identified synergies derive from the core science cases of one of the observatories, which can be enhanced by combining with the other. We refer the reader to the "*Advanced Astrophysics with the Square Kilometre Array*" document[2] for a broad overview of SKA science cases, updated at 2014. For a more detailed description of the science cases that will be enabled by *Athena*, we refer the reader to the White and Supporting Papers written in support of the mission proposal[3]. Other synergies are genuinely new topics which can be opened up only by using both facilities in tandem. We stress, however, that this White Paper can only represent a partial view of the vast astrophysical landscape that will be explored and revealed by these future facilities. This undoubtedly includes many phenomena that have yet to be discovered, by these or other future observatories, in which joint Athena-SKA observations will bring new insight.

Table 1: *Athena* main scientific requirements

| Parameter | Requirements | Scientific driver |
|---|---|---|
| Effective Area at 1keV | ≥1.4 m² | Early groups, cluster entropy and metal evolution, WHIM, first stars |
| Effective Area at 6keV | 0.25 m² | Cluster bulk motions and turbulence, AGN winds and outflows, BH spin |
| HEW (spatial resolution) | 5" on-axis, 10" off-axis | High-z AGN, early groups, AGN feedback on cluster scales |
| WFI point source sensitivity | $2.4 \times 10^{-17}$ erg/s/cm² (in 450 ks) | AGN evolution, first stars, high-z groups |
| X-IFU spectral resolution | 2.5 eV | WHIM, cluster hot gas energetics and AGN feedback, AGN outflows |
| WFI spectral resolution | 150 eV (at 6 keV) | GBH spin, reverberation mapping |
| WFI Field of View | 40' × 40' square | High-z AGN, census AGN, early groups, cluster entropy evolution |
| X-IFU Field of View | 5' effective diameter | AGN feedback in clusters, IGM physics |
| ToO Trigger efficiency | 50% | WHIM, GW |
| ToO Reaction time | ≤ 4 hrs | WHIM, first generation of stars, GW |

Notes – See Athena Science Requirements document (Lumb and den Herder 2017).

Table 2: Imaging sensitivity of SKA1-LOW (top) and SKA1-MID (bottom) at a central frequency $\nu_c$ for spectral line ($\sigma_L$) and continuum ($\sigma_C$) observations. For SKA1-LOW, values of the confusion noise, $\sigma_{conf}$, at $\nu_c$ and $\theta_{min}$ are listed according to Condon et al. 2012. Numbers in the table are taken from Tab. 1 and Tab. 2 in the *Anticipated SKA1 Science Performance document*, by Braun et al. 2017.

| $\nu_c$ MHz | $\sigma_L$ $\mu$Jy/beam | $\sigma_C$ $\mu$Jy/beam | $\sigma_{conf}$ $\mu$Jy/beam | $\theta_{min} - \theta_{max}$ arcsec |
|---|---|---|---|---|
| 82 | 3261 | 47 | 183 | 17−850 |
| 158 | 1258 | 18 | 13 | 8.9−444 |
| 218 | 973 | 14 | 4 | 6.4−321 |
| 302 | 794 | 11 | 1 | 4.6−232 |
| 770 | 303 | 4.4 | - | 1.092−145.6 |
| 1430 | 137 | 2.0 | - | 0.587−78 |
| 4940 | 95 | 1.4 | - | 0.17−22.7 |
| 12530 | 85 | 1.2 | - | 0.067−8.9 |

Notes – $\sigma_L$ is derived for $\Delta \nu / \nu_c = 10^{-4}$ and $\sigma_c$ for $\Delta \nu / \nu_c = 0.3$, both for an observation of 1 hour. The range of Gaussian FWHM beam sizes for which the approximate sensitivity value applies is given by $\theta_{min}$ to $\theta_{max}$.

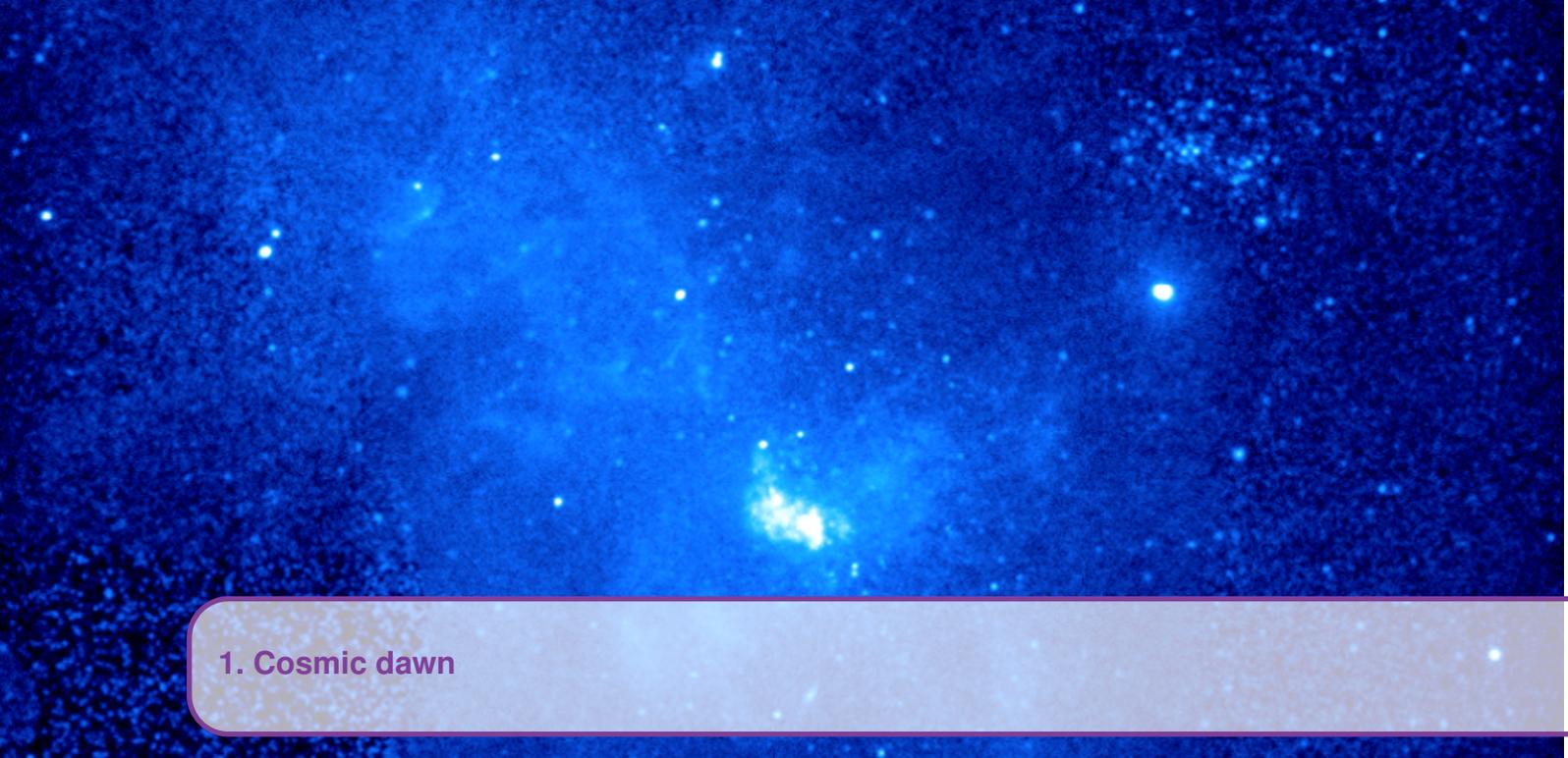

# 1. Cosmic dawn

## 1.1 The epoch of reionization

### 1.1.1 Introduction

Following recombination at $z \sim 1100$, when the Universe was only ~300 000 years old, cosmic gas began to collapse under gravity and condense onto dark matter structures. Around a hundred million years later, this gravitational collapse culminated in the birth of the first stars and galaxies. During this Cosmic Dawn (CD), the light from the first galaxies spread out, influencing their surroundings through complicated feedback processes. Eventually, at some point during the first billion years, the radiation from the first galaxies heated and ionized virtually every atom in our Universe, resulting in its last major phase change: the Epoch of Reionization (EoR).

The latest observations of the EoR suggest that the bulk of it occurred sometime between $6 \lesssim z \lesssim 10$. Despite recent progress on understanding the timing of the EoR (driven by observations of the optical depth to the Cosmic Microwave Background (CMB) and the damping wing in Quasi Stellar Objects (QSO) spectra; see Greig and Mesinger 2017b for a recent, detailed summary), we know very little about *how* this cosmic milestone unfolded. Was it driven by abundant faint galaxies or rare bright ones? To what extent did AGN contribute? Did radiative or SNe feedback play a large role in shaping the star formation history of these galaxies? When did recombining, Lyman limit systems begin to regulate the ionizing background?

Moreover, the CD is more than just the EoR. Following thermal decoupling from the CMB at $z \sim 200$, the Intergalactic Medium (IGM) adiabatically cooled with the expansion of the Universe. Barring the influence of radiation fields, this meant that the average temperature of the IGM at $z \sim 10$ could have been as low as $\sim$ few K. As the cosmic ionization fronts originating from the first galaxies swept through the IGM, gas was impulsively heated to temperatures of $\sim 10^4$ K (e.g., Hui and Gnedin 1997). If the EoR did proceed inside an adiabatically cold IGM, the contrast provided by the cosmic HI and HII patches would produce a very large 21-cm signal. Such extreme, "cold reionization" models have already been ruled out with current observations (Pober et al. 2015; Greig, Mesinger, and Pober 2016).

Instead, empirical relations based on local, star-forming galaxies suggest that the X-rays from the first galaxies heated the IGM to temperatures above the CMB before the EoR. *Virtually nothing is currently known about this Epoch of Heating (EoH).* X-rays, with their long mean free paths, can easily escape early galaxies, partially pre-ionizing the IGM and heating it through free-free interactions of the ejected electrons. Which sources dominated the early X-ray background: high-mass X-ray binaries, the hot interstellar medium (ISM), faint AGN, etc.? Did cosmic rays or even more exotic processes like dark matter annihilation play a role?

Our hope for answering these and many more "Cosmic Dawn" questions will come in the near future, enabled by the spin-flip transition of neutral hydrogen. This transition produces a photon with a wavelength of 21 cm in the rest frame. The strength of the cosmic 21-cm emission depends on both the ionization and the thermal state of the IGM. By observing the intensity of this emission against the CMB, the SKA will map out the 3D structure of the EoR and EoH (see Fig. 1.1). *Encoded in these patterns are the properties of the*

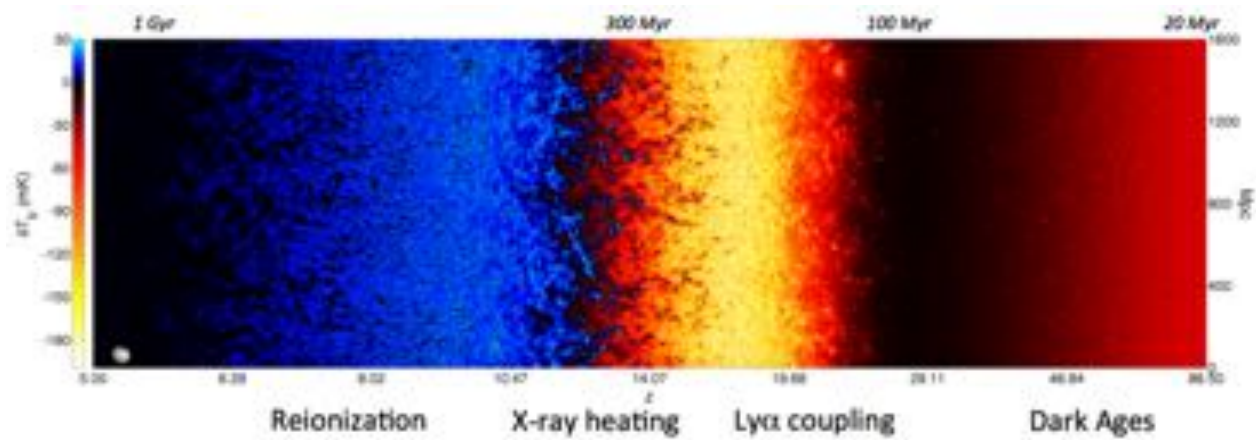

Figure 1.1: Light-cone slice through the 21-cm brightness temperature offset from the CMB. The main cosmic milestones can be seen from right to left: (i) first stars (Wouthuysen-Field coupling in black→yellow), (ii) first black holes (X-ray heating in yellow→blue), (iii) first galaxies (cosmic reionization in blue→black). The SKA will be able to detect reionization and X-ray heating with a very high signal to noise ratio, perhaps even pushing into earlier epochs with marginal detections (Mesinger, Greig, and Sobacchi 2016). *The UV and X-ray properties of the first galaxies are encoded in the timing and the patterns of the signal.*

*first structures of our Universe.*

To take advantage of the wealth of information in the signal, the SKA CD working group is proposing a three-tier survey strategy, starting with the first phase of SKA1-LOW in Australia (e.g., Koopmans et al. 2015): (i) a *shallow* field with ∼ 10 h integration over the whole sky (i.e., the visible 3-π steradians) (minimizing the cosmic variance of the signal and enabling overlap with other instruments); (ii) a *deep* field with ∼ 1000 h over ∼100 deg² (allowing for power spectrum estimation at the highest redshifts and imaging at the lowest redshifts); and (iii) a *medium* field with ∼ 100 h over ∼ 1000 deg² (as a compromise between the two extremes). The second phase of construction is expected to begin in the mid-2020s, and would increase the collecting area of the instrument by an order of magnitude.

*Athena* will *offer a complementary view of the early Universe.* The *Athena*/WFI, with its large collecting area at 1 keV, can carry out surveys one to two orders of magnitude faster than *XMM-Newton* and *Chandra*. A four-tier survey is envisioned with the *Athena*/WFI. Both SKA and *Athena* will allow us to study the first sources, but at very different energy regimes. In the following section, we list some areas of overlap between the two efforts, suggesting strategies to best answer cosmic dawn science questions.

### 1.1.2 Did galaxies or AGN drive reionization? When does the transition occur?

There has been recent controversy surrounding the relative contribution of AGN and star-forming galaxies to the cosmic radiation fields. At lower redshifts, the ionizing background (UVB) is dominated by AGN, while empirical extrapolations of AGN and galaxy luminosity functions (LFs) predict that above some redshift, star-forming galaxies dominate the UVB (e.g., Haardt and Madau 2012). These predictions depend on the relative abundance of faint AGN and galaxies (e.g., Giallongo et al. 2015; Parsa, Dunlop, and McLure 2017).

- *Conducting high-z surveys of faint AGN*: The planned *Athena*/WFI deep survey will push the X-ray LFs at high redshifts to over an order of magnitude fainter than current limits (e.g. Gilli, Comastri, and Hasinger 2007). Of the order of 20 AGN are expected to be discovered at luminosities of $L_X \lesssim 10^{45}$ erg s⁻¹ at $8 \lesssim z \lesssim 10$ (e.g., Aird et al. 2013).
- *Improving our understanding of the $L_X - L_{UV}$ relation*: Estimating the contribution of X-ray detected high-z AGN to the UVB requires a knowledge of the corresponding ionizing luminosities. While near-IR (NIR) coverage should be present with other instruments, allowing us to measure the non-ionizing UV magnitudes of the AGN, it will be necessary to understand the relative contributions of the AGN and the stars within the host galaxy (e.g., Lusso et al. 2011). Indeed, contamination by massive stars has been flagged as an issue in interpreting the UV luminosities of high-z faint AGN recently detected with *Chandra* (e.g., Ricci et al. 2017).

### 1.1.3 What is the large-scale environment of high-z AGN?

Do high-$z$ AGN live in special environments? There is evidence that bright QSOs live in overdense regions (e.g., Garcia-Vergara et al. 2017). Does this translate to the EoR?

- *Cross-correlating the high-$z$ AGN with the 21-cm signal: are AGN environments typically over-ionized?* If AGN during the EoR also live in overdense environments, we would expect them to be surrounded by a larger-than-average cosmic HII region. Cross-correlating the high-$z$ AGN detected by *Athena* with the 21-cm signal detected with the SKA can give us the typical sizes of the surrounding HII regions.
- *Imaging high-z, bright AGN*: One of the main science goals of SKA1-LOW is imaging the CD. If indeed AGN locations are over-ionized, SKA will be able to identify potential sites for *Athena* follow-up. This can be done either blindly using a matched filter technique to boost signal-to-noise (e.g. Datta et al. 2016), or using phase 2 of SKA-LOW, which should be operational on the same timescale as *Athena* and will have an order of magnitude more collecting area than phase 1.

### 1.1.4 Which source(s) of X-rays drove the epoch of heating in the early Universe?

As mentioned above, we also do not know what the dominant sources of X-rays were in the early Universe, i.e. those responsible for the EoH. Empirical relations based on local star-forming galaxies (e.g., Mineo, Gilfanov, and Sunyaev 2012a) and population synthesis models (e.g., Fragos et al. 2013) point to high-mass X-ray binaries as likely sources. However, locally the hot ISM has a comparable soft-band luminosity (Mineo, Gilfanov, and Sunyaev 2012b). More exotic sources, such as cosmic rays (Leite et al. 2017) or dark matter annihilation (e.g. Evoli, Mesinger, and Ferrara 2014; Lopez-Honorez et al. 2016) have also been proposed. Strategies to address this question include:

- *Characterizing the soft-band Spectral Energy Distributions (SEDs) of local star-forming galaxies*: Because photons with energies $\gtrsim 2$ keV have a mean free path longer than the Hubble length, it is the soft X-rays that are responsible for the EoH. *Athena* will have over an order of magnitude larger effective area in the soft band compared to *XMM-Newton* and *Chandra* (Willingale et al. 2013). This should allow us to better characterize the soft-band luminosities of local and moderate redshift star-forming galaxies, and how this scales with the star formation rate (e.g., Lehmer et al. 2016). This will be important in interpreting EoH observations with SKA, since the intrinsic shape of the soft-band SED is degenerate with the unknown level of self-absorption by the ISM of the host galaxy (Greig and Mesinger 2017a).
- *Cross-correlating the unresolved X-ray background with the 21-cm signal*: The soft X-rays are responsible for heating the IGM during the first billion years. However, the hard X-rays from the same sources should still show up in the present-day X-ray background. Although likely too faint to be resolved by *Athena*, these EoH sources could show up in a cross-correlation of the unresolved X-ray background with the 21-cm signal during the EoH. This would isolate the cosmic origin of part of the unresolved X-ray background.

## 1.2 The first stars

### 1.2.1 Introduction: Population III Stars

Population III stars (Pop-III) are assumed to form from a gas of primordial composition in dark matter minihalos, and represent the first sources of light, heat and metals in the early Universe. Due to their complete lack of heavy elements, they are believed to be massive ($\sim 40\,M_\odot$) to very massive (in the range $100 - 1000\,M_\odot$), producing an order of magnitude more ionizing photons than present-day stars (e.g., Tumlinson and Shull 2000). However, their imprint on the EoR is degenerate with the influence of the unknown star-formation efficiencies and ionizing escape fractions of the first galaxies. Breaking this degeneracy and isolating the role of Pop-III stars would be very helpful in interpreting the 21-cm signal from the EoR, as discussed in the previous section.

The relevance of primeval stars, however, is manifold, as they also represent a way to produce the seeds of massive black holes (powering quasars at $z\sim6$), and they provide the feedback that regulates the assembly process of the first galaxies (Ciardi and Ferrara 2005). Directly probing Pop-III stars might be possible by catching them at the time of their explosion as hypernovae, pair-instability supernovae or Gamma Ray Bursts (GRBs). Indeed, at least a fraction of Pop-III stars, if their Initial Mass Function is biased towards large masses, could collapse into massive black holes (Abel, Bryan, and Norman 2002; Bromm et al. 2009). If such stars are able to launch a relativistic jet that successfully breaks out of the massive/extended progenitor's envelope, then they should be recognizable as extremely powerful GRBs (Komissarov and Barkov 2010). Pop-III GRBs would be unique probes of primordial cosmic star formation and of the intergalactic medium (Salvaterra 2015, and references therein).

### 1.2.2 *Athena* and SKA: revealing GRB-PopIII progenitors

Although the mass range for Pop-III progenitors is still uncertain, recent estimates for the extreme super-collapsar case (e.g. Mészáros and Rees 2010) and numerical simulations (Suwa and Ioka 2011; Nagakura, Suwa, and Ioka 2012) suggest that the jet can break out from a Pop-III massive progenitor if energy injection lasts $\sim 10^{3-4}$ s. The $\gamma$–ray burst event should be extremely long and energetic, with an isotropic equivalent kinetic energy $> 10^{56}$ erg, corresponding to a typical GRB isotropic luminosity of $10^{52-53}$ erg/s. GRBs produced by Pop-III stars might already be within reach of current and forthcoming gamma-ray satellites (e.g. Nakauchi et al. 2012). The most difficult issue is how to recognise them. The large energy and duration should set them apart from the distribution of long and short GRBs in the $\gamma$-ray diagnostic planes (Burlon et al. 2016). While the measurement of a GRB at z~20 would be the most direct evidence of a Pop-III progenitor (e.g., Bromm et al. 2009; Abel, Bryan, and Norman 2002), Pop-III GRBs could also occur, and be more numerous, at much lower redshifts (down to z~5, Tornatore, Ferrara, and Schneider 2007; Campisi et al. 2011) due to the relatively long and patchy transition from primordial stars to normal Population II (Pop-II)/Population I (Pop-I) stars. For these lower redshift events, the identification of a genuine Pop-III progenitor would require complementary (and multi-wavelength) observations.

*Athena* can uniquely trace the closest environment of GRBs through X-ray absorption spectroscopy of the ionised medium (Jonker et al. 2013). A pristine metal-free environment can be established via deep (down to a few percent solar) upper limits on the metallicity. These should be provided by early (< few days) GRB follow-up by *Athena*/X-IFU. Abundance-ratio diagnostics could then indirectly probe the effect of metal pollution by Pop-III in the environment back–illuminated by Pop-II/I GRBs at intermediate redshifts (Ma et al. 2015; Heger and Woosley 2010). The long-lasting interaction of the jet with the progenitor's envelope, before the former can break out, could deposit a considerable amount of energy in a cocoon surrounding the jet. Similarly to the interpretation proposed for the long duration GRB 130925a (Piro et al. 2014), *Athena* could detect the X-ray thermal signature of the cocoon emission as it breaks out of the star.

Distinctive features (energy and duration) of the prompt $\gamma$–ray emission and X-ray observations alone would, however, be insufficient to identify GRBs from Pop-III progenitors. Valuable orthogonal information can be obtained from SKA observations of the radio afterglow: the extreme kinetic power of the jet is expected to produce bright radio afterglows peaking at late times (Ghirlanda et al. 2013). While the X-ray afterglow light curve decay of a Pop-III GRB is similar–to that of a Pop-II event, it has been shown (de Souza, Yoshida, and Ioka 2011; Ghirlanda et al. 2013; Burlon et al. 2016; Meiksin and Whalen 2013) that flux densities of 1-10 mJy at 5 GHz could be reached 100 days after the prompt $\gamma$–ray event by a Pop-III prototypical burst (Nakauchi et al. 2012; Mesler et al. 2014). At these epochs, a typical Pop-II GRB would be a factor 10-100 dimmer. Therefore, late time detection of a bright radio afterglow would add the last piece of the puzzle. This final confirmation can be obtained only by the SKA a few years after the Pop-III GRB, when the flux has faded below the $\mu$Jy level and the jet has become non-relativistic: at this stage, the measurement of the jet's total energy can prove the extraordinary progenitor, and so provide the definitive evidence for the discovery of a Pop-III GRB.

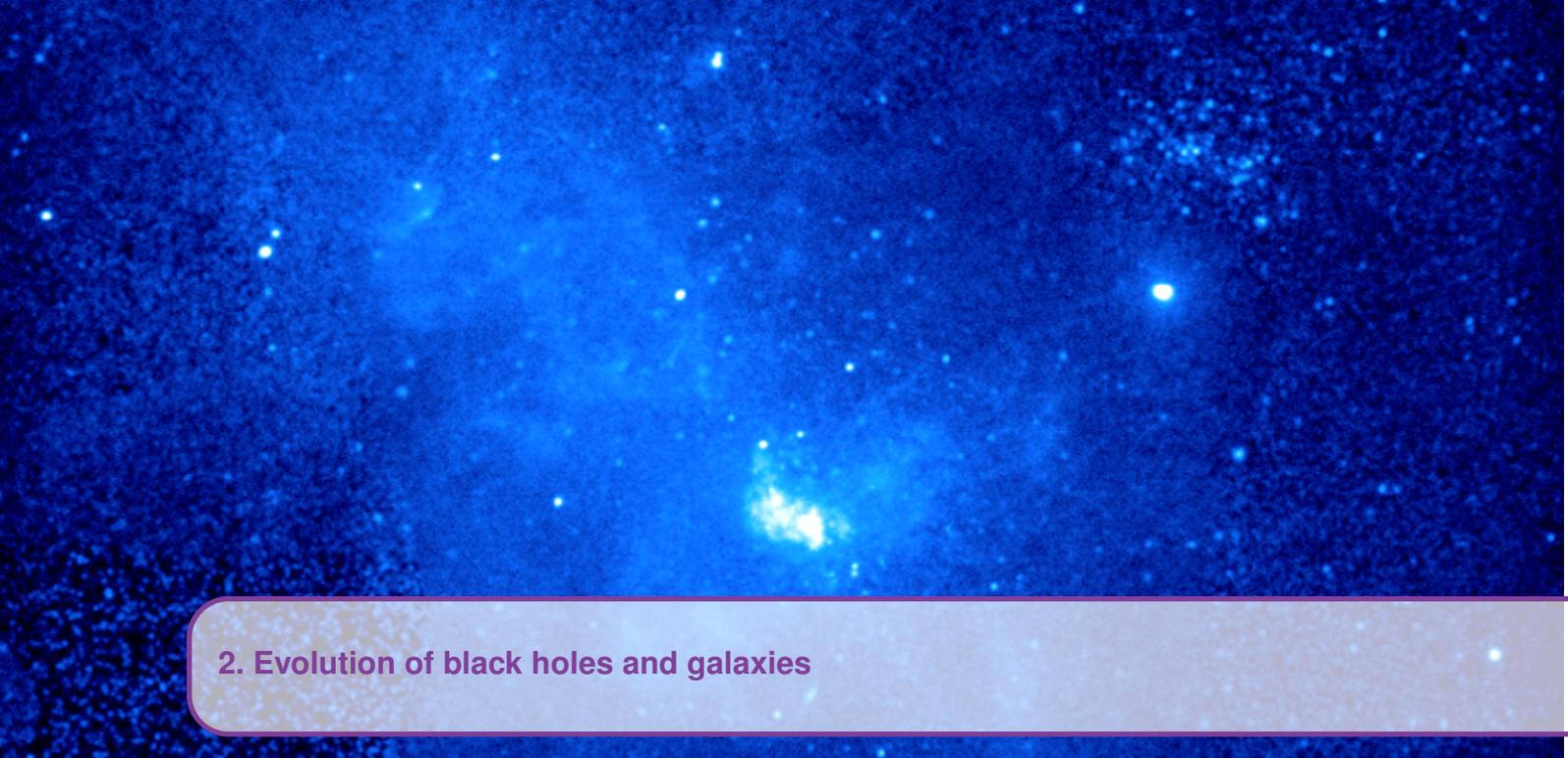

## 2.1 Introduction

The $M_{BH} - \sigma$ relation seems to point to a synergistic relationship between the growth of galaxies and their central super-massive black holes (SMBHs) (e.g., Batiste et al. 2017). Indeed, integrated over many galaxies and AGN, there is a broad agreement in the redshift evolution of the star-formation-rate density and the black-hole accretion rate density (Madau and Dickinson 2014). They both show a rapid increase from high redshifts, reach a peak at $z \sim 2$, before declining by an order of magnitude to their current values. However, determining the exact physics of the star formation/accretion connection is exceedingly difficult beyond the local Universe. In nearby AGN, there does appear to be a relationship between nuclear star formation and accretion activity, but it only appears when examining star formation within a few hundred pc of the nucleus. Investigating nuclear processes at the peak of the star-forming and accretion activity at $z \sim 1$–2 will require a new generation of instruments with exquisite sensitivity and resolution.

A further complication in studying the physics on scales of $\sim 100$ pc around AGN is that the majority are obscured behind a significant column of gas and dust. Therefore, UV and optical tracers of star formation in the nuclear environments are heavily extinguished, and IR signatures may be confused with emission from dust heated directly by the AGN. In contrast, radiation at both X-ray and radio wavelengths is relatively immune to both obscuring gas and confusion from other sources, and so studying AGN with radio and X-ray telescopes provides a powerful view of the AGN/star formation relationship. As shown below, deep AGN surveys by *Athena* and SKA will be able to trace the star-formation history of multiple types of AGN, solving many of the outstanding questions about the influence of black-hole growth in the build-up of galaxies.

### 2.1.1 X-ray background and the associated radio sources: key to AGN evolution

Surveys making use of the large collecting area of *Athena* are expected to yield about 180,000 obscured AGN at $z = 1$–4 (Georgakakis et al. 2013), including Compton-thick objects required for a complete census of accretion activity. SKA imaging of these AGN survey fields may be able to achieve detection of Star Formation Rates (SFRs) as low as $\sim 10\ M_\odot$ yr$^{-1}$, depending on the depth of the radio observations (Murphy et al. 2015). With such a large dataset to draw from, computing the radio number counts of AGN will provide significant constraints on the evolution of star formation in AGN host galaxies. Such an experiment was first proposed by Ballantyne 2009, who showed that using empirical relationships between AGN radio and X-ray luminosities, X-ray background models can be used to predict the number counts of AGN at radio frequencies down to sub-$\mu$Jy flux levels. Ballantyne 2009 used this procedure to show that additional radio flux due to star formation within AGN host galaxies may be necessary in order to match the observed AGN radio counts at a flux density of $\sim 50\ \mu$Jy. Indeed, the $\mu$Jy AGN population will provide a very clean sample to trace the accretion and galactic star formation histories of Seyfert galaxies over a significant fraction of cosmic time.

These predictions have been updated here using a recent X-ray background model (calibrated to fit the latest *NuSTAR* data; Harrison et al. 2016) that employs the Ueda et al. 2014 hard X-ray luminosity function, the Burlon et al. 2011 $N_H$ distribution, and the Ballantyne 2014 relationship between the obscured AGN fraction and the X-ray luminosity. In predicting the AGN radio number counts, the radio flux from the AGN core must be included. This calculation makes use of the Panessa et al. 2015 relationship between the core 1.4 GHz luminosity and the 2–10 keV luminosity and assumes an $\alpha = 0.2$ spectrum for the compact cores (Massardi et al. 2011). If an SFR is added to the AGN population then it produces a radio luminosity given by the Murphy et al. 2011 relationship (assuming $\alpha = 0.7$). Finally, the most recent radio AGN number counts from Padovani et al. 2015 and Smolcic et al. 2017 are included to constrain the model.

The design-driving science cases of both SKA and *Athena* placed a strong emphasis on comprehensive inventories of sources and structures in the cosmos by means of large surveys, to which a substantial fraction of the observing time will be dedicated at both facilities.

The advances in survey capabilities provided by SKA and *Athena* (encoded in metrics such as survey speed, areal coverage at given depth (or grasp), signal-to-noise × area, etc.; see figure 2.1) should deliver orders of magnitude improvements compared to currently operating survey instruments at radio and X-ray wavelengths, respectively. Thus, a large discovery space will be opened up by those surveys, and in this chapter we aim to present a few topical areas where major progress is anticipated.

By their very nature, surveys are usually designed to serve multiple scientific purposes, and so, unsurprisingly, the science enabled by the SKA and *Athena* surveys spans a very large range of astrophysical topics, from the very first stars and galaxies (sections 1.1 and 1.2), to the overall evolution of accreting black holes and star-forming galaxies (sections 2 and 2.2) to the detailed properties of the star-forming gas (2.3) and inter-stellar medium (2.4) in galaxies with or without nuclear SMBH.

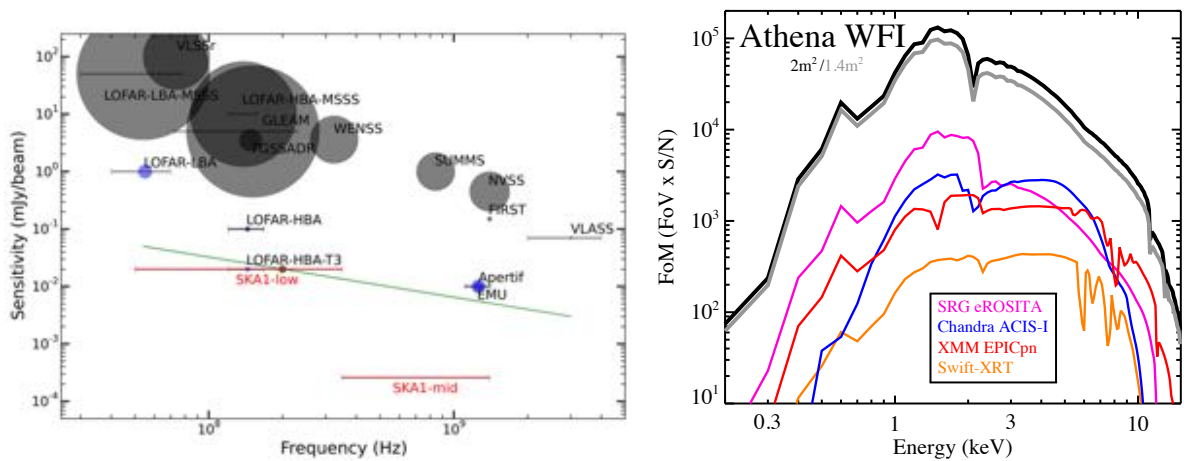

Figure 2.1: Figures of merit for surveys. *Left:* Past, current and future radio surveys sensitivity vs. frequency. The size of the symbols scales with the beam size (spatial resolution); SKA-1 surveys are marked in red (adapted from Shimwell et al. 2017). *Right:* Signal-to-noise (S/N) integrated over the field of view of current or future X-ray telescopes vs. energy. S/N is here proportional to the effective area divided by the spatial resolution (adapted from Wik et al., SPIE Proc. Vol. 10709 (2018), in press).

The left panel of Figure 2.2 shows the predicted AGN radio number counts at 1.4 GHz when no star formation flux is added to the AGN emission. The solid line is the total counts from all AGN, including both radio-loud and radio-quiet AGN, and shows that obscured AGN dominate the radio counts at all flux densities. The radio-loud AGN fraction is determined empirically by ensuring the total counts pass through the observed data at flux densities > 1 mJy, and the fainter radio-loud AGN counts measured in the Extended *Chandra* Deep Field-South (CDFS) (green data points; Panessa et al. 2015).

The 1.4 GHz radio luminosity of the radio-loud objects is always $10^{-2} \times$ the X-ray luminosity (Terashima and Wilson 2003). The radio-loud fraction never rises above 15% and the remaining AGN are radio-quiet with a core radio flux computed as described above. The figure clearly shows that the predicted AGN radio counts fall far below the observed data at flux densities < 1 mJy, and this mis-match is caused by the radio-quiet AGN.

The right panel of Figure 2.2 shows the predicted 1.4 GHz AGN counts after including a redshift and luminosity dependent star-formation law for AGN host galaxies: SFR$\propto [(1+z)/(1+((1+z)/2.9)^{5.6})](\log L -$ $40)^{1.75}$. This redshift evolution is similar to, but rises slower than, the overall growth in star-formation-rate density determined by Madau and Dickinson 2014. The plot shows that these predictions provide a good description of the observed AGN radio counts. The calculation included a flux limit of $5 \times 10^{-17}$ erg cm$^{-2}$ s$^{-1}$, easily within the grasp of *Athena* (Georgakakis et al. 2013); thus, these predictions should describe the expected SKA counts from targeting *Athena* survey fields.

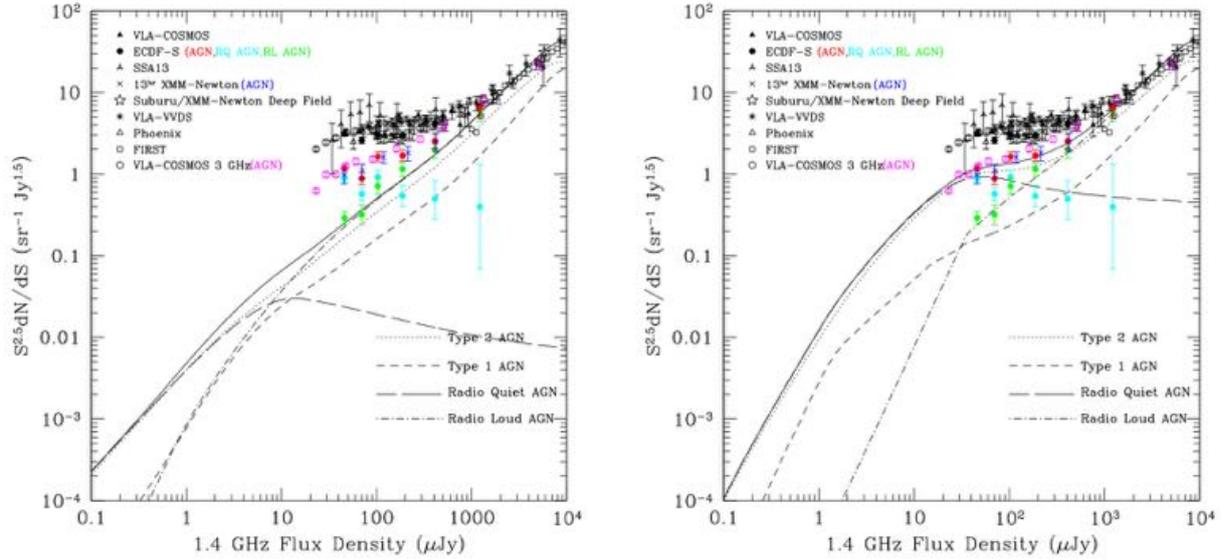

Figure 2.2: *Left:* The solid line plots the predicted Euclidean-normalized 1.4 GHz radio counts from 0.1 µJy to 10 mJy obtained from the X-ray background modeling described in the text. No contribution from star formation in the AGN host galaxy is included in the predictions. The dotted and short-dashed lines plot the contribution to the counts from obscured and unobscured AGN, respectively. These curves include AGN of all radio powers. Type 2 AGN dominate both the radio-loud and radio-quiet populations at these flux levels. The contribution from radio-quiet AGN is plotted as the long-dashed line, while radio-loud AGN are shown as the dot-dashed line. These curves include both obscured and unobscured AGN. The points indicate the observed counts obtained from various surveys: Very Large Array (VLA)-COSMOS (Bondi et al. 2008), CDFS (Kellermann et al. 2008), SSA13 (Fomalont et al. 2006), 13hr *XMM-Newton* (Seymour, McHardy, and Gunn 2004), Subaru/*XMM-Newton* Deep Field (Simpson et al. 2006), VLA-VVDS (Bondi et al. 2003), Phoenix (Hopkins et al. 2003) and FIRST (White et al. 1997). The blue and red data plot the estimated AGN contribution to the radio counts from the 13hr *XMM-Newton* (Seymour et al. 2008) and Extended CDFS (Padovani et al. 2015) surveys, respectively. The cyan/green data points are estimates of the radio-quiet/loud AGN contribution in the Extended CDFS (Padovani et al. 2015). The magenta data points are the AGN radio counts recently determined by Smolcic et al. 2017 at 3 GHz and then converted to 1.4 GHz. *Right:* As in the left panel, but now including radio-emission from star formation in every AGN host galaxy. The SFR varies with both redshift and AGN luminosity: SFR$\propto [(1+z)/(1+((1+z)/2.9)^{5.6})](\log L - 40)^{1.75}$. The model now takes into account the expected *Athena* flux limit of $5 \times 10^{-17}$ erg cm$^{-2}$ s$^{-1}$ (Georgakakis et al. 2013).

The sensitivity of SKA observations, combined with the large yield of AGN from *Athena* surveys, will allow the star formation history of AGN host galaxies to be precisely traced. For example, Figure 2.3 shows the predicted AGN radio counts (from the model shown in Figure 2.2, right panel) broken down into 3 redshift bins. The model predicts that star formation from radio-quiet obscured Seyfert galaxies at $1 \leq z \leq 3$ will dominate the AGN radio counts for flux densities $< 10$ µJy. Constructing AGN radio counts for different redshift and luminosity bins, and then comparing the results to models such as this one, will provide measurements of the star formation rate history of AGN host galaxies. Similar experiments could also be performed for obscured and unobscured AGN to determine how the star formation rate may or may not be related to the nuclear obscuration in AGN at $z \sim 1$.

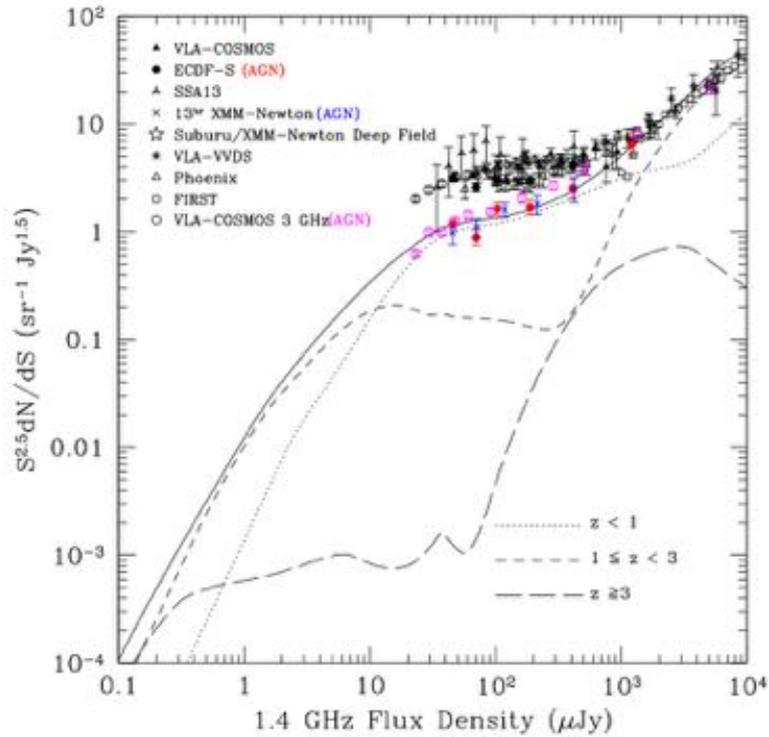

Figure 2.3: As in the right panel of Figure 2.2, but the contributions to the total AGN number counts from sources at difference redshifts are now indicated.

## 2.2 *Athena* and the faint radio sky

Radio astronomy has changed. For years its focus was relatively rare sources, which emit mostly non-thermal radiation across the entire electromagnetic spectrum, i.e., radio quasars and radio galaxies. Radio surveys have now reached such faint flux densities that the dominant populations are the star-forming galaxies (SFGs) and the more common non-jetted (radio-quiet) AGN (see Padovani 2016, for a review). These sources make up the bulk of the extragalactic sky, which has been studied for decades in the infrared (IR), optical, and X-ray bands. This change in the dominant radio-source population will be more extreme with the SKA: for example, for $S_{1.4GHz} \geq 1$ $\mu$Jy, SFGs and non-jetted AGN are expected to make up $\approx 80\%$ and $10\%$ of the sources, respectively (Padovani 2016, and references therein).

It took more than thirty years to figure out the source population of the $\lesssim 1$ mJy radio sky, partly because the classification of radio sources at these flux density limits is complex, but most crucially because the relevant, and necessary, data at other wavelengths were not available. Only by combining radio, IR, optical, and X-ray data can one accurately classify faint radio sources. The availability of deep X-ray data is therefore a vital prerequisite for tackling various hot issues that can be addressed in the radio band, such as the co-evolution of SMBHs and their host galaxies, the demographics of jetted and non-jetted AGN and their relationship, and the cosmic star formation history of the Universe.

For the $\lesssim 1$ $\mu$Jy radio sky, Padovani 2011 has derived order of magnitude estimates of the X-ray, optical/NIR, and mid-IR (MIR)$-$far-IR (FIR) fluxes of sub-$\mu$Jy radio sources, and then evaluated the current and, particularly, future availability of the relevant multi-wavelength data. Fig. 2.4 (an updated version of Padovani 2016, Fig. 2) shows that in the X-ray band, sources with $S_{1.4GHz} \sim 1$ $\mu$Jy should have $f_{0.5-2keV} \approx 10^{-17}, \approx 10^{-18}$, and well below $10^{-18}$ erg cm$^{-2}$ s$^{-1}$ for non-jetted AGN, SFGs, and jetted AGN respectively. These values are beyond the reach of eROSITA, which will provide large area X-ray surveys and detect millions of (X-ray brighter) AGN. The deepest X-ray survey currently available is the 7 Ms CDFS, which reaches $f_{0.5-2keV} \sim 6.6 \times 10^{-17}$ erg cm$^{-2}$ s$^{-1}$ over $\sim 0.07$ deg$^2$ (50% completeness level; Luo et al. 2017). *Athena* will reach $f_{0.5-2keV} \sim 2 \times 10^{-17}$ erg cm$^{-2}$ s$^{-1}$ in 1 Ms[1], and given its survey speed will cover larger areas much more efficiently than *Chandra*. Fig. 2.4 shows that *Athena* should be able to detect the bulk of non-jetted AGN down to $S_{1.4GHz} \approx 1$ $\mu$Jy, and provide upper limits for SFGs and jetted AGN. Since only AGN can have hard X-ray powers $(2 - 10$ keV$)$ $L_x \gtrsim 10^{42}$ erg s$^{-1}$, *Athena* will differentiate AGN from SFGs. Based on Fig. 16 of Luo et al. 2017 many AGN will be below this value and therefore will need to be identified by using ancillary X-ray information (e.g., intrinsic X-ray absorption, X-ray variability, the presence of a K-shell Fe line at 6.4 keV) and data at other wavelengths. Below this radio flux density, very few radio sources will have an X-ray counterpart in the foreseeable future.

---
[1]At these levels *Athena* is not only background limited, but also confusion limited, and integrating further will not improve its sensitivity (Aird et al. 2013).

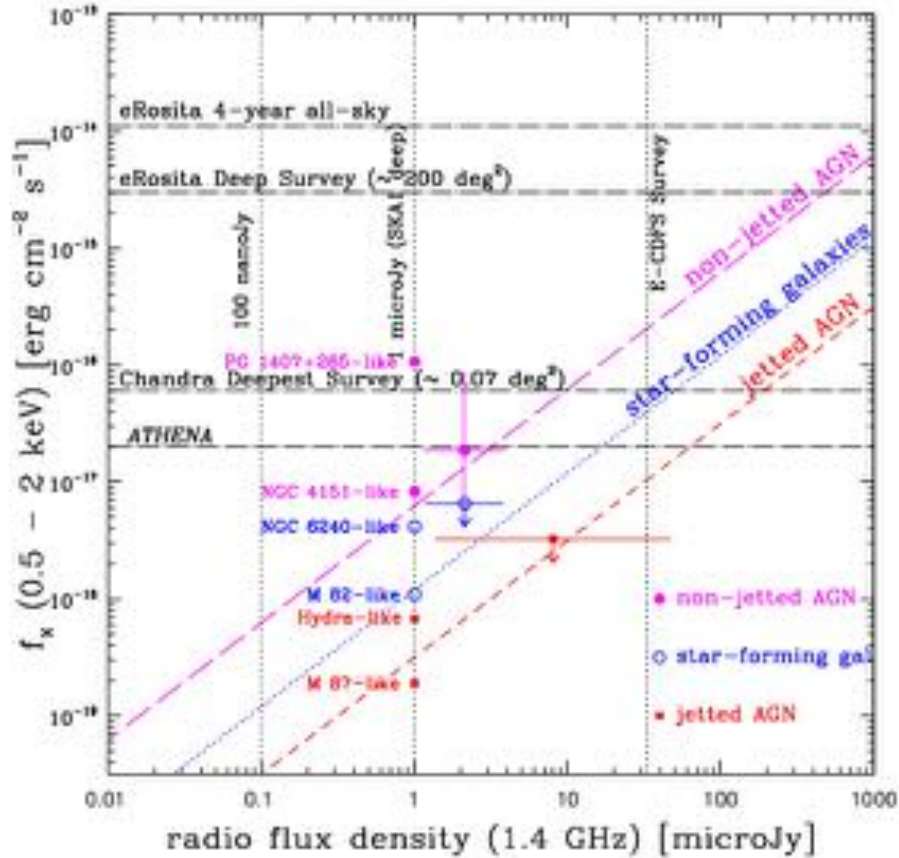

Figure 2.4: $0.5 - 2$ keV X-ray flux vs. the 1.4 GHz radio flux density for faint radio sources. The loci of radio-selected, non-jetted AGN (magenta long-dashed line), SFGs (blue dotted line), and jetted AGN (mostly of the low-excitation type: red short-dashed line) are indicated. The scaled X-ray fluxes of prototypical representatives of the three classes at $S_{1.4GHz} = 1$ $\mu$Jy are also shown. Finally, the mean radio and X-ray flux values, or upper limits, for sources belonging to a hypothetical sample characterised by $S_{1.4GHz} \geq 1$ $\mu$Jy, as extrapolated from the Extended CDFS sample, with error bars indicating the standard deviation, are also marked. The horizontal dot-dashed lines indicate the limits of (from top to bottom): the eROSITA all-sky and deep surveys, *Chandra*'s deepest survey (Luo et al. 2017; this limit refers to the 50% completeness level: the faintest flux limit is about one order of magnitude smaller over 1 arcmin²) and *Athena*. Updated from Fig. 2 of Padovani 2011.

Proper exploitation of the $\lesssim 1$ $\mu$Jy radio sky will *require* synergy with many other contemporaneous astronomical facilities. *SPICA* should easily detect the bulk of the $\gtrsim 1$ $\mu$Jy population in the MIR (and perhaps FIR) band, while deep *SPICA* exposures should be able to detect many radio sources at $S_{1.4GHz} \lesssim 0.1$ $\mu$Jy (Padovani 2016). However, while the far-IR–radio correlation is useful to identify jetted AGN, it cannot be used to identify the non-jetted ones, which fall on the same locus as SFGs.

Many optical/NIR facilities (including the Wide Field Infrared Survey Telescope (*WFIRST*), the Large Synoptic Survey Telescope (LSST), the next generation giant telescopes, and [possibly] the James Webb Space Telescope (*JWST*)) will also be available by the time *Athena* flies. For example, the bulk of the $\gtrsim 1$ $\mu$Jy radio sky should be detected in the optical band by the LSST and will be well within the sensitivity of *WFIRST* (Padovani 2011; Padovani 2016). Sources having $R_{mag} > 27.5$ will be within reach of the *JWST* and the next generation giant telescopes[2], with diameters between 25 and 39 m and "first light" expected in the mid-2020s, which however will be covering relatively small fields of view (up to $\approx 0.2$ deg$^2$ for the smallest one). *JWST* and the ELTs may be the main facilities to secure redshifts of $\mu$Jy radio sources but even they could have problems in the nJy regime. We stress, however, that the optical band does not give the full story, being strongly affected by absorption and/or dilution by the host galaxy. For example, there are many cases of sources that show no sign of nuclear activity in their optical spectra but are strong ($L_x \gtrsim 10^{43}$ erg s$^{-1}$) X-ray sources.

The SKA will provide radio morphology information, but this will not necessarily allow the separation of SFGs and non-jetted AGN if radio emission is dominated by star-formation processes in both populations (Padovani 2016). The full SKA will detect AGN cores, and therefore will provide direct evidence of AGN activity, by combining high sensitivity with milliarcsecond resolution (McAlpine et al. 2015).

In summary, radio astronomy is now a fully multi-wavelength enterprise. The proper exploitation of SKA surveys will require very strong synergy with other contemporaneous astronomical facilities. These will include, among others, *WFIRST*, the LSST, the next generation giant telescopes, *JWST*, *SPICA*. *Athena*, in particular, will play a very important role in differentiating AGN from SFGs in the SKA era.

## 2.3 A multi-wavelength view of star formation in galaxies

### 2.3.1 Introduction

Galaxies are known to evolve through a mixture of internal processes (e.g. conversion of gas into stars, chemical enrichment) and external processes (e.g. gas accretion, galaxy encounters), whose relative importance (the so-called "nature versus nurture" debate) is still one of the major open questions in the field of extragalactic astrophysics. Much effort is also devoted to investigating whether global properties (e.g. mass, morphology, environment) or local properties (e.g. density, chemical composition) are the main drivers of galaxy formation and evolution.

From an observational point of view, optical spectroscopy of large galaxy samples – for example SDSS (Abazajian et al. 2009) and GAMA (Driver et al. 2011) – has provided robust evidence that many, if not most, of their physical and/or observable properties are strongly correlated (e.g. Disney et al. 2008; Ascasibar and Sánchez Almeida 2011). The advent of large integral-field spectroscopic surveys, such as CALIFA (Sánchez et al. 2012; Sánchez et al. 2016), MaNGA (Bundy et al. 2015), and SAMI (Croom et al. 2012; Bryant et al. 2015) has revealed that analogous "scaling relations" hold on resolved (of the order of $\sim$kpc) scales (e.g. Rosales-Ortega et al. 2012; Cano-Díaz et al. 2016; González Delgado et al. 2016), and the combination of optical data with radio observations of the 21-cm line in atomic hydrogen, such as THINGS (Walter et al. 2008), VLA-ANGST (Ott et al. 2012), LVHIS (Koribalski 2008), as well as other emission lines from molecular transitions in CO (e.g. HERACLES Leroy et al. 2009) has opened up the possibility of investigating the connection between stellar and gas surface density, chemical composition, and star formation efficiency on similarly local scales (e.g. Bigiel et al. 2008; Bigiel et al. 2011; Leroy et al. 2008; Leroy et al. 2013; Lara-López et al. 2013; Ascasibar et al. 2015; Kudritzki et al. 2015; Schinnerer et al. 2016).

Among these correlations, star-forming galaxies (the so-called "blue cloud") are neatly segregated from passively-evolving systems (the "red sequence") in a colour-magnitude diagram (e.g. Tully, Mould, and Aaronson 1982; Strateva et al. 2001; Baldry et al. 2004; Baldry et al. 2006). Objects in the intermediate region, known as the "green valley", are typically considered to be a transition population (e.g. Bell et al. 2004; Faber et al. 2007; Martin et al. 2007; Schiminovich et al. 2007; Wyder et al. 2007; Mendez et al. 2011; Gonçalves et al. 2012), and their colours are often interpreted as evidence for a putative "quenching" process that rapidly interrupts any star formation activity (see e.g. Peng et al. 2010). On resolved scales, the situation is less straightforward, and a detailed analysis of the mean stellar age (Zibetti et al. 2017) shows a clear correlation between local surface density, as well as with the internal structure of the galaxies (e.g. bulge, arm, and inter-arm regions).

It has long been proposed (e.g. Dekel and Silk 1986) that supernova explosions are powerful enough to drive large-scale outflows that expel a significant fraction of the gas reservoir, at least in relatively low-mass systems, and thus regulate the stellar-to-gas mass fraction and the instantaneous star formation rate. Models of galaxy evolution often invoke an as-yet-undetermined physical mechanism to quench all star formation activity within a very short time scale. The signatures of AGN in the optical spectra of many of the galaxies in the "green valley" (e.g. Salim et al. 2007; Schawinski et al. 2007; Schawinski et al. 2010), the correlation between the estimated properties of the central SMBH and those of the host galaxy (e.g. Magorrian et al. 1998; Ferrarese and Merritt 2000; Gebhardt et al. 2000), and the observation of high-speed winds in the immediate vicinity of AGN (e.g. Ganguly and Brotherton 2008; Tombesi et al. 2010) have led to the hypothesis that AGN-driven outflows may blow away a significant fraction of the gas reservoir in the host galaxy and hence simultaneously prevent further star formation and black-hole growth (e.g. Silk and Rees 1998; Di Matteo, Springel, and Hernquist 2005; Menci et al. 2006; Fabian 2012; Zubovas and King 2012). Cosmological simulations lend support to this scenario, and there is widespread agreement that feedback from both supernovae and AGN play a key role in regulating the conversion of gas into stars, although the details are strongly dependent on the adopted phenomenological prescriptions for star formation and feedback, as well as on their precise numerical implementation (e.g. Scannapieco et al. 2012; Hopkins et al. 2014; Crain et al. 2015; Guidi, Scannapieco, and Walcher 2015; Peters et al. 2017).

---

[2]These include, in order of decreasing diameter size, the Extremely Large Telescope (ELT); the Thirty Meter Telescope (TMT); and the Giant Magellan Telescope (GMT).

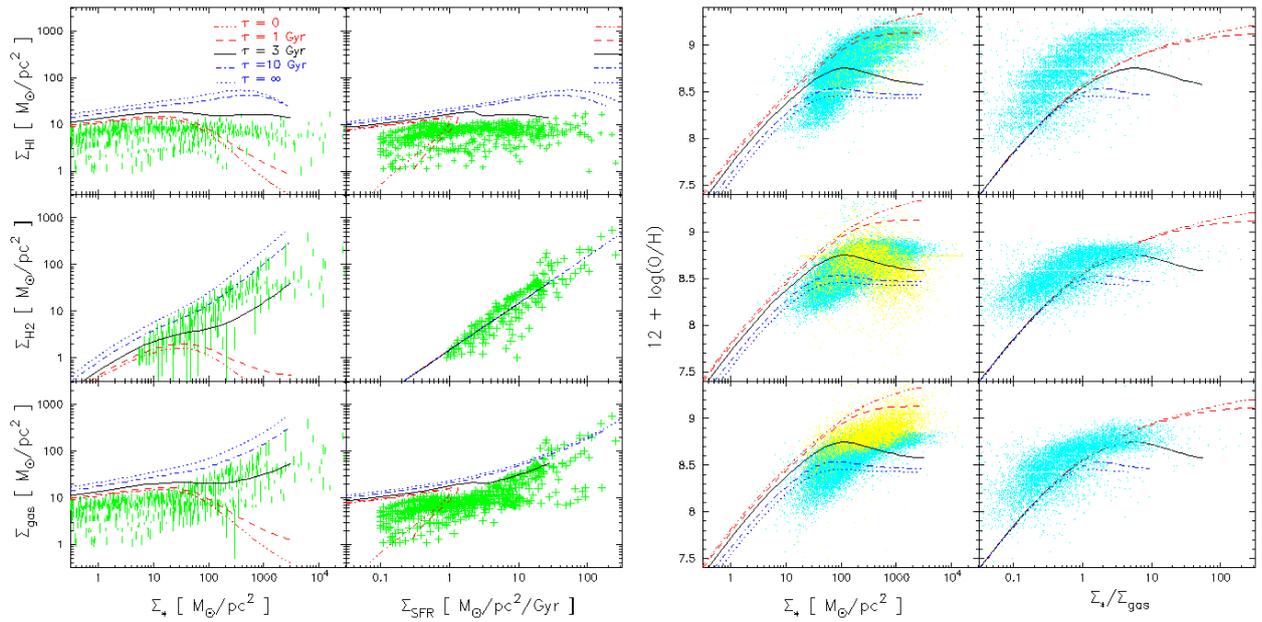

Figure 2.5: Local scaling relations between the surface densities of stars ($\Sigma_*$) and star formation rate ($\Sigma_{SFR}$), and (*left*) those of atomic ($\Sigma_{HI}$), molecular ($\Sigma_{H2}$), and total gas ($\Sigma_{gas}$) and (*right*) gas-phase oxygen abundance. Points represent different observations, and lines illustrate the predictions of chemical evolution models (Ascasibar et al., in preparation).

Nevertheless, this mainstream view is not beyond question. It has been argued that many of the observed properties of dwarf galaxies (Tassis, Kravtsov, and Gnedin 2008; Gavilán et al. 2013), the Milky Way (Mollá et al. 2017), and nearby spiral galaxies (Figure 2.5) can be understood in terms of the efficiency of the conversion of gas into stars, regulated not by supernova explosions driving significant galactic outflows, but mainly by the availability of atomic hydrogen and the formation and destruction of $H_2$ molecules. Regarding AGN feedback, different analyses of optical spectra (e.g. Ascasibar and Sánchez Almeida 2011; Casado et al. 2015) cast serious doubts on the idea that there is a sharp transition between "star-forming" and "passive" galaxies, suggesting instead that galaxies are distributed along a gradual "ageing" sequence. Once again, spatially resolved observations are indicating a more complex picture, but with no strong evidence for sudden "quenching" (or any other AGN feedback) affecting the star formation activity of the whole galaxy (Figure 2.6).

### 2.3.2 The key role of *Athena* and SKA

The availability of multi-wavelength (spatially-resolved spectroscopic) data for a statistically representative sample of galaxies is of the utmost importance in order to identify the main mechanisms that govern galaxy formation and chemical evolution. More specifically, the combination of radio, optical, and X-ray observations will enable the effects on the host galaxy of molecule formation and photodissociation, supernova explosions, and AGN-driven outflows to be determined, disentangling the contribution of internal and external processes on local and global scales, and discriminating between different theoretical scenarios.

Optical spectra provide information on the stellar population and the warm ($T \sim 10^4$ K) ionised gas, such as the total mass, local surface densities, resolved star formation history, detailed chemical composition, and kinematics. The radio continuum traces the presence of supernovae and warm ionised gas through their synchrotron and bremsstrahlung emission, respectively, whereas radio emission lines probe the cold ($T \sim 100$ K) gas phases. Supernovae and AGN can be identified as X-ray point sources, while extended emission traces the hot ($T > 10^6$ K) gas responsible for driving large-scale outflows. Combining measurements over such a broad range of the electromagnetic spectrum is the best (if not the only) way to overcome model degeneracies, constraining the physical properties of all gas phases and the total energy budget associated with different feedback processes.

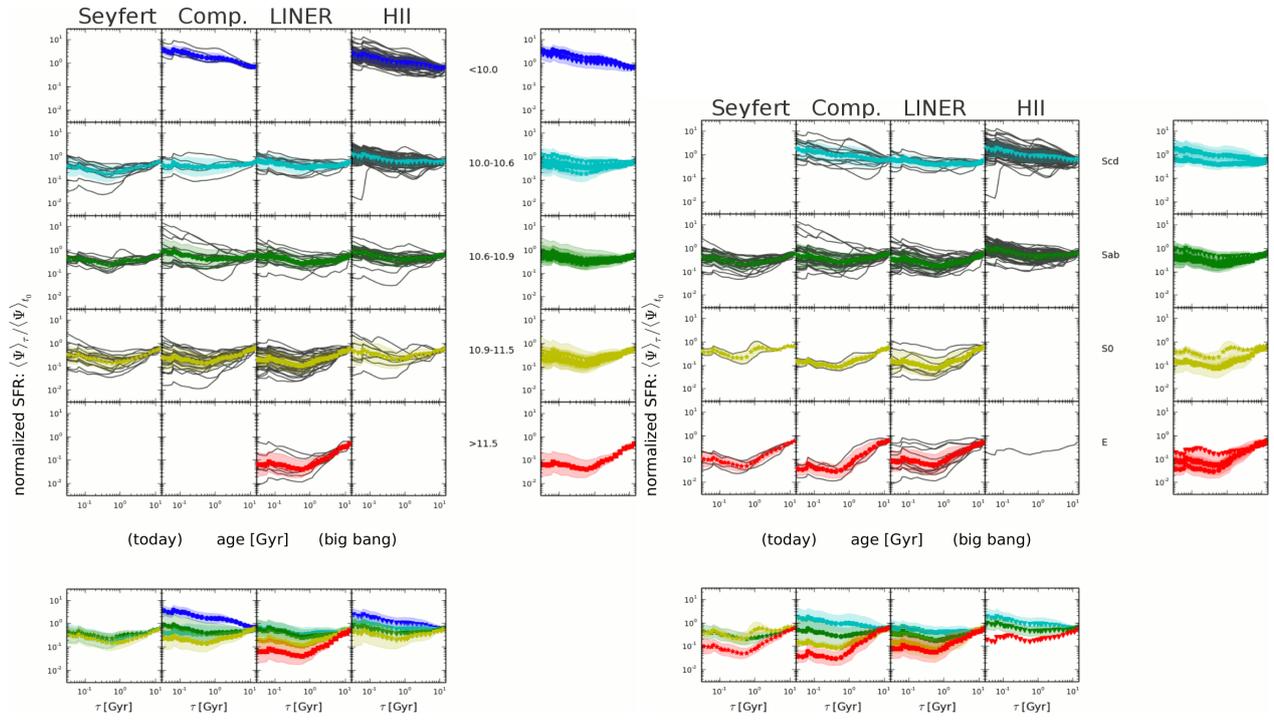

Figure 2.6: Star formation histories of CALIFA galaxies, classified according to their optical emission line ratios into different "activity" classes (columns) in different mass (rows on the left panel) and morphology (rows on the right panel) bins (Ascasibar et al., in preparation).

Despite the large quantity of observations that has become publicly available in recent years, assembling all the necessary data is far from a trivial task, which is severely hampered by the different observational biases that are characteristic of each wavelength regime (e.g. sensitivities, spatial and spectral resolutions, field of view). While optical integral-field spectroscopic surveys typically reach (sub-)arcsecond resolution over a $\sim 1$ arcmin field of view, the highest resolutions attained by current interferometric 21-cm observations of nearby galaxies (e.g. THINGS; Walter et al. 2008) are of the order of $6''$ or larger, over much more extended regions in the sky. As a result, surveys at different wavelengths usually target systems with different angular diameters, and therefore there is a surprisingly small number of galaxies that have multi-wavelength spatially-resolved spectroscopic data. The spatial resolution of current X-ray instrumentation is somewhat in between those of current optical IFU surveys and 21-cm radio surveys, but the limited spectral resolution available for all but the brightest sources makes it difficult to carry out detailed analyses of the chemical composition and kinematics of the hot gas.

This situation will change in the near future, thanks to the advent of SKA and, to a lesser extent, its precursors. In the *Athena* era, SKA1-MID will observe (during Phase 1) a substantial number of galaxies to a column density limit of the order of $10^{20}$ cm$^{-2}$ resolved on $\sim 3''$ scales (increasing depth and/or spatial resolution further will only be feasible with the full SKA). Complementary to 21-cm emission, which maps the distribution of atomic hydrogen throughout the galaxy, absorption studies towards radio-loud AGN will also offer detailed information on the distribution and kinematics of the gas around the central AGN. The addition of high-resolution X-ray spectroscopic data will for the first time make it possible to compile all the necessary information about the different phases of the gas duty cycle, connecting them with the main properties of the host galaxies.

## 2.4  The gas content of distant active galaxies with SKA and *Athena*

### 2.4.1  Introduction

The highly penetrating capabilities of both radio and X-ray radiation allow us to probe the interstellar medium of galaxies across a wide range of density and temperature. Recent studies of bright radio galaxies have revealed the presence of dense and dusty gas near their active nuclei that can be traced both by 21-cm

HɪI absorption and soft X-ray absorption (Allison et al. 2015; Glowacki et al. 2017; Moss et al. 2017). A key ambiguity in these studies is whether the absorption revealed in HɪI (along a censored sight-line to the background radio source) is produced by the same dense gas as seen in the absorption of soft X-rays, and additionally whether both of these absorbers are measuring gas near the torus, in the nearby circum-nuclear medium, or on larger galactic scales. A particular focus of previous work has been on the Gigahertz-Peaked Spectrum galaxy population, which is thought to represent AGN at their youngest, and so to offer insights into the early stages of radio-galaxy evolution (e.g. Vink et al. 2006; Ostorero et al. 2010). These studies have found evidence for a correlation between $N_{HI}$ measured from HɪI absorption and $N_H$ measured from soft X-ray absorption; however, $N_{HI}$ is typically an order of magnitude smaller than $N_H$. Since $N_H$ measures the total hydrogen column (including neutral, ionised and molecular hydrogen), we expect $N_H$ to be greater than $N_{HI}$ in general. However, it remains an open question whether this difference is being driven by the effects of ionisation fraction or elemental abundance assumptions on the X-ray side, or spin temperature and covering fraction assumptions on the radio side. It is thus important to understand the relationship between HɪI and X-ray absorption in more detail, to better understand the role that neutral gas plays in AGN evolution and its relationship to the high-energy emission traced by X-rays.

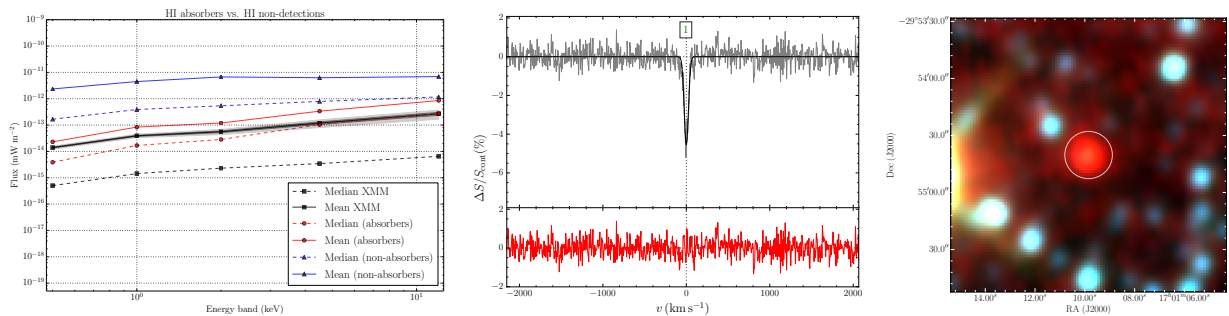

Figure 2.7: *Left*: Comparison of the average X-ray properties of HɪI absorbers (red circles) and non-absorbers (blue triangles) showing increased steepness for absorbers indicative of correlated soft X-ray absorption. *Middle*: detection of HɪI absorption in the BETA pilot sample towards PKS 1657−298, with a single Gaussian component identified by the adopted Bayesian approach. *Right*: Wide-Field Infrared Survey Explorer (Wright et al. 2010) three-colour image (W1/W2/W3) of PKS 1657−298 showing a reddened source that is strongly detected in all bands, and in good agreement with the Very Long Baseline Interferometry (VLBI) position of the galaxy (white circle). All images from Moss et al. 2017

### 2.4.2 Connecting neutral hydrogen and soft X-ray absorption in AGN

Using a catalogue of known HɪI absorbers/non-absorbers, Moss et al. (submitted; Fig 2.7 left panel) have compared HɪI and X-ray properties and identified a statistically-significant correlation between HɪI absorption optical depth (or corresponding upper limit) and the hardness ratio in X-rays indicated by the ratio of $f_5$ to $f_1$ in *XMM-Newton*. In order to test this correlation between HɪI absorption and X-ray absorption, pilot observations have been carried out during commissioning with the Australian Square Kilometre Array Pathfinder (ASKAP) precursor, the Boolardy Engineering Test Array (Hotan et al. 2014; McConnell et al. 2016), towards bright radio galaxies selected on the basis of evident absorption in archival *XMM-Newton* X-ray data. Of the five galaxies observed so far, a new detection of HɪI absorption has been obtained towards the most-absorbed object (Fig 2.7 centre and right panels). These early results emphasise a connection between the HɪI and X-ray absorption, which is being investigated further using data from current and upcoming telescopes including ASKAP, *XMM-Newton*, and eROSITA, in preparation for science with next-generation telescopes.

These pilot observations reflect positively on the future prospects of radio/X-ray complementarity in this area, particularly in the context of synergy between *Athena* and the SKA in the next decade. With SKA1-MID, we will be able to trace HɪI absorption in low-column density systems ($\sim 10^{20}$ cm$^{-2}$) out to $z = 3$ at 2″ resolution. When combined with the expected sensitivity of *Athena* (an order of magnitude more than *XMM-Newton*/*Chandra*) and high-resolution spectral coverage, this opens up a completely new window for spatial investigations of the multiphase distribution of hot and cold gas near the nucleus of the most massive and luminous active galaxies. *Athena* is predicted to identify low-luminosity obscured AGN up to a redshift $z \sim 6$, which is highly complementary to the SKA1-MID range and in fact, can be supplemented further by the inclusion of SKA1-LOW, which will cover a frequency range of 50-350 MHz ($3 \lesssim z \lesssim 25$). This high-redshift parameter space is particularly unexplored in the context of existing facilities for both HɪI absorption and X-ray absorption studies.

Combining SKA and *Athena*, we open up our window on the Universe not only to the period in which star formation is in decline from $z \sim 2$ to present (e.g. Madau and Dickinson 2014), but also to the period leading up to this decline, potentially allowing us to probe the changes in available gas reservoir that altered the content of the Universe ($\sim$12 billion years of evolution). We will be able to trace this evolution through the observable changes in H<sub>I</sub> absorption fraction, H<sub>I</sub> column density, radio luminosity, X-ray absorption fraction, $N_H$ column density, and X-ray luminosity. Moreover, higher frequency SKA (Band 5) observations will allow us to trace the dense molecular gas from which stars form, thanks to the availability of low-J CO, CS and HCN transitions at $z >$2-3 (depending on the final Band 5 frequency range; see Wagg et al. 2015).

In addition, for the lower redshift Universe, we anticipate being able to detect H<sub>I</sub> emission in the same galaxies, which, together with population studies of H<sub>I</sub> absorbers to determine the likely spin temperature distribution (Allison et al. 2016), will help with conclusively resolving the ambiguity between $N_H$ and $N_{HI}$. The specific combination of sensitivity in both SKA and *Athena* with angular resolution in SKA and spectral resolution in *Athena* is particularly beneficial for detailed studies of distant faint AGN (including complex physical modelling of their X-ray spectra), sources that are currently absent from our samples due to limitations of existing instruments. These transformational next-generation instruments will be key in determining how the properties of obscuration of AGN change from low-luminosity to high-luminosity radio/X-ray galaxies, and whether the multi-wavelength properties of galaxies across orders of magnitude in column density are consistent.

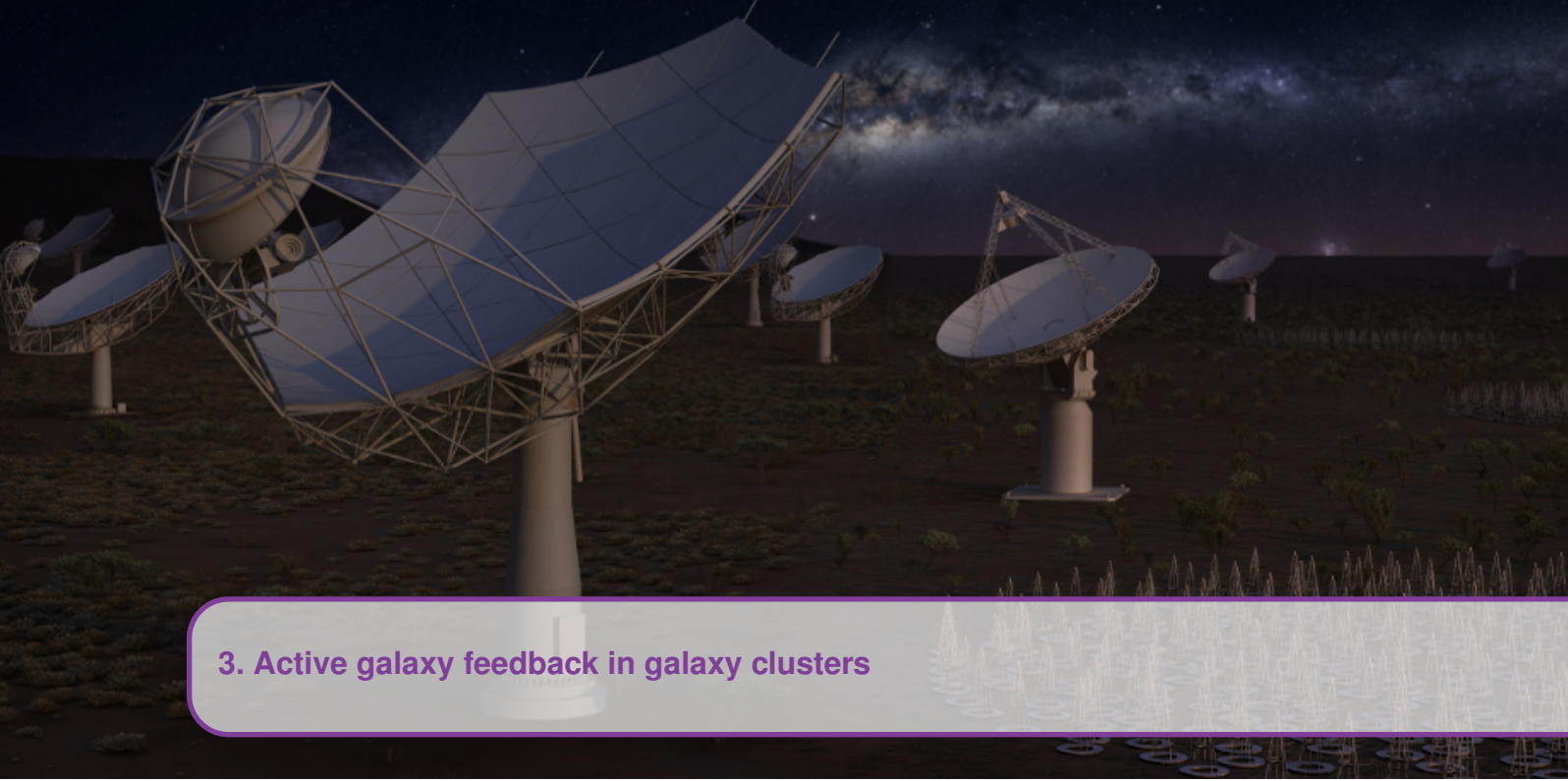

## 3.1   Introduction

It is well known that the baryonic component of galaxy clusters is dominated by a hot and diffuse gas (the intracluster medium, ICM), which permeates the space between cluster galaxies. One of the most important scientific advances enabled by joint studies with the current generation of X-ray and radio facilities has been a step-change in our understanding of how radio-loud AGN interact with the ICM. This has been achieved via spectroscopic measurements of ICM cooling (e.g. Peterson et al. 2003), observations of ICM cavities and ripples (e.g. Fabian et al. 2003) (see e.g. Fig. 3.1), and of shocks surrounding lobes of radio galaxies (e.g. Kraft et al. 2003; Croston et al. 2009). At the same time, increasing evidence suggests that the environment, and in particular the possible interactions with a surrounding ICM, play an important role in shaping the apparent dichotomy in accretion mode (spectral classification) of radio-loud AGN (High and Low Excitation Radio Galaxies, HERGs and LERGs respectively, e.g. Ineson et al. 2015). In addition to tracing the role of AGN jets in the evolution of massive galaxies, *Athena* and the SKA will also allow us to determine how AGN feedback and environmental influence on radio-loud AGN populations evolve to high redshifts.

The interplay between AGN jets and the ICM has wider implications for *Athena*'s objective of characterizing the forming and evolving ICM. Specific *Athena* science objectives in this area are set out by Croston et al. 2013; Pointecouteau et al. 2013 and Ettori et al. 2013, while SKA programs in related areas are discussed by Prandoni and Seymour 2015; Kapinska et al. 2015 and Smolcic et al. 2015. In the following sections, we will focus on specific areas of synergy between *Athena* and the SKA, which include not only the feedback processes of AGN in galaxy clusters, but also wider topics at the intersection of radio-loud AGN and cluster physics, such as the detection and associated physical studies of inverse Compton X-ray emission related to active and/or remnant radio lobes, as well as the discovery of high-redshift groups using radio-loud AGN as pointers to rich environments.

## 3.2   The impact of jets on the thermodynamic evolution of the ICM

Joint radio and X-ray analyses have shown that Fanaroff and Riley Class I radio galaxies are present in about 70% of the central dominant (cD) galaxies of cool-core clusters (see e.g. Gitti et al. 2015, and references therein), with bubbles of relativistic plasma excavating cavities in the ICM distribution (Fig. 3.2). Mechanical energy from radio-loud AGN is expected to be responsible for driving turbulence in the ICM, which dissipates into heat contributing to offset radiative cooling (e.g. McNamara and Nulsen 2012). It is, however, still unknown how the energy generated from the AGN on sub-parsec scales couples so efficiently to gas distributed over tens of kpc, and how this energy is dissipated into the hot ICM. Understanding the role of radio-loud AGN feedback is one of the key requirements for characterizing the non-gravitational physics acting during cluster evolution, which causes cluster thermodynamical properties, e.g. ICM entropy profiles, to deviate from the predictions of a pure gravitational collapse model (see, e.g., left panels of Fig. 3.3).

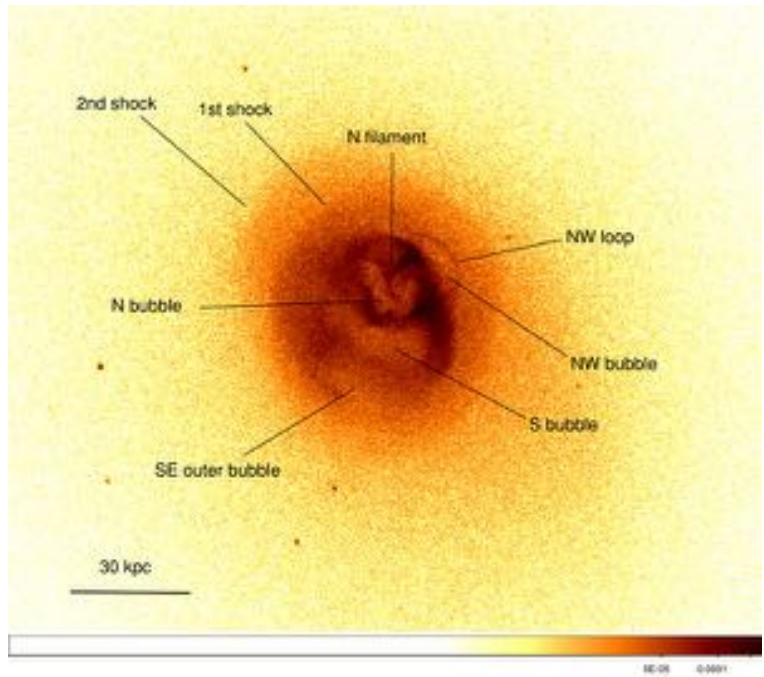

Figure 3.1: Structures related to the interaction between the ICM and the AGN at the center of A2052, as revealed by a *Chandra* image of the galaxy cluster in the 0.3–2.0 keV band. Extracted from Blanton et al. 2011.

An important aim of *Athena* is to determine the thermodynamic and chemical evolution of the ICM from the epoch of cluster formation to the present day (Pointecouteau et al. 2013). The energy input from AGN is a crucial ingredient in this evolution, and the relative contributions of energy deposition from jets and from AGN winds are not well constrained. While *Athena* itself will make advances in quantifying these contributions (Cappi et al. 2013), if we want to determine the cumulative impact of jet energy injection on the ICM this will require input from the SKA. Specifically, this must include the breakdown of high-redshift AGN populations by morphology, source size and spectrum, to enable a robust assessment of energy input, and the mapping of this energy input to evolving dark matter haloes. For a jet of constant power, radio luminosity at a particular observed frequency evolves substantially over the source lifetime (e.g. Hardcastle and Krause 2014), while systematic variations in non-radiating particle content also affect the translation from radio luminosity to jet power and injected energy (e.g. Dunn and Fabian 2004; Croston, Ineson, and Hardcastle 2018).

The SKA1-MID continuum surveys will enable the radio-loud AGN population to be catalogued down to the lowest luminosities at $z > 3$. Fully determined radio luminosity functions compiled from wide sky areas out beyond the peak of star formation and the epoch of cluster formation will be an invaluable resource for evaluating the role of AGN feedback in galaxy evolution. Preliminary feasibility studies investigating the detection and characterization of the radio lobes associated with X-ray cavities carried out by Gitti et al. 2015 indicate that SKA1-MID will allow us to unveil radio-mode feedback at any level in clusters out to the maximum redshift at which virialised clusters have so far been detected in the X-ray band ($z \sim 1.7$), while we may be limited to lower redshifts ($z < 0.6$) for galaxy groups. It is important to stress that complementary lower frequency observations with SKA1-LOW can identify remnants of older AGN outbursts, invisible at GHz frequencies due to their very steep synchrotron spectrum (see, e.g., Fig. 3.2, where the 74 MHz radio emission maps the more external X-ray cavities, most likely containing the oldest population of relativistic electrons, not detected at 1.4 GHz).

Over the next few years, LOw-Frequency ARray (LOFAR) surveys are already expected to advance substantially our understanding of radio-loud AGN life cycles, and comparisons of survey populations with hydrodynamical simulations of jet evolution in realistic environments should enable the development of better radio diagnostics of jet power and energy input that use the combination of observables such as luminosity, morphology, source size and spectrum and polarisation (e.g. Kapinska et al. 2015). To extend these methods to the redshift range of most interest for *Athena*, investigations of ICM evolution will require SKA1-MID continuum surveys to provide sources sizes and morphological information across the luminosity range at $z \gtrsim 3$, so that the cumulative energy input from radio jets into the forming ICM can be assessed. By fully constraining the physics and energetic impact of jets from $z \gtrsim 3$ to the present, the combination of SKA and *Athena* will greatly enhance the separate capabilities of both observatories to transform our understanding of the evolution of galaxies and their environments over the crucial periods of cluster formation and the peak of AGN activity.

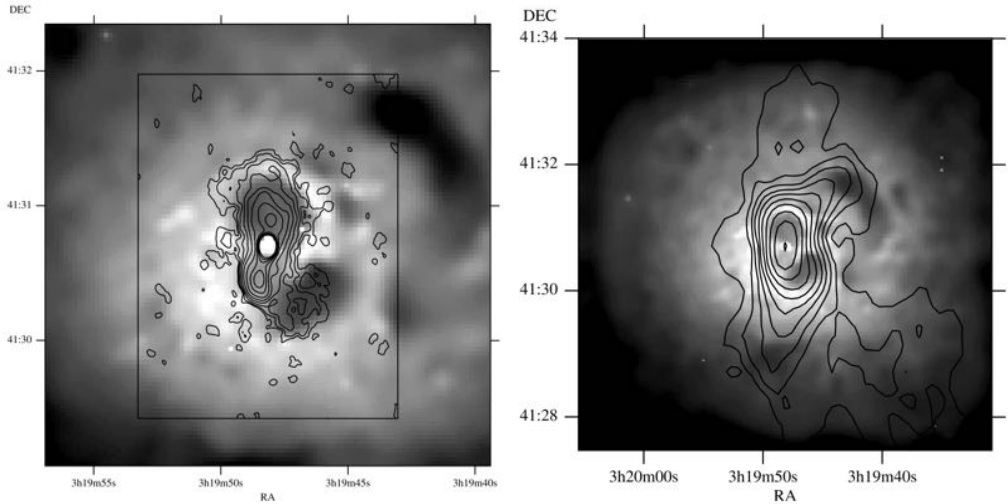

Figure 3.2: 1.4 GHz (*left*) and 74 MHz (*right*) radio contours overlaid on the *Chandra* image of the core of the Perseus cluster. Please notice the different spatial scales of the two images. From Fabian et al. 2002.

To summarize, AGN feedback is a crucial ingredient to understand the non-gravitational heating processes that drive the thermodynamical evolution of the ICM. The combination of X-ray and radio observations provide a powerful tool to trace both thermal and non-thermal components. While *Athena* will be able to characterize the heating and cooling history of the ICM and to measure AGN-outburst-driven outflows directly, at the same time SKA1-LOW will trace the long-term variability of the feedback process over several hundreds of million years, and SKA1-MID will be able to detect outflows and constrain AGN fueling up to $z \gtrsim 3$.

## 3.3 Inverse Compton X-ray emission of radio lobes

### 3.3.1 Tracing the contribution of AGN magnetic fields and relativistic particle injection to the ICM

It is well known (see e.g. Govoni and Feretti 2004; Ferrari et al. 2008) that non-thermal X-ray photons are produced via inverse Compton scattering of the CMB by the same population of relativistic electrons that are responsible for the synchrotron emission of radio galaxies (Fig. 3.4). By jointly modelling the radio spectrum and X-ray emission it is possible to directly measure the electron energy distribution and infer the magnetic field strength within the radio lobes.

*Chandra* and *XMM-Newton* have enabled us for the first time to measure magnetic field strengths reliably for the population of Fanaroff and Riley Class II (FRII) radio galaxies, leading to some of the most robust estimates to date of extragalactic magnetic fields (e.g. Croston et al. 2005; Kataoka and Stawarz 2005; Ineson et al. 2017); however, *Athena* will build on this legacy to allow direct measurements of magnetic field strengths in FRIIs and radio-lobe remnants at $z > 2$, via efficient X-ray inverse-Compton follow up of SKA-selected samples of hundreds of objects spanning a range of radio and environmental properties. Better constraints on the physical conditions within lobes and remnants embedded in groups and clusters at high redshift will provide important input for investigating the evolution of cluster magnetic fields. This method will nicely complement magnetic field measurements through Rotation Measure studies (see Sect. 4.4).

Particle acceleration models for diffuse radio sources in clusters suggest that a seed population of relativistic particles is required to produce the observed extended radio structures (further details in e.g. Brunetti and Jones 2014 and Sect. 4), with radio-loud AGN providing an obvious source. Additionally, the mixing of radio-lobe plasma into the ICM is expected to be important for the evolution of cluster magnetic fields (e.g. Xu et al. 2010). Remnant lobes of previous cycles of nuclear activity (see Sect. 3.3.2) are undoubtedly an important link in this evolution, with sensitive low-frequency radio observations beginning to confirm the important connections between radio galaxies and other extended cluster radio sources (e.g. Bonafede et al. 2014b; van Weeren et al. 2017).

SKA will build on LOFAR studies of AGN life cycles (e.g. Kapinska et al. 2015) and will enable a range of methods for studying cluster magnetism (e.g. Bonafede et al. 2015; Johnston-Hollitt et al. 2015; Govoni et al. 2015, see also Sect. 4.4). *Athena* inverse-Compton studies will allow us direct measurements of magnetic field strength and relativistic particle densities in embedded lobes and bright remnants, providing valuable input in the development of observationally calibrated models for the evolution of cluster magnetic fields.

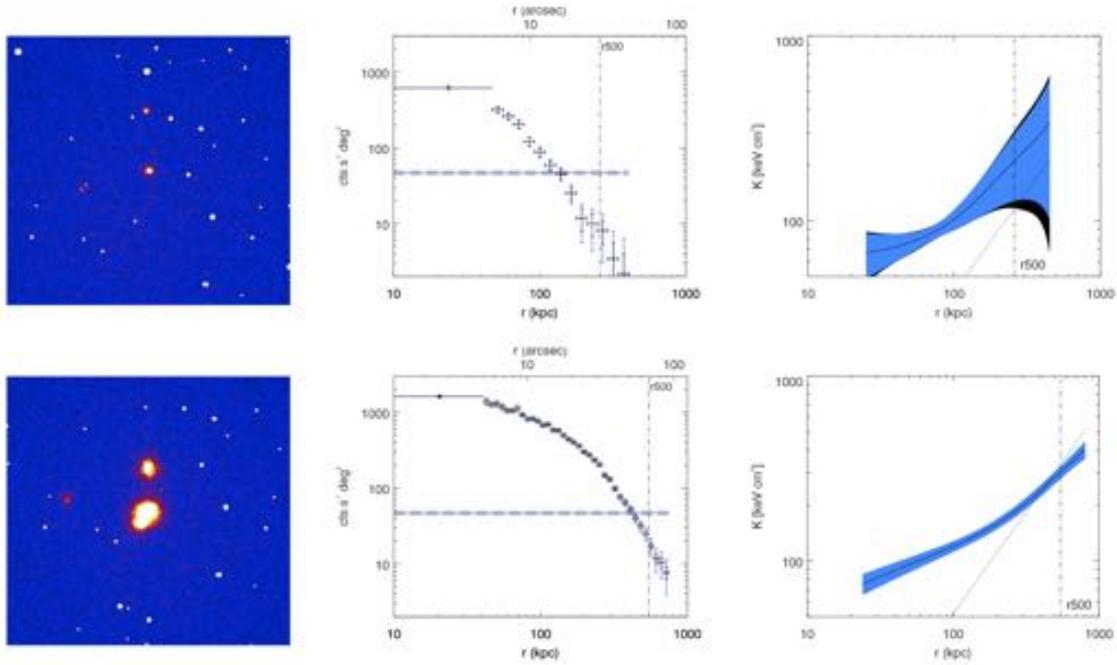

Figure 3.3: *Left column*: 100 ks *Athena*/WFI predicted images of a simulated group of galaxies (and surrounding individual AGN sources) at a redshift of $z = 2$ (top) and evolved to $z = 1$ (bottom). *Middle column*: corresponding surface brightness profiles, extracted in the $[0.5 - 2.5]$ keV energy band. The vertical dot-dashed lines indicate $R_{500}$. *Right column*: evolution of the entropy profiles, obtained from an inversion of analytical density and temperature models, given the X-ray surface brightness and projected temperature. The dotted lines show the profile derived from numerical simulations only implementing gravitational heating processes (Voit, Kay, and Bryan 2005). For more details, see Pointecouteau et al. 2013, from which the plots are extracted.

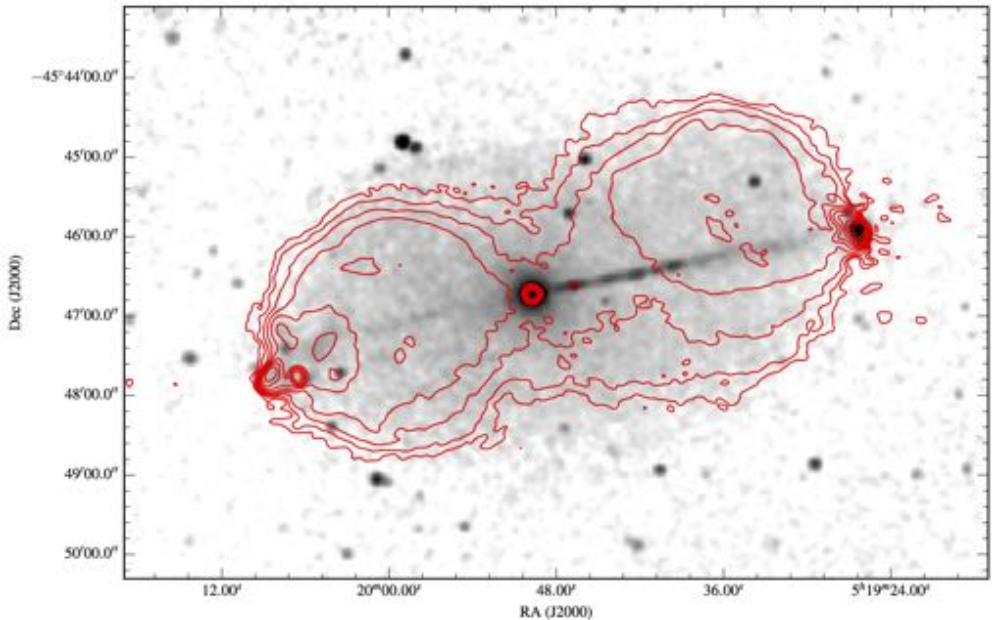

Figure 3.4: *Chandra* X-ray (0.5-5.0 kev) image of the nearby ($z = 0.035$) radio galaxy Pictor A. Superposed are radio Australia Telescope Compact Array (ATCA) contours (5.5 GHz). While the X-ray emission in the jet is synchrotron emission from accelerated particles, the extended X-ray emission associated with the radio lobes is attributed to inverse Compton scattering of the CMB. Extracted from Hardcastle et al. 2016a.

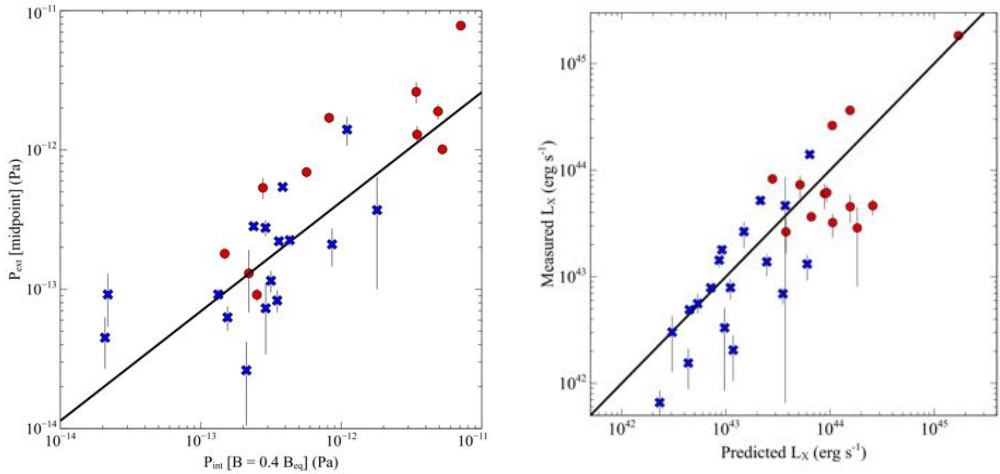

Figure 3.5: *Left*: internal lobe pressure, calculated assuming magnetic field strengths a factor of a few less than the equipartition value (Croston et al. 2005; Ineson et al. 2017), for 27 FRII radio galaxies vs. measured external ICM distribution (Ineson et al. 2017). Results suggest that radio-only pressure estimates can predict the external mid-lobe pressure both at low-$z$ (blue crosses, $z < 2$) and at intermediate-$z$ (red circles, $0.4 < z < 0.6$). *Right*: the predicted group X-ray luminosities determined from the radio-measured pressures of FRII (x-axis) are compared with the measured X-ray luminosities of the surrounding ICM (y-axis). The black line represents the best fit in the left panel, the line of unity on the right. Extracted from Croston et al. 2017.

### 3.3.2 Inverse Compton emission from radio galaxies: an important component of the X-ray sky

The possibility of measuring magnetic fields through the powerful combination of radio and X-ray emission has important astrophysical applications, but it has also been suggested that X-ray emission from both active radio galaxies and remnant lobes (i.e. those no longer being actively fed by a jet) could be an important component of the X-ray sky at high redshifts (e.g. Mocz, Fabian, and Blundell 2011). This should be taken into account carefully when looking for X-ray sources associated with clusters and groups, in particular beyond $z > 0.5$, when they are expected to be extended but small objects (from a few tens of arcseconds to a few arcminutes; Croston et al. 2013, ; Fig. 3.3).

LOFAR surveys indeed suggest that such remnants may be common (e.g. Hardcastle et al. 2016b). It is therefore important for high-redshift group/cluster studies with *Athena* to establish whether group catalogs obtained via an *Athena*/WFI survey could suffer significant contamination by non-thermal X-rays from active or remnant lobes, and whether the *Athena*/X-IFU characterization of such groups can be carried out in the presence of such non-thermal contributions. Ongoing LOFAR surveys (e.g. Shimwell et al. 2017; Hardcastle et al. 2016b; Williams et al. 2016) should enable both active and remnant sources of intermediate luminosities at $z > 2$ to be detected in large numbers; however, the spatial resolution achieved by these surveys ($\gtrsim 5$ arcsec) will mean that only sources larger than a few hundred kpc can be fully characterized.

While SKA1-MID's higher frequency range is less suited to steep spectrum remnant emission than LOFAR, it is sources of radio luminosities at or above the canonical Fanaroff & Riley luminosity break (Fanaroff and Riley 1974) that are predicted to have X-ray inverse-Compton emission at a level comparable to thermal group emission. A SKA1-MID continuum survey at $\sim 1$ GHz with $\sim 0.5$ arcsec resolution will enable the geometry of moderately luminous remnant sources, as well as of active FRIIs, to be estimated with the accuracy needed to model the synchrotron emission and to predict the expected X-ray inverse Compton emission. SKA1-LOW will also be useful for this work, being more sensitive to steep spectrum remnants, but, similarly to LOFAR, its expected spatial resolution will limit its value for characterizing high-redshift remnants (Kapinska et al. 2015).

## 3.4 Finding the earliest galaxy groups and clusters efficiently with radio surveys

It has long been argued that radio-loud AGN at high redshift are pointers to rich environments, and many known protoclusters have been found by targeting luminous high-redshift radio-loud AGN (e.g. Wylezalek et al. 2013; Castignani et al. 2014). Despite this common association, recent work at X-ray and optical wavelengths has established that most radio galaxies at $z < 0.6$, across a wide luminosity range, inhabit galaxy groups (e.g. $L_X < 10^{44}$ erg s$^{-1}$), rather than rich clusters (Ineson et al. 2015; Best 2004).

Simple considerations of lobe hydrodynamics suggest that this will remain true at $z > 2$, and in particular radio-loud AGN are likely to trace evolved groups with a hot ICM. Wide-field, high-resolution continuum radio surveys therefore offer considerable potential to assist with *Athena*'s objective of finding and characterizing the earliest galaxy groups (Pointecouteau et al. 2013). In particular, it has recently been shown that, due to the relatively narrow magnetic field strength distribution of the FRII (classical double) population (e.g. Ineson et al. 2017), *radio-only* estimates of FRII internal pressures appear to be a reliable estimator of the mass and X-ray luminosity of the surrounding galaxy group or cluster to within a factor of $\sim 2$ (Croston et al. 2017) (see Fig. 3.5). This method offers considerable potential for identifying large samples of candidate evolved galaxy groups: unlike optical/IR selection, it preferentially selects for groups with a hot ICM, enabling efficient *Athena* follow-up of the most evolved systems at $z > 2$.

Radio polarization information is also expected to prove to be a powerful environmental diagnostic (e.g. O'Sullivan et al. 2015), while tailed sources are expected to act as cluster indicators (Blanton et al. 2001; Johnston-Hollitt, Dehghan, and Pratley 2015). Exploiting these methods efficiently requires reliable morphological classification and size estimates for catalogued AGN populations; the expected spatial resolution of an all-sky SKA1-MID continuum survey at $\sim 1$ GHz (corresponding to a few kpc at $z = 3$) is ideally matched to these requirements at the redshifts where the first groups with a hot ICM are expected to occur.

While further development (including testing and calibration) of these methods is needed – e.g. using LOFAR surveys for lower redshift samples, combined with *Chandra*/*XMM-Newton* data and optical surveys – SKA continuum surveys have the potential to provide large catalogs of candidate evolved galaxy groups, clusters and proto-clusters at $z > 2$, which can be targeted efficiently for characterization with *Athena*. The comparison with serendipitous samples obtained through other methods will enable careful investigation of selection effects at $z > 2$.

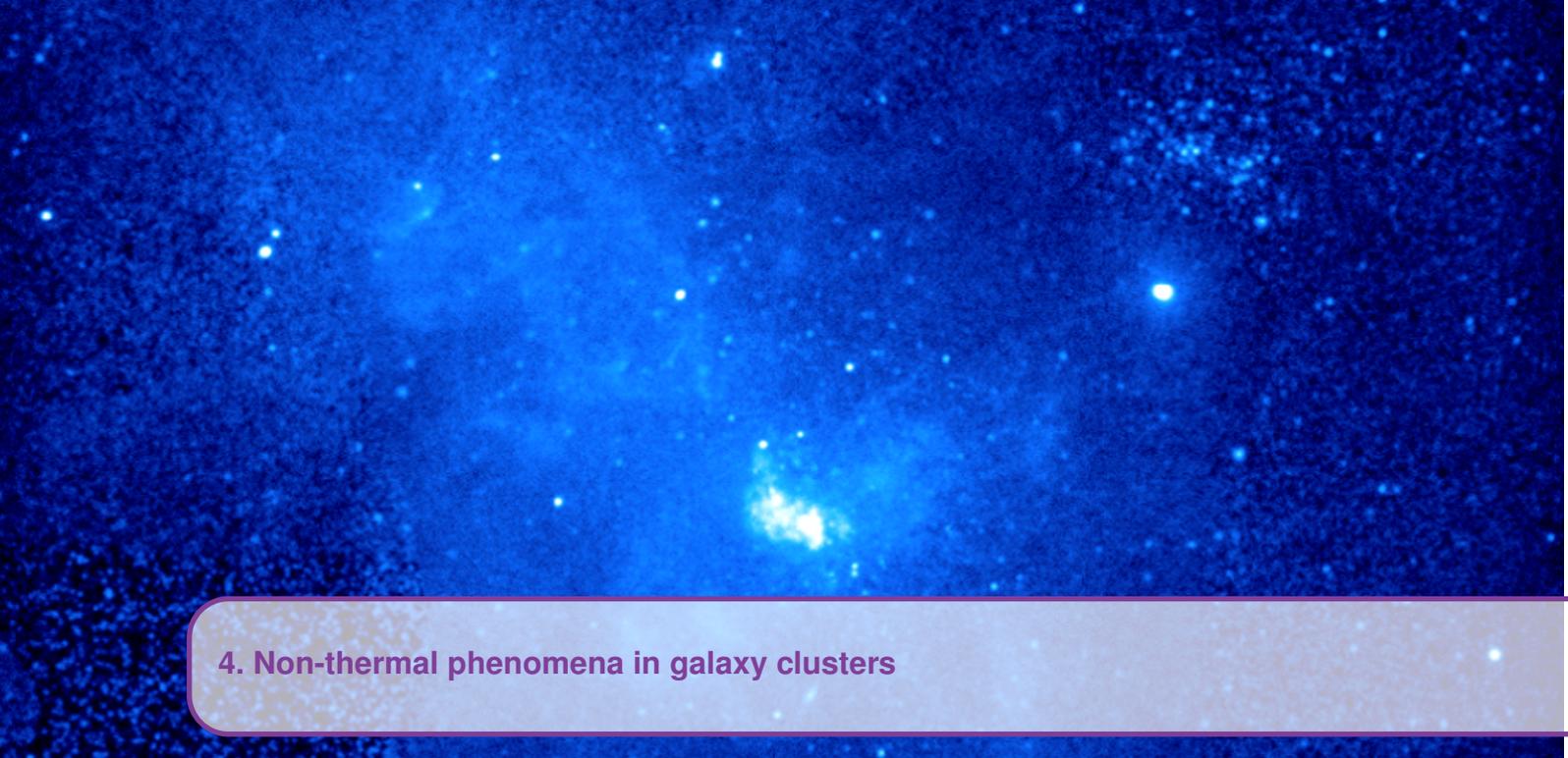

## 4. Non-thermal phenomena in galaxy clusters

### 4.1 Introduction

Radio observations prove the existence of relativistic particles and magnetic fields associated with the ICM, through the presence of Mpc-scale diffuse synchrotron emission in the form of giant radio halos and peripheral relics (for reviews see Ferrari et al. 2008; Feretti et al. 2012; Brunetti and Jones 2014).

Radio relics are Mpc-size, polarized sources located on the outskirts of merging galaxy clusters. Relics have elongated shapes and display filamentary morphologies (e.g., Owen et al. 2014; van Weeren et al. 2016). Radio halos are unpolarized sources that display smooth morphologies (e.g., Brown and Rudnick 2011), with the radio emission approximately following the X-ray emission from the thermal gas. Large Mpc-size halos are found in merging galaxy clusters (e.g., Cassano et al. 2010). Mini-halos with sizes of $\lesssim 500$ kpc are found at the centre of relaxed cool-core clusters, and they surround the central radio source associated with the brightest cluster galaxy (e.g., Gitti, Brunetti, and Setti 2002; Giacintucci et al. 2014b).

Over the last decade, significant progress has been made in our understanding of halos and relics, primarily through a combination of radio and X-ray studies. It has become clear that cluster merger shocks, and probably turbulence, play an important role in accelerating particles to high energies, as required to produce the observed emission (e.g., Brunetti et al. 2008; Finoguenov et al. 2010; van Weeren et al. 2010; Shimwell et al. 2015). In addition, recent work has found evidence for the re-acceleration of AGN fossil radio plasma at galaxy cluster shocks (e.g., Bonafede et al. 2014b; van Weeren et al. 2017). This indicates that to understand relics, and possibly also radio halos, the presence and distribution of radio galaxies needs to be taken into account, in addition to particle acceleration produced by shocks and turbulence. Furthermore, significant progress has been made thanks to the *Fermi* gamma-ray non-detections of galaxy clusters (e.g., Ackermann et al. 2010; Ackermann et al. 2014), which have disfavoured a pure hadronic origin for radio halos by limiting the content of cosmic ray protons in the ICM (see e.g., Brunetti and Jones 2014, for a critical discussion).

While great progress has been made, there are still key open questions and problems. For relics, it is still unclear by which mechanism the particles are re-accelerated, and whether re-acceleration is required for all radio relics. For halos and mini-halos, so far only indirect evidence has been found to support a connection with ICM turbulence, and many details of the turbulent re-acceleration mechanism are poorly understood. In addition, the relative contributions of fossil electrons and cosmic ray protons in the generation of diffuse cluster-scale emission remains unclear. In general, both the evolution of the non-thermal component of the ICM over cosmic time and its impact on the formation and evolution of galaxy clusters remain poorly constrained. In this chapter, we will show how combined SKA-*Athena* observations have the potential to greatly increase our understanding of the physics of shocks, turbulence, and particle acceleration in the ICM.

Major breakthroughs can be achieved by combining the survey capability and deep imaging of SKA1 (MID and LOW) with the high-resolution spectroscopy of *Athena*/X-IFU to investigate the presence of a connection between diffuse radio emission from the ICM and its turbulent state. Furthermore, the superb spectral resolution of *Athena*/X-IFU will make it possible to explore the presence of deviations from a Maxwellian distribution in the ICM of the host clusters of both radio halos and relics. *Athena* will allow accurate measurements of temperature jumps in the cluster outskirts where relics reside (difficult with present facilities), and this information combined with SKA1 measurements of the spectral and polarization properties of diffuse radio emission will provide crucial constraints for theoretical models of radio relics. More details are provided in the following Sections.

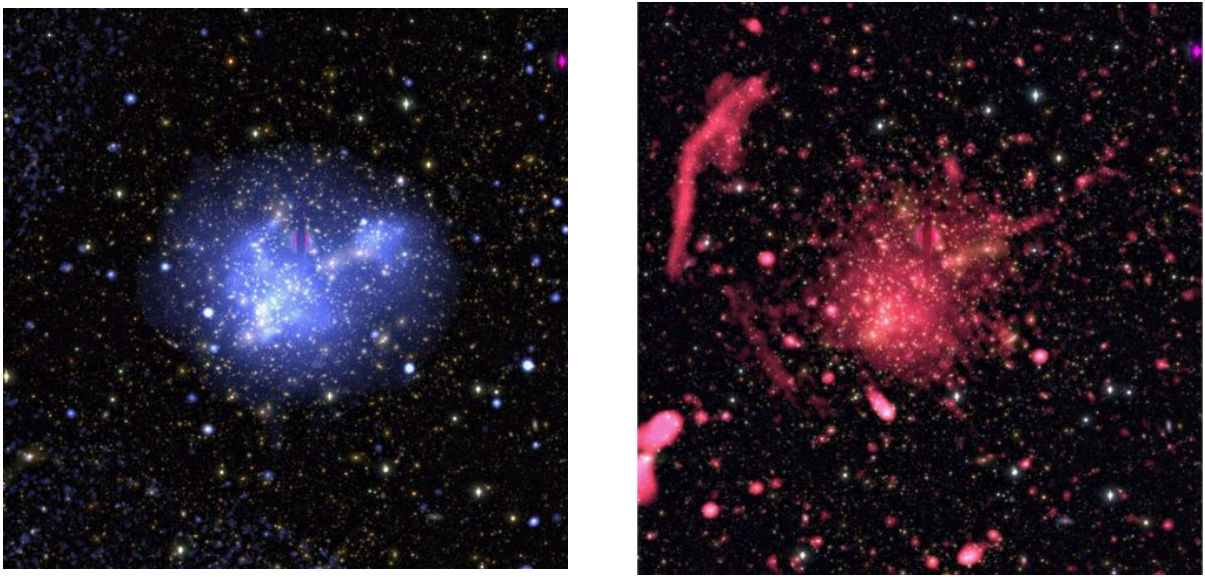

Figure 4.1: A2744: *Left: Chandra* 0.5-2.0 KeV X-ray (in blue) overlaid on the optical BRz Subaru color image from Medezinski et al. 2016. *Right:* radio image from the 1–2 GHz JVLA wide-band 15″ uv-tapered map on the same Subaru color image. Figures adapted from Pearce et al. 2017.

## 4.2 Giant radio halos and the physics of the ICM

Giant radio halos are the largest non-thermal sources in galaxy clusters. They are Mpc-sized emission regions with a morphology similar to that of the thermal X-ray emission (see e.g.; Fig. 4.1). The currently favoured scenario for the origin of radio halos is the reacceleration model, in which the relativistic particles producing the radio emission are reaccelerated by turbulence generated in the ICM during cluster-cluster mergers (Brunetti et al. 2001; Petrosian 2001; Brunetti and Lazarian 2007). The seed electrons are expected to be injected into the ICM by the activity of AGN and starburst galaxies (primary electrons; e.g. Völk, Aharonian, and Breitschwerdt 1996; Berezinsky, Blasi, and Ptuskin 1997; Völk and Atoyan 1999), and/or by the hadronic interactions between relativistic protons and thermal protons (secondary electrons; e.g. Dennison 1980; Blasi and Colafrancesco 1999). Reacceleration models are supported by the statistical connection that is observed between radio halos and cluster mergers (e.g. Cassano et al. 2013; Cuciti et al. 2015), and predict that clusters hosting radio halos are more turbulent than clusters without Mpc-scale radio emission (see Eckert et al. 2017, for some first observational support for this picture). Despite the physical details of acceleration mechanisms remaining poorly understood, calculations suggest that radio halos can be powered in massive clusters if turbulence is energetically significant, requiring typical turbulent-eddy velocities of $\geq 300$ km s$^{-1}$. Numerical simulations demonstrate that these turbulent velocities are possible in the ICM (Vazza et al. 2009; Miniati 2014), and can be measured very well by *Athena* (Ettori et al. 2013).

In fact, the reacceleration scenario requires a significant change in our understanding of the ICM, and it has important implications for the physics of the ICM more broadly. In this model, a hierarchy of complex mechanisms drains a fraction of the energy of the large-scale motions that are generated by the process of cluster formation into electromagnetic fluctuations and collisionless particle acceleration mechanisms on much smaller scales (Brunetti and Jones 2014, for review). This is possible because the ICM is a weakly collisional plasma (with a Coulomb mean free path very much larger than the Larmor radius of particles) with high-beta plasma value (the ratio of thermal and magnetic pressure), and in this case, plasma instabilities and kinetic effects play important roles in regulating micro-physical properties (e.g. Santos-Lima et al. 2014, and references therein). In fact, the collisional parameter in the ICM (namely the ratio of the Coulomb collision frequency and plasma frequency) is $R_C \sim 10^{-16}$ and it is well known that in $R_C << 1$ media wave-particle interactions become more important than Coulomb collisions.

Non-thermal phenomena in galaxy clusters are probes of this complex physics and provide an important research area for the SKA and its precursors and pathfinders [LOFAR, Murchison Widefield Array (MWA), ASKAP]. ICM turbulence is the key factor for these processes. It drives both the amplification of the ICM magnetic fields and the mechanisms of particle reacceleration that are probed by giant radio halos; moreover, it is a unique probe of ICM physics, because the generation, dissipation and cascading of the different turbulent branches depend on (and affect) ICM microphysics (Brunetti and Jones 2014; Schekochihin and Cowley 2006). Therefore the combination of adequate radio observations of non-thermal components in the ICM with direct measurements of cluster turbulence from X-ray spectroscopic measurements will allow a major advance in our understanding of the physics of non-thermal and thermal ICM.

### 4.2.1 Turbulence and radio halos

*Athena*'s micro-calorimeter (*Athena*/X-IFU) will provide direct constraints on turbulence in the ICM, by resolving spectral features on $\sim 5$ arcsec scales, with a spectral resolution of 2.5 eV, and over a field of view of 5 arcmin. These measurements, combined with radio observations, will allow tests of the prediction that clusters hosting radio halos are more turbulent than clusters without Mpc-scale radio emission (see also Eckert et al. 2017). This is a crucial step towards understanding non-thermal phenomena in galaxy clusters. Moreover, the combination of spatially resolved measurements with *Athena* and SKA will allow further advances. In fact, the complex interplay between particles and turbulence governs the acceleration and diffusion of particles, which in turn, together with the magnetic field distribution, determine the properties of the synchrotron emission (e.g. Lazarian and Pogosyan 2012; Lazarian and Pogosyan 2016). The SKA will measure the synchrotron brightness, spectrum and polarization of radio halos with unprecedented detail, exploring for the first time the statistics and the relationships between these quantities on 5-15 arcsec scales. The ICM turbulence is hydrodynamic on large scales, and becomes MHD on smaller scales. The transition between the two regimes occurs at the MHD scale, namely the scale at which the turbulent velocity equals the Alfvén speed,

$$l_A = L_0 M_A^{-3} \simeq 1.7 \frac{L_0}{(\sqrt{\beta_{pl}} M_0)^3} \qquad (4.1)$$

where $M_A$ and $M_0$ are the Alfvénic and total turbulent Mach number, $L_0$ is the turbulent driving scale and $\beta_{pl}$ is the beta–plasma number. The scale $l_A$ is also where the magnetic-field power spectrum peaks, reaching quasi-equipartition with the kinetic energy of turbulence. As a consequence, this scale plays a key role in determining the polarization properties of the synchrotron radio emission produced in the turbulent ICM. In practice, in super-Alfvénic turbulence the intrinsic polarization scales as $1/\sqrt{N}$, where $N$ is the number of magnetic field cells intercepted by the observer. This scale depends on the combination of observational beam size (decreasing for lower resolution, similarly to beam depolarization) and turbulent properties (decreasing for larger Mach numbers). At the same time, the turbulent acceleration rate increases with turbulent Mach number, allowing the acceleration of electrons at increasingly higher energies, and the generation of radio synchrotron emission at increasingly higher frequencies. In Fig. 4.2 we show the maximum steepening synchrotron frequency of radio halos produced by turbulent reacceleration models, as a function of the turbulent broadening that is induced by the same turbulence in the ICM (see caption). In the same plot, we show the percentage of intrinsic polarization that is predicted by the models, assuming an observational beam of 5 arcsec and the redshift of the Coma cluster. The key prediction of models is that radio halos with steeper spectra are generated in a less turbulent ICM, and have higher intrinsic polarization. We also note that – in general – the intrinsic polarization of radio halos generated by turbulent reacceleration is expected to increase at lower radio frequencies.

### 4.2.2 Collective plasma effects and ICM microphysics

According to current models, plasma effects are important for radio halos, as they determine the efficiency of the chain of mechanisms that generate these radio sources (Miniati 2015; Brunetti and Lazarian 2011; Brunetti and Lazarian 2016). These effects are in general subtle and intermittent in nature; however, measuring their imprint at the sites of particle acceleration will be important for our understanding of ICM physics. One key consequence of collective plasma effects is a possible departure of the energy distribution of ICM particles from the Maxwellian. k-distribution functions are expected to be generated in the cases where Coulomb collisions are largely sub-dominant with respect to particle-wave interactions in the plasma (e.g., Yoon 2014; Livadiotis 2015). Under these conditions ionisation equilibrium in the ICM is different from that calculated using Maxwellian distributions, which has observable consequences for X-ray emission lines. Markevitch et al. (in prep) carried out calculations of X-ray spectra in the case of Maxwellian and k-distribution functions and conclude that the superb spectral resolution of *Athena* will allow disentangling the two situations . In particular, according to their calculations, the strongest effects are observed in the case of dielectric (satellite) recombination lines. Departures from the Maxwellian distribution are expected to be found in particular at the sites of turbulent acceleration that can be best identified with SKA observations at 5-10 arcsec resolution (regions of flatter local synchrotron spectra, see also Fig. 4.2), and so the combination of *Athena* and SKA observations provides the most efficient observational approach.

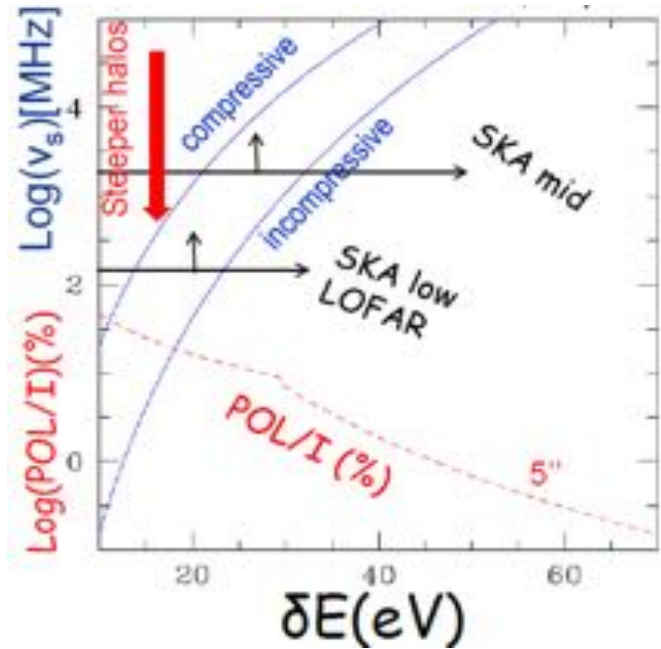

Figure 4.2: Maximum synchrotron frequency from radio halo models as a function of the turbulent broadening measured at 6.7 keV (blue lines). Calculations assume compressible (Brunetti and Lazarian 2007) and incompressible (Brunetti and Lazarian 2016) turbulent acceleration physics, a sound speed =1500 km/s, a value of the beta plasma =100 and a turbulent injection scale $L_0 = 300$ kpc. Percentage of intrinsic linear polarization as a function of the turbulent broadening (dashed red line). Calculations assume an observational beam =5 arcsec at the distance of the Coma cluster. Black arrows provide the lower bound to the maximum synchrotron frequencies of radio halos that can be observed with the SKA1-MID and the SKA1-LOW.

### 4.2.3   Statistical investigation of radio halos

A statistical link between cluster dynamical state and the presence of extended radio halo emission has now been demonstrated in a number of publications (Fig.4.3, right panel; e.g. Cassano et al. 2010; Cassano et al. 2013; Rossetti et al. 2011; Wen and Han 2013; Cuciti et al. 2015; Parekh et al. 2015; Mantz et al. 2015; Yuan, Han, and Wen 2015). These works have shown that Mpc-scale radio halo emission is associated almost exclusively with systems that present highly perturbed X-ray images, indicating significant dynamical activity, with a few possible outliers (Bonafede et al. 2014b; Kale and Parekh 2016; Sommer et al. 2017), while objects with a relaxed X-ray morphology do not appear to host radio emission. When plotted in the radio power-X-ray luminosity (or equivalently, Sunyaev-Zel'dovich – SZ – signal or mass) plane, there is a clear bimodality in the cluster population: an empirical correlation exists for systems with detected radio emission, while objects with regular X-ray morphology are undetected in radio (see Fig.4.3). One can expect clusters to evolve cyclically in this plane as major merger events generate turbulence that lights up the radio halo, which then evolves through a progressive steepening of the spectral index and a dimming of the overall radio power over time (Donnert et al. 2013). A recent attempt to link the presence of radio halo in clusters to ICM turbulent motions can be found in Eckert et al. 2017. These authors used the amplitude of density fluctuations in the ICM derived from *Chandra* X-ray data of ~ 50 galaxy clusters as a proxy for the ICM turbulent velocity. They found a segregation in the turbulent velocity of radio halo and radio-quiet clusters, with the turbulent velocity of the former being on average higher by about a factor of two. These results, although convincing, are still indirect (not a true measurement of ICM turbulence), and limited by the sensitivity of radio observations to the most luminous and massive clusters.

In addition, there is increasing evidence for the existence of merging clusters that do not show the presence of radio halos at $\sim 1$ GHz frequency (e.g. Cassano et al. 2010; Cassano et al. 2013; Cuciti et al. 2015). This is not surprising in the context of the turbulent re-acceleration scenario since it is expected that less turbulent systems should host radio halos characterized by ultra steep radio spectra (ultra-steep spectrum radio halos, USSRH; Cassano, Brunetti, and Setti 2006; Brunetti et al. 2008), which can be unveiled at lower frequency (i.e., less than 600 MHz). These systems are expected to be generated by relatively less efficient mergers (i.e., mergers with small mass ratios — minor mergers—, or mergers between less massive clusters) or, less frequently, they can represent the rising or fading phase of a giant radio halo. Low-frequency SKA precursors and pathfinders, such as MWA and LOFAR, have already started to unveil the existence of USSRH (e.g. Wilber et al. 2017; Duchesne et al. 2017); however, observations with SKA1-LOW have the potential to unveil the bulk of this population (Cassano et al. 2015).

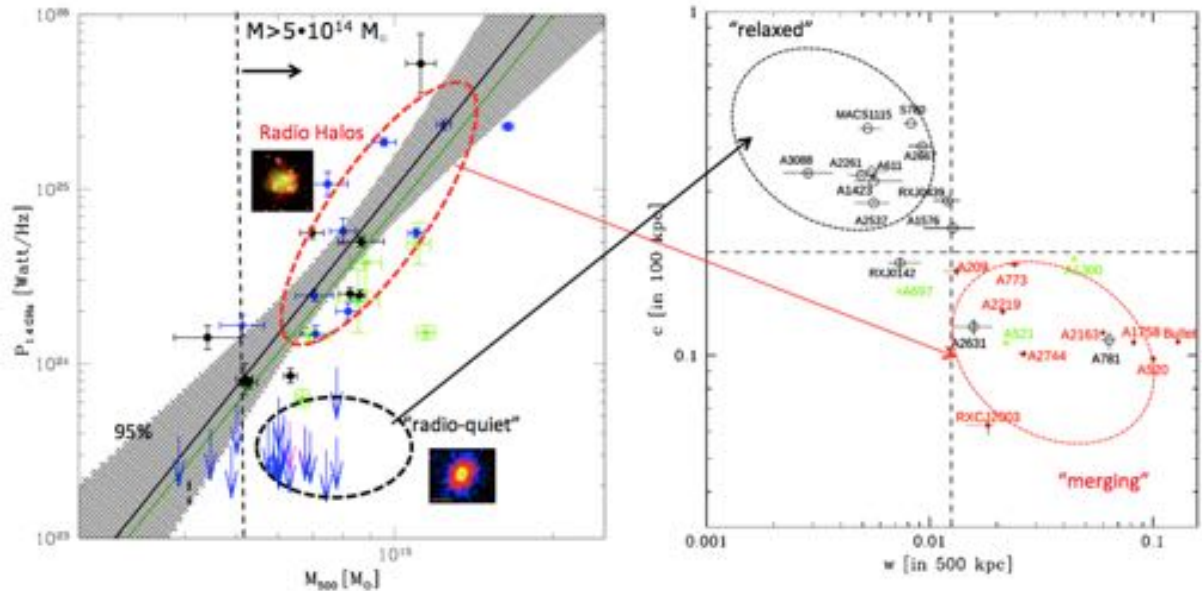

Figure 4.3: *Left:* Radio Halos (RH) power $P_{1.4}$ vs cluster mass $M_{500}$ (adapted from Cassano et al. 2013). Symbols are clusters belonging to the GMRT RH Survey (Venturi et al. 2008, , and ref. therein): RH and upper limits (in blue), RH with ultra-steep spectra (USSRH, in green) and clusters with RH from the literature (in black). The best-fit relation to giant RH (black line) is reported together with its 95% confidence region (shadowed regions). *Right:* distribution of clusters with $M_{500} \geq 5.5 \times 10^{14} M_\odot$ in the merger diagnostic diagram: concentration parameter, $c$, versus the emission centroid shift, $w$ (Cassano et al. 2013, for details). Clusters with RH (red points and green asterisks) and clusters without RH (black points) clearly occupy different regions, the upper left region is the place of relaxed clusters, whereas the lower right region is the place of merging clusters.

To make progress on this front, it will be necessary to define an unbiased sample of clusters that extends down to low masses and out to high redshifts, such as will be available from other multi-wavelength (optical, SZ, and X-ray) surveys in the coming years. In this context, radio detections and non-detections with SKA1 will allow us to construct a statistical view of the full radio-halo population. Systematic follow-up with *Athena* would allow us to build a statistical picture of the level and occurrence rate of turbulence in the full cluster population, and to determine its relation to the presence or absence of radio-halo emission. In fact, the synergistic use of SKA and *Athena* will *for the first time* enable a direct, objective measurement of the gas velocity field in clusters with and without diffuse radio halos, thus probing the scenario described above.

In particular, with SKA1 data we will be able to:

- detect unprecedented numbers of extended radio sources, classify them both morphologically and spectrally to provide a physically robust taxonomy, and relate them to clusters detected in other wavelengths (e.g. Cassano et al. 2015);
- detect extended radio emission down to lower mass and higher redshift ($z \sim 1$) than is currently possible, allowing us to test the hypothesis that radio-halo generation is not limited to the highest-mass systems, and extending the empirical scaling between radio power and mass to currently-inaccessible regimes;
- investigate the bimodality in the empirical scaling between radio power and mass, by probing the low radio power region, and potentially discovering a population of steep-spectrum sources that are currently undetected;

- using the SKA All-Sky Rotation Measure Survey (Johnston-Hollitt et al. 2015), which will provide rotation measures from both background sources and embedded cluster sources, including resolved AGN (Johnston-Hollitt, Dehghan, and Pratley 2015) and relics (Bonafede et al. 2015), we will be able to probe the magnetic field in clusters to an unprecedented level, determining the field strengths in and around shocks in the ICM and even potentially detecting polarised emission from radio halos themselves (Govoni et al. 2015)

With *Athena*, it will be possible to:

- use high throughput *Athena*/WFI imaging observations to investigate the ICM turbulent power spectrum through fluctuations in *multiple* tracers (e.g. density, pressure, entropy; see e.g. Gaspari et al. 2014), from highly perturbed to relaxed systems, and from low masses to high masses;
- use the *Athena*/X-IFU to make measurements of bulk and turbulent ICM velocities (either averaged over the cluster extent, or in regions; Ettori et al. 2013), and for the first time, to probe the projected turbulent velocity power spectrum.

### 4.2.4  The case of radio mini-halos

Diffuse radio emission on the scale of cluster cores ($< 300$ kpc), known as radio mini-halo emission, is found in a large fraction of massive ($M_{500} > 6 \times 10^{14} M_\odot$) cool-core clusters (Giacintucci et al. 2017) surrounding the central active radio galaxy that is nearly always found at the centre of such clusters (e.g., Mittal et al. 2009). Similar to giant radio halos, their presence requires the *in situ* generation of relativistic particles on the cluster core scale, via hadronic interaction or particle re-acceleration by turbulence (e.g. Brunetti and Jones 2014, for a review). Mini-halos often appear bounded by one or two X-ray cold fronts (Mazzotta and Giacintucci 2008; Giacintucci et al. 2014b; Giacintucci et al. 2014a) that result from sloshing of the cool gas in the central core (e.g., Ascasibar and Markevitch 2006). This coincidence supports the idea that mini-halos arise from the reacceleration of seed relativistic electrons in the magnetized cool core (Gitti, Brunetti, and Setti 2002; Gitti et al. 2004) as a consequence of sloshing-induced turbulence (e.g., Mazzotta and Giacintucci 2008; ZuHone et al. 2013). Numerical simulations show that sloshing motions can indeed amplify magnetic fields and develop turbulence in the area enclosed by the cold fronts, which may lead to the generation of mini-halos (ZuHone et al. 2013). Interestingly, the *Hitomi* satellite has provided an unprecedented view of the Perseus cluster, hosting a famous mini-halo (e.g. Gendron-Marsolais et al. 2017), showing for the first time what spatially resolved high-resolution X-ray spectroscopy can deliver (thanks to its SXS calorimeter spectrometer). These observations have shown that the level of turbulence close to the core of Perseus is of the order of $\sim 164 \pm 10$ km s$^{-1}$ on scale of $\sim 20 - 30$ kpc (Hitomi Collaboration et al. 2016). This turbulence is already sufficient to explain the radio mini-halo in the Perseus core, and possibly to balance the radiative cooling in the core (Hitomi Collaboration et al. 2016). This is just a foretaste of what the *Athena*/X-IFU will actually deliver, thanks to its combined spatial (5 arcsec) and spectral (2.5 eV) resolution, by mapping the velocity field of the hot gas in groups and clusters of galaxies down to a precision of 10-20 km s$^{-1}$ for velocities ranging between 100-1000 km s$^{-1}$ (Barret et al. 2016). On the radio side, SKA1 surveys are expected to unveil the bulk of radio mini-halos in the Universe (Gitti et al. 2015), furnishing the most complete sample of cluster cores to follow up with *Athena*/X-IFU.

An important question still to be answered is whether or not there is a physical connection between giant radio halos and mini-halos. Although it is clear that they are hosted by clusters in a different dynamical state, one might argue that during mergers there is likely to be a transition between the presence of a central mini-halo to a more giant halo, since it is known that ICM turbulent motions can destroy cluster cores and also transport and re-accelerate relativistic electrons on larger scales (e.g. Brunetti and Jones 2014; Bonafede et al. 2014a; Venturi et al. 2017). Alternatively, it has been proposed that mini-halos are of hadronic origin, while giant radio halos experience a transition from the central hadronic emission to a leptonic external component due to turbulent re-acceleration (Zandanel, Pfrommer, and Prada 2014). Future observations with the SKA of the external region of cluster mini-halos are necessary to discriminate among the possible scenarios, and their combination with *Athena*/X-IFU spectroscopic characterization of the cluster cores will really help to solve this puzzle.

Finally, it remains unclear what role the central AGN plays in the origin of mini-halos. In addition to secondary electrons that can be furnished by the hadronic interaction between Cosmic Ray (CR) protons and thermal protons (Pfrommer and Enßlin 2004), cluster central AGN cyclically feed the ICM with relativistic electrons that are then observed at low frequencies as radio lobes filling X-ray cavities in the ICM (see also Chapter 2.4.2). These electrons can be the seeds to be re-accelerated by the turbulence active in mini-halos (e.g. Cassano, Gitti, and Brunetti 2008). SKA1-LOW observations of cluster central AGN, in combination with *Athena* observations and future observations in the gamma-ray regime (e.g.; with *Fermi*, Cherenkov Telescope Array – CTA) are crucial to disentangle the contribution of the different sources of relativistic electrons in cluster mini-halos.

## 4.3 Giant radio relics and shocks in the ICM

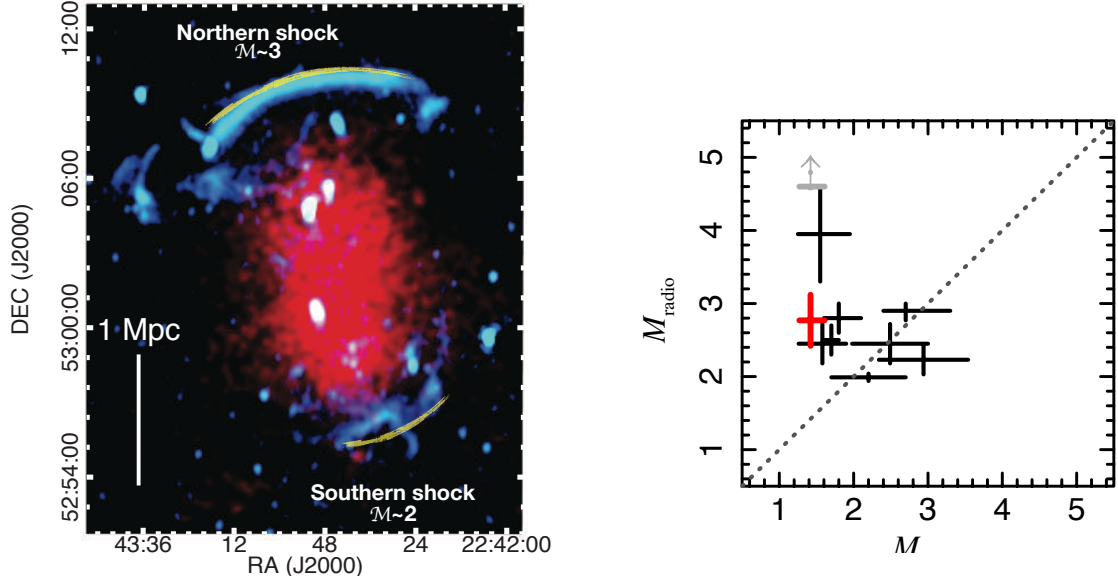

Figure 4.4: *Left:* Smoothed 0.5-2.0 keV *Suzaku* X-ray image (red) and WSRT 1.4 GHz radio image of CIZA J2242.8+5301 (cyan). The yellow lines depict the approximate locations of the shock fronts confirmed by *Suzaku* from Akamatsu et al. 2015. *Right:* The Mach number derived from radio ($\mathscr{M}_{\text{radio}}$) plotted against that from the ICM temperature determined from *Suzaku*.

Radio relics are diffuse radio emission features typically located at the periphery of clusters ($> 1$ Mpc from the cluster centre); they have elongated shapes and are strongly polarized (up to 60%–70% levels). Because of their morphology and location, it is believed that radio relics trace shock structures in the ICM caused by cluster mergers/structure formation activity, where the emitting CR electrons can be accelerated or re-accelerated (e.g. Miniati et al. 2001; Brüggen et al. 2012). The elongated filamentary morphology of radio relics, their polarization properties, and the observed spectral index gradients (e.g. van Weeren et al. 2010) strongly support a shock-relic connection. In fact, in some relics, a shock front has been identified in the X-rays in coincidence with the radio relic position (e.g. Markevitch 2010; Giacintucci et al. 2008; Finoguenov et al. 2010; Akamatsu and Kawahara 2013; Botteon et al. 2016b; Botteon et al. 2016a, see Fig. 4.4 left).

In the simplest picture, shock acceleration in radio relics is described according to diffusive shock acceleration (DSA) theory (e.g. Bell 1978; Drury 1983). In the DSA regime, the steady state spectrum of test-particle CRs at a shock is a power law in momentum, $f(p) = K p^{-(\delta_{inj}+2)}$, where the slope is related to the shock Mach number via:

$$\delta_{inj} = 2 \frac{\mathscr{M}^2 + 1}{\mathscr{M}^2 - 1} \tag{4.2}$$

where $\mathscr{M} = V_{sh}/c_s$ is the Mach number of the shock and the synchrotron spectral index $\alpha_{\text{inj}} = (\delta_{inj} - 1)/2$ (the observed particle and synchrotron spectra are steeper, $\delta_{\text{obs}} = \delta_{inj} + 1$ and $\alpha_{\text{obs}} = \alpha_{inj} + 0.5$, respectively, due to particle energy losses). In this simplified picture, observations of radio relics can therefore be used to constrain the properties of merger shocks.

Shocks discovered in the X-rays are apparently weak, with Mach numbers estimated from the Rankine-Hugoniot condition (via temperature and/or density jumps across the shock region) $\mathscr{M}_X \sim 1.5 - 3.0$ (e.g. Markevitch and Vikhlinin 2001; Markevitch and Vikhlinin 2007; Botteon, Gastaldello, and Brunetti 2018). The properties of X-ray detected shocks are in line with what is found in high-resolution cosmological simulations, where shocks with $\mathscr{M} > 3$ are rare inside the cluster virial radius (e.g. Ryu et al. 2003; Pfrommer et al. 2006; Vazza et al. 2009). However, this poses a problem for the origin of radio relics, since weak shocks are expected to be relatively inefficient as particle accelerators (e.g. Kang and Ryu 2013; Hong et al. 2014), especially if they are accelerating CR from the thermal pool. This problem becomes dramatic when one has to explain very luminous radio relics (such as the ones in CIZA2242.8+55301 and A3667) and relics with synchrotron spectra flatter than those expected from DSA, if the X-ray derived shock Mach number is assumed (see discussion in Brunetti and Jones 2014). There are indeed cases in which a discrepancy between the radio-derived ($\mathscr{M}_{\text{radio}}$) and X-ray-derived ($\mathscr{M}_X$) Mach numbers has been found for some relics, typically $\mathcal{M}_{\text{radio}} > \mathcal{M}_{\text{X}}$, (see Fig.4.4 right). Another issue is that explaining radio relics with standard DSA overpredicts the expected gamma-ray flux (e.g., Vazza and Brüggen 2014; Brunetti and Jones 2014; Griffin, Dai, and Kochanek 2014; Vazza et al. 2015a).

Many of the challenges discussed above can be mitigated assuming a scenario in which merger shocks in radio relics are re-accelerating a pre-existing population of mildly relativistic electrons (e.g. Markevitch et al. 2005; Kang, Ryu, and Jones 2012; Pinzke, Oh, and Pfrommer 2013). Alternative mechanisms, such as shock drift acceleration (e.g. Guo, Sironi, and Narayan 2014a; Guo, Sironi, and Narayan 2014b) have been also proposed, since these are efficient for electron acceleration, and easily solve the problem of the overproduction of the gamma rays. Other issues related to observational biases in the radio and X-ray observations are also discussed as potentially contributing to the observed discrepancy between ($\mathcal{M}_{\text{radio}}$) and ($\mathcal{M}_{\text{X}}$) (see Sect 4.3 in Akamatsu et al. 2017 for details). Some possible issues include: projection effects, which can lead to the underestimation of the Mach numbers from X-ray observations (Skillman et al. 2013; Hong, Kang, and Ryu 2015); clumpiness and inhomogeneities in the ICM (Nagai and Lau 2011; Simionescu et al. 2011), which will lead to nonlinearity of the shock-acceleration efficiency (Hoeft and Brüggen 2007); a nonuniform Mach number as a result of inhomogeneities in the ICM, which are expected in the periphery of the cluster (Nagai and Lau 2011; Simionescu et al. 2011; Mazzotta et al. 2011); an underestimation of the post-shock temperature, with electrons not reaching thermal equilibrium (such a phenomenon is observed in SNRs); van Adelsberg et al. 2008; Yamaguchi et al. 2014; Vink et al. 2015). The limitations of current observatories prevent a firm conclusion being reached.

### 4.3.1 SKA-*Athena* synergy for the study of giant radio relics

Future observations with SKA and *Athena* are expected to shed light on the mechanisms of particle acceleration in the cluster merger shocks responsible for the formation of giant radio relics. In addition, the unprecedented capabilites of *Athena* and SKA will allow us to discriminate between the effects of possible systematics and new physical discoveries. Here we introduce some examples based on the *Athena*/X-IFU instrument. The high spatial ($\sim 5''$) and spectral resolution ($E/\Delta E$=2400 @ 6 keV) of *Athena*/X-IFU will enable us to disentangle the influence of projection effects, to measure ion-electron non-equilibrium in shock-heated regions, and to apply new diagnostics for non-thermal electrons.

#### The projection effect in radio relics

The morphology of radio relics indicates that the underlying shocks are expected to have a curved three-dimensional structure, which can lead to a two-temperature structure in the post-shock region (even without any inclination angle). The high spectral resolution of *Athena*/X-IFU enables us to disentangle two-temperature structure (mixing of pre- and post-shock temperatures along the line-of-sight) via the spectroscopic approach. Based on spectral temperature diagnostics (such as line fluxes, Bremsstrahlung exponential cut-off), *Athena*/X-IFU can reconstruct the original temperature structure. We refer here to Kaastra et al. 1996; Kaastra et al. 2004; Akamatsu et al. 2012 for more details and applications to clusters of galaxies.

#### The ion-electron non-equilibrium

In regions affected by shock heating, an electron–ion two-temperature structure is predicted based on numerical simulations (e.g. Takizawa 1999). This could lead to an underestimation of the post-shock temperature, because it is hard to constrain the ion temperature with current X-ray satellites and instruments and so the main observable is the electron temperature of the ICM, which will reach temperature equilibrium with the ions on a timescale corresponding to the ion-electron relaxation time after shock heating. Ion-electron non-equilibrium could be a possible cause of the Mach number discrepancy between radio and X-ray shock speed estimates. There are indeed several claims that electrons can be more rapidly energized (so-called instant equilibration) than by Coulomb collisions (Markevitch 2010; Yamaguchi et al. 2014). However, it is difficult to investigate this phenomenon without better spectra than those from the current X-ray spectrometers. The upcoming *Athena* satellite can shed new light on this problem by measuring the ion and electron temperatures from the line broadening and ratios of each ion.

#### Diagnostic approach for non-thermal electrons

It is usually assumed that the ICM is in collisional ionisation equilibrium, and that it can be well described by a Maxwellian electron distribution. However, when shocks and turbulent flows are present, deviations from a Maxwellian distribution may occur (see also Sect. 6.2.2 for the role of turbulence), associated with the temperature gradients or caused by particles accelerated by the shocks (e.g. Kaastra, Bykov, and Werner 2009). As the Coulomb thermal relaxation time increases with electron energy proportional to $E^{3/2}$, supra-thermal electrons are relatively long-lived, and may cause an additional pressure contribution. The presence of shock-accelerated non-thermal electrons in the post-shock region changes the relative intensities of satellite lines, for example in the $Fe$xxv $j$-line (Kaastra, Bykov, and Werner 2009). To reveal observationally the non-Maxwellian tails in the electron distributions requires the high-resolution spectra that can be obtained by *Athena*/X-IFU. In combination with detailed spectral modeling, this would allow a determination of the non-thermal electron energy distribution that could then be compared to the relativistic CR population observed with SKA. This information tells us how thermal and non-thermal energy transfers (injection and distribution) take place in the shock region.

### The shock-acceleration mechanism

To determine which particle acceleration mechanism(s) operate in the ICM, and to establish the origin of the ICM cosmic rays, detailed radio and X-ray measurements are required. *Athena* will for the first time provide accurate shock measurements via the ICM temperature jump. Currently, precise measurements cannot be obtained, since relics are located in the faint cluster outskirts where the X-ray count rates are very low. The SKA can obtain detailed spectral shape and polarization measurements, thanks to its excellent uv-coverage and sensitivity. Such measurements can then be compared with shock acceleration models (Kang and Ryu 2016; Fujita et al. 2015; Kang, Ryu, and Jones 2012). A key comparison will be that between the radio and X-ray derived Mach numbers (e.g., Akamatsu et al. 2017).

Finally, deep SKA1-LOW observations will enable the detection of steep-spectrum AGN fossil radio plasma, to determine the role of this plasma in the formation of radio relics.

## 4.4   Magnetic fields in galaxy clusters

Magnetic fields are an essential ingredient of many astrophysical phenomena, but important questions remain about their structure, origin and evolution. The most promising technique to derive a detailed view of the magnetic field in clusters, and on larger scales in the intergalactic medium, is the analysis of Faraday rotation of radio galaxies located inside and behind the cluster magneto-ionic medium (Fig. 4.5).

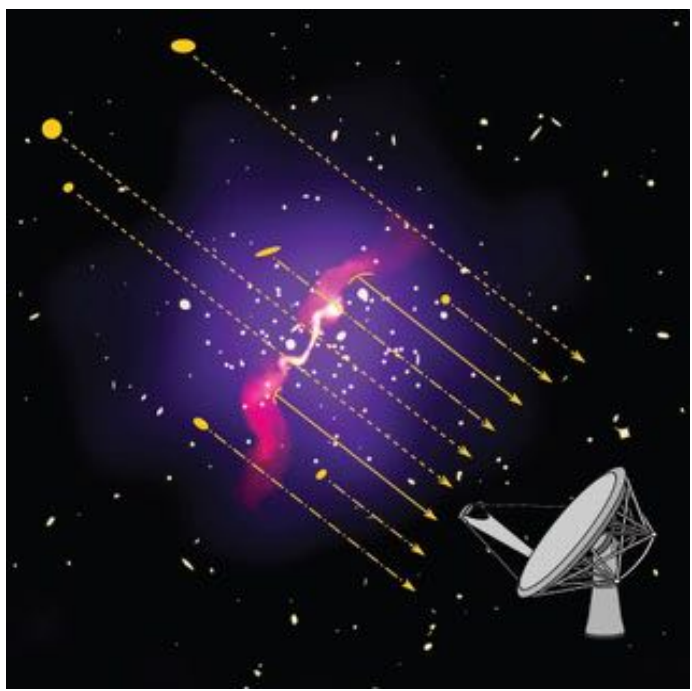

Figure 4.5:  Schematic of a nearby cluster showing X-ray emission in purple, and radio galaxies both in the cluster and in the background. The Faraday rotation of the polarized sources can be used to examine the cluster magnetic field (from Johnston-Hollitt et al. 2015).

Faraday rotation has a simple dependence on $\lambda^2$, assuming a uniform Faraday screen. The observed parameter is the Rotation Measure (RM), which is linked to the value of magnetic field along the line of sight and to the thermal electron density (e.g. Govoni and Feretti 2004). The strength and structure of cluster magnetic fields can be obtained by numerical techniques, given a 3-D magnetic field model and the density distribution of the intracluster gas. Simulated RM images of radio galaxies located inside or behind a cluster are then calculated by numerically integrating the product of magnetic field strength and gas density along the line of sight, and by applying to the results a filter that takes into account the observational beam and noise. The comparison between simulated and observed RMs allows us to constrain the magnetic field parameters (Murgia et al. 2004).

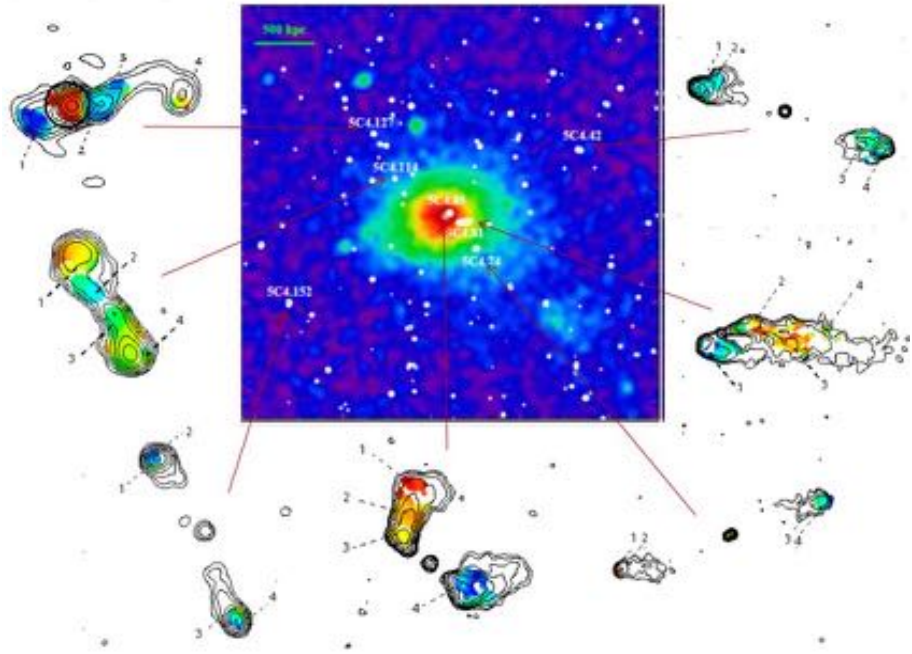

Figure 4.6: Coma cluster X-ray emission, and RM images of the polarized radio galaxies used to derive the magnetic field intensity, structure and profile, adapted from Bonafede et al. 2010.

Information about the magnetic field can thus be derived only if the density of the ionized medium along the line of sight is known or assumed. To date, magnetic field investigations are performed by assuming a spherically symmetric model (beta model) for the thermal gas, and so any possible substructures are completely neglected. The best studied cluster so far is Coma, whose magnetic field has been obtained with RM information on 7 radio galaxies in the cluster central region (Bonafede et al. 2010, see Fig. 4.6). The magnetic field of the peripheral Coma SW region, where the NGC 4869 infalling group and the relic are located, has been derived with 7 additional radio galaxies in that area (Bonafede et al. 2013).

Good estimates of the intensity of the magnetic field at the centre of clusters have been obtained so far, whereas the magnetic field structure (profile, coherence scale, minimum and maximum scales, power spectrum) is poorly known. These parameters are of paramount importance, since they are linked to the magnetic field formation, amplification and degree of ordering during the cloud collapse.

Moreover, deviations of the Faraday rotation from the simple $\lambda^2$ law are expected in the case of non uniform screen, and/or a mixed thermal and relativistic plasma.

### 4.4.1 SKA-*Athena* synergies

The SKA can deliver stunning new data sets that will address the above issues. The foundation for these experiments will be an all-sky (i.e. the visible 3-$\pi$ steradians) SKA survey of RM, in which Faraday rotation toward $\sim 10^7$ background sources will provide a dense *RM grid* for probing magnetism in the Milky Way, nearby galaxies, distant galaxies, clusters and protogalaxies (Johnston-Hollitt et al. 2015). According to the current polarized source counts to 4 $\mu$Jy, it is expected to obtain 60 - 115 RMs per cluster at z=0.05, 15 - 30 RMs at z=0.1, 1 - 2 RMs out to a redshift of 0.5 (Johnston-Hollitt et al. 2015). Using these data, we can map out the evolution of magnetized structures and can thus reveal what cosmic magnets look like, how they formed, and what role they have played in the evolving Universe, provided we have accurate knowledge of the gas density distribution.

From simulations of large-scale structure formation in the Universe, gas perturbations are expected on a range of spatial scales, and complex gas patterns can survive for long timescales (Kravtsov and Borgani 2012). These fluctuations have been detected on scales of 20 to 500 kpc, at levels of 5 - 10 % (e.g. Schuecker et al. 2004; Churazov et al. 2012). Detailed knowledge of the power spectrum of the gas-density perturbations is crucial to obtain the accurate and correct interpretation of the sensitive RM measurements. The presence of substructure and gas clumpiness within a galaxy cluster will lead to variation in the line-of-sight electron densities towards the polarized radio sources.

With *Athena*'s sensitivity, it will be possible to trace the diffuse X-ray emission to the outer cluster regions, leading to improved knowledge of the thermal gas density up to large cluster radii. These observations will provide a completely new window on the processes responsible for the assembly of galaxy clusters (Ettori et al. 2013).

The combined radio – X-ray studies possible with the SKA and *Athena* will provide detailed information on the magnetic field strength and structure in merging and relaxed clusters, allowing any deviations of Faraday rotation from the $\lambda^2$ law to be disentangled. They will also allow us to: probe the magnetization of low-density environments at the cluster outskirts and in the intergalactic medium; investigate the link between magnetic field and cluster temperature, the gas density (see e.g. Govoni et al. 2017), and the dynamical state of the system; derive the effects of mergers and other sources of magnetization, such as AGN, on magnetic field evolution; constrain models of magnetic-field formation and amplification; and ultimately shed light on the origin of cosmic magnetism.

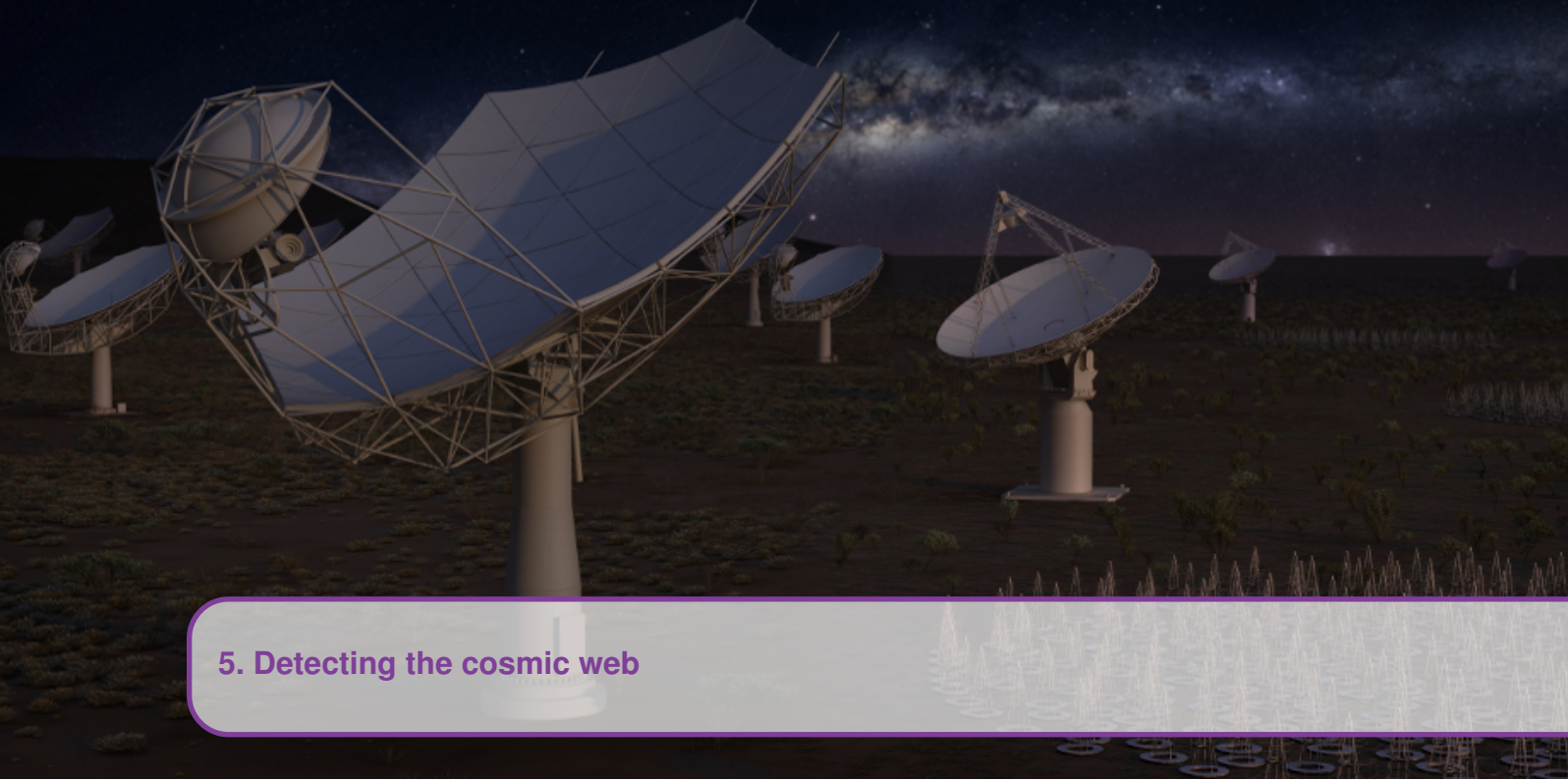

## 5.1 Introduction

The majority of the baryons in the Universe lie in intergalactic space, as predicted by numerical simulations of structure formation in a Lambda Cold Dark Matter ($\Lambda$CDM) cosmology (Cen and Ostriker 1999). Part of this baryonic component has been observed via the absorption of hydrogen and metal lines against bright background sources (particularly those at high redshift), where intergalactic medium (IGM) photoionised filaments with T$\simeq 10^4 - 10^5$ K have been detected in optical surveys. Below $z = 2$, accretion of these baryons onto the deep dark matter potential wells of massive halos induces shocks that heat the Warm-Hot Intergalactic Medium (WHIM) in the cosmic web to temperatures above $10^5$ K, and almost completely ionize the H atoms. In addition, galactic outflows powered by stellar and AGN feedback enrich the IGM. These processes severely hamper the absorption method for the detection of this important baryonic component, which is expected to contain up to ∼40% of the baryons in the Universe (see e.g. Nicastro 2016, and references therein). The presence of this hotter phase of baryonic material has only recently been firmly detected via absorption features in the X-ray spectrum of a bright quasar (Nicastro et al. 2018).

An alternative direct detection of the expected thermal (Bremsstrahlung) and non-thermal (synchrotron) radiation from the WHIM is at the moment extremely challenging, due to to the low densities ($\sim 10^{-5}$ cm$^{-3}$), temperatures ($10^5 - 10^7$ K) and magnetic fields ($10^{-9} - 10^{-7}$ G) expected within the cosmic web. In the following sections we describe the SKA-*Athena* prospects for detecting the WHIM through its expected thermal and non-thermal continuum emission (Sect. 5.2), and through joint radio and X-ray spectral studies (Sect. 5.3).

## 5.2 Continuum emission from the cosmic web

In the current $\Lambda$CDM scenario, the peripheral regions of galaxy clusters are still subject to high-speed matter accretion, which should drive the continuous conversion of infall kinetic energy into thermal and non-thermal energy components (e.g. Reiprich et al. 2013). The consequent presence of both warm thermal gas and magnetized plasma within the WHIM provides three direct observables in radio and in X-rays: diffuse synchrotron radiation, thermal bremsstrahlung emission (Sect. 5.2.1) and Faraday RM (Sect. 5.2.2).

### 5.2.1 Detecting the WHIM through its shocked gas

The exciting possibility of detecting both thermal and non-thermal emission with *Athena* and SKA will allow us to better constrain the (as yet unknown) plasma conditions at strong accretion shocks, in a hitherto poorly known environment. Moreover, this will enable the typical magnetization of the WHIM to be constrained, which according to several scenarios may encode information about the origin of extragalactic magnetic fields (e.g. Vazza 2016). In this context, we outline below the results of large cosmological simulations of extragalactic magnetic fields, recently produced by Vazza and collaborators using the *ENZO* code (Bryan et al. 2014) and tailored to predicting the chances of observing the shocked WHIM around galaxy clusters with SKA1-LOW and *Athena*.

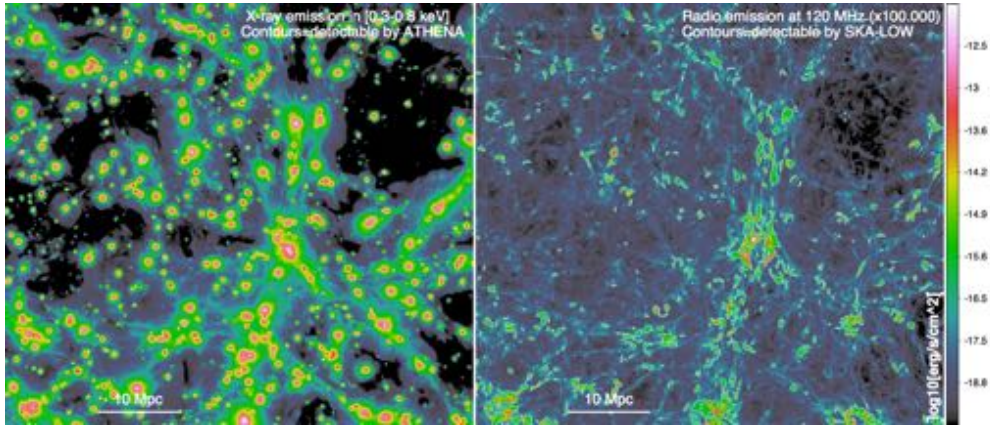

Figure 5.1: $50^2$ Mpc$^2$ maps of the intrinsic X-ray (*left*) and radio (*right*) emission obtained through the simulations of Vazza and collaborators. Contours indicate the regions detectable by SKA and *Athena*.

Table 5.1: Percentage of detected area covered by different simulated objects.

|  | *Athena* | SKA1-LOW | *Athena* ∩ SKA1-LOW |
|---|---|---|---|
| galaxy clusters | 50% | 19% | 11% |
| outskirts | 10% | 19% | 1.4% |
| filaments | 0.1 % | 1.8% | 0.01% |

We focus here on a 200 Mpc$^3$ volume simulated using $2400^3$ cells and dark-matter particles, run on the Piz-Daint supercluster at CSCS-ETHZ (Lugano). A simple uniform initial magnetic field seed of $B_0 = 1$ nG (comoving) at $z = 38$ was assumed, which is at the level of the upper limits on primordial magnetogenesis inferred from the analysis of the CMB (e.g. Subramanian 2016, and references therein).

The radio emission from shock-accelerated relativistic electrons was computed as in Vazza et al. 2015b, while for the X-ray emission a single temperature plasma in ionization equilibrium within each 3D cell was assumed, adopting the APEC emission model (e.g. Smith et al. 2001). As this simulation does not contain any treatment of chemistry, a uniform metallicity of $Z = 0.3 Z_\odot$ was assumed everywhere, and both the continuum and line emission from each simulated cell were considered. After computing the radio and X-ray emission in the rest frame, redshift corrections were applied and the emission maps were processed further in order to simulate mock observations sensitive to the low-$z$ cosmic web with SKA and *Athena* ($d_L \approx 200$ Mpc, i.e. $z = 0.045$). In detail, the simulations included:

- a mock observation with SKA1-LOW at 120 MHz as in Vazza et al. 2015b, including the effects of baseline sampling, finite beam size and considering the (confusion-limited) sensitivity of $0.03 \times 10^{-6}$ Jy/arcsec$^2$;
- mock long X-ray exposures (1 Megasecond) with *Athena*/X-IFU in the $0.3 - 0.8$ keV energy range, with a typical binning into $5' \times 5'$ pixels, in order to maximize the signal to noise ratio over the entire FOV of *Athena*/X-IFU. Background counts from the instrument, as well as from astrophysical background, and X-ray absorption relative to a high galactic latitude ($n_H = 2 \times 10^{20}$cm$^{-2}$) were included.

The resulting maps of the intrinsic X-ray and radio emission from $50^2$ Mpc$^2$ projected are shown in Fig. 5.1.

This approach allows us to assess the fraction of galaxy clusters, cluster outskirts and filaments expected to be detectable by SKA1-LOW and *Athena*. The identification of the virial volume of clusters, of their outskirts (defined as the projected area between $R_{100}$ and $2R_{100}$, with the additional masking of gas clumps), and of filaments, is based on techniques presented elsewhere (Gheller et al. 2016). Those regions whose radio flux is $\geq 3$ times the confusion level of SKA, or whose X-ray flux is $\geq 3$ the S/N ratio for *Athena*, were considered as "detectable" and are shown by the white contours in Fig. 5.1.

Table 5.1 gives the surface fraction of the simulated maps that is detectable using *Athena*, SKA1-LOW, or both. While a large fraction of the virial region of clusters will be detected with 1Ms exposures with *Athena*/X-IFU, SKA1-LOW will detect only a lower fraction of the virial volume, where it is crossed by powerful enough shocks. On the other hand, SKA1-LOW surveys should detect a much larger fraction of filaments compared to *Athena*, provided that the magnetic field intensity is of the order $\sim 10 - 100$ nG.

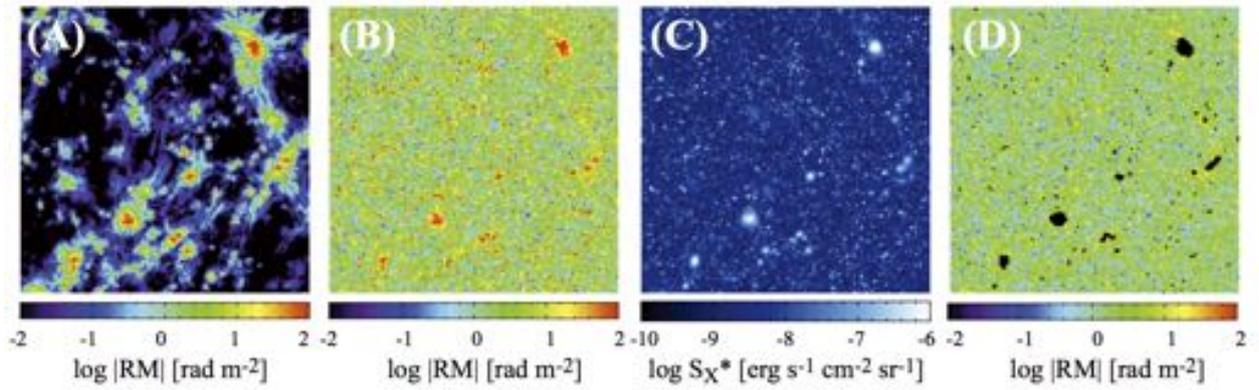

Figure 5.2: From left to right: (A) RM map of 200 deg² determined for the region up to $z = 0.1$, (B) As for (A), but integrated out to $z \sim 5$, (C) As for (B) but plotting the X-ray surface brightness $S_X^*$, (D) As for (B), but excluding pixels with $S_X^* \geq 10^{-10}$ erg/s/cm²/sr.

These results indicate that cluster outskirts represent an important case of synergy between the two instruments, with an estimated $\sim 1.4\%$ of detected regions by both instruments. These multiple detections should often occur in the wake of dense and hot large-scale accretions, along the collision axis of ongoing merger events. Despite the low percentage of multiple detections, this suggests the possibility of revealing the shocked WHIM gas beyond the virial volume of $\sim 2-3$ galaxy clusters for every Gpc³ volume.

This important SKA/*Athena* synergy will allow us to perform WHIM science with *Athena* not only using absorption lines towards high-z powerful sources (which will provide only integrated information about the low density WHIM; Sect. 5.3), but also using *Athena*/X-IFU measurements of emission line properties to characterize the thermal and chemical properties of regions undergoing powerful enough accretion. In particular, robust detections will allow us to constrain the Mach number of these strong shocks, and in combination with SKA detections (in continuum emission with SKA1-LOW and also possibly in polarisation with SKA1-MID) enable measurement of the electron acceleration efficiency in the WHIM phase.

This possibility makes it essential to plan for the synergistic investment of long ($\sim$ Ms) exposures with *Athena* of a few promising targets, based on the guidance of radio detections expected to be delivered several years in advance by SKA1-LOW.

### 5.2.2 Detecting the WHIM through Faraday rotation

Using a model of the intergalactic magnetic field based on MHD turbulence simulations (Ryu et al. 2008), Akahori and Ryu 2010 studied RMs through filaments of galaxies and found that RM is sensitive to dense parts of filaments (thus having a bias similar to the X-ray emission). The Faraday rotation of emission passing through a filament is a random walk process. The root mean square (rms) value of RMs for the local Universe is $\sim 1$ rad m⁻² (panel A of Fig. 5.2). Akahori and Ryu 2011 extended this work by integrating over filaments up to a redshift of 5, and found that the rms value of RMs reaches $\sim$ several -10 rad m⁻² (panel B of Fig. 5.2). They also found that the structure function of the RMs is $\sim 100–200$ rad² m⁻⁴ down to 0.2° scales.

The study of RM benefits from exploiting the presence of bright background polarized sources, and the SKA's unprecedented sensitivity will deliver dense RM grids on the sky (Sect. 4.4). The grid density will be further increased by including linearly-polarized Fast Radio Bursts (FRBs), which will additionally provide the column density of free electrons via their dispersion measure (Akahori, Ryu, and Gaensler 2016; Ravi et al. 2016). A longstanding issue in exploring the RM of the WHIM is to extract its contribution from the total observed RM, which includes multiple magnetized components along the line of sight: in addition to the WHIM, this includes the magnetized plasma associated with the background radio source itself, intervening extragalactic sources, and the Milky Way. The SKA's ultra-wide-band polarimetry will allow us to partly distinguish such components by depolarization studies (Akahori, Gaensler, and Ryu 2014) and/or Faraday tomography (Akahori et al. 2014). Dense RM grids will provide the accurate structure functions needed to reduce the Galactic foreground contribution, and so unveil the characteristic structures of the intergalactic magnetic field (Akahori et al. 2013).

In an extracted RM map of the cosmic web with sufficiently large field-of-view [1], sightlines through galaxy clusters contribute significantly to the statistics of the RMs. It has been confirmed that the pixels influenced by clusters can be satisfactory removed if we identify X-ray clusters above $10^{-10}$–$10^{-8}$ erg/s/cm$^2$/sr, which is roughly close to the detection limit of current X-ray facilities (Akahori and Ryu 2011) (C and D panels of Fig. 5.2). Nevertheless, unidentified X-ray-faint clusters are a potential limitation for the detection of RM associated with the WHIM. Therefore, *Athena*'s high-sensitivity observations will be very important to achieve successful cluster subtraction so as to enable discovery of the WHIM in RM maps.

---

[1] Roughly 100 deg² may be required so as to overcome cosmic variance. The North and South Galactic caps are ideal target fields, since the Milky Way foreground RM is expected to be minimum.

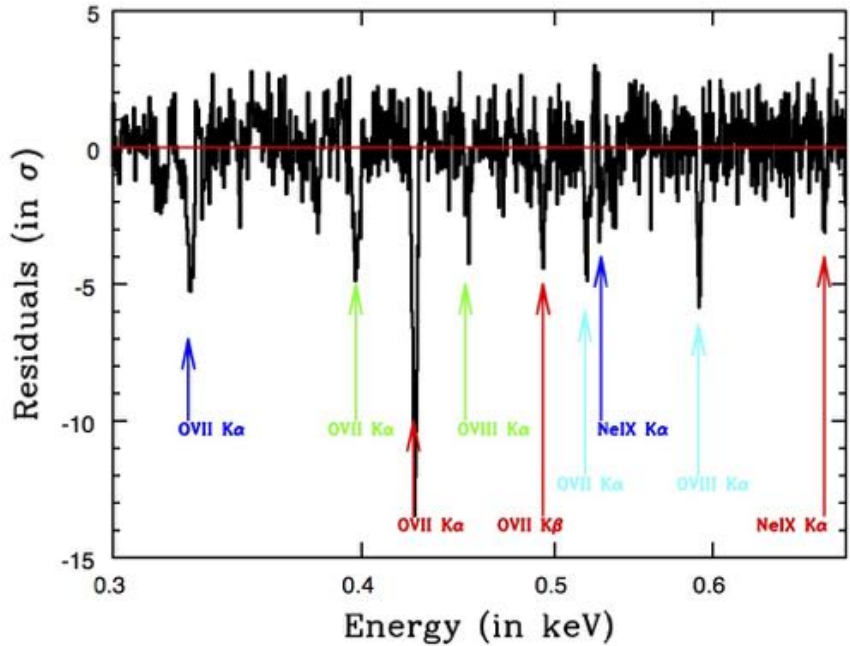

Figure 5.3: 100 ks *Athena*/X-IFU mock spectrum of the $z = 0.851$ blazar 3C 454.3: 4 different WHIM systems are clearly detected, all in multiple metal lines.

The SKA's unique survey speed and wide frequency coverage will allow us to achieve an all-sky polarimetric survey and all-sky RM grids (which are the highest priority science objectives for cosmic magnetism, see Johnston-Hollitt et al. 2015). From this, according to previous work (Akahori, Gaensler, and Ryu 2014; Akahori et al. 2014), we will be able to build an added-value catalog of WHIM candidates that could be high-priority targets to be observed with *Athena*, unveiling the density of the WHIM, which can be compared to the density derived from FRBs. Such an analysis would enable direct derivation of the WHIM magnetic field strength from RM measurements.

## 5.3 The WHIM and the "roaming" baryons

As detailed above, below $z \sim 2$ the IGM is so hot as to be highly ionized, and the sensitivity of current UV and X-ray spectrometers is not sufficient to detect the broad and shallow HI-Ly$\alpha$ line or the dominant OVII-OVIII metal lines, respectively, against bright background AGN. However, very recently Nicastro et al. 2018 obtained a high quality X-ray spectrum of the bright quasar 1ES 1553+113 at z>0.4, by using the longest (1.75 Mega-seconds) observation performed so far on a single target with the XMM-Newton spectrometer (RGS). Two main absorption lines of the O VII He$\alpha$ were detected with high significance, and their spatial position within a galaxy overdense region and their number density agree well with theoretical expectations, suggesting they are originating from the hot-phase of the WHIM (T$\approx 10^{5.7} - 10^{6.3}$ K). The derived mass density of baryons is in the range 9-40% of the total baryon density, potentially accounting for the missing baryons. This important discovery opens the door to the study of a range of background sources across the sky to confirm whether these findings are truly universal and to further investigate the physical state of the WHIM, which will be possible thanks to the high sensitivity of *Athena* .

The ubiquity of metal lines detected so far in IGM filaments at higher $z$, and the deficit of baryons (compared to the universal $\Omega_b/\Omega_{DM}$ ratio) found in today's galaxies, suggest that in the past the gas in galaxies has been enriched and then injected into the IGM by powerful AGN and star formation (SF) events. Yet, direct evidence for baryon removal from galaxies is scarce and it is unclear when and how the enrichment of the IGM took place. The other missing piece of the cosmic puzzle is the recent SF activity of galaxies along the blue sequence, which can be explained only if the IGM filaments can cool, flow into the galaxy and feed SF. Hence, we expect to find cold condensations of neutral gas inside hot filaments as these approach and enter the potential wells of small- or medium-sized galaxies.

Numerical simulations have shown that streams of cold gas are unable to form around massive halos, where a hot corona is more likely to develop and SF is quenched. To date HI surveys have been able to image only low-mass clouds in the proximity of Local Group spirals (M31 and M33), or massive clumps in galaxy outskirts, leaving open the possibility of tidal interaction remnants or disk-halo circulation. To test if the 40-50% of currently missing baryons are indeed trapped in a warm-hot phase of the cosmic web, as suggested by the recent discovery (Nicastro et al. 2018), and are continuously roaming in and out of galaxy potential wells (and so possibly feeding SF), X-ray spectroscopy with *Athena*, and radio imaging-spectroscopy with the SKA are both crucial.

Below we summarize the major benefits of exploiting this synergy for our understanding of the roaming baryon phenomenon in the local Universe ($z < 1$).

- Current hydrodynamical simulations predict that about 10 OVII absorbers per unit redshift are expected along a random line of sight, down to the *Athena* equivalent-width sensitivity of EW(OVII)$> 0.14$ eV. Following *Athena* detections, SKA1 imaging at 21 cm is expected to allow the identification of galaxies in the neighborhood of the filaments down to HI masses of order $2 \times 10^9 \, M_\odot$ (M33 like) up to z=0.2.

- The surroundings of the nearest galaxies, expected to be associated with the WHIM, can be imaged with high 21-cm sensitivity to search for cold streams connecting the disk to the hotter filaments. The expected HI column densities of this gas are across the HI-HII transition region i.e. between $10^{17}$ and $10^{19}$ cm$^{-2}$ and can be imaged with SKA1-MID at a resolution of 1.5 arcmin in 10 hours. For massive halos, where a hot accretion mode is more likely, the HI counterpart to the *Athena*'s metal signatures can be acquired with future FUV instrumentation, through the detection of broad and faint Ly$\alpha$ lines.

- For bright background targets, *Athena* will be able to detect several OVII as well as lower/higher ionization oxygen and neon lines (e.g. Fig. 5.3), which will allow temperature and ion abundance determinations up to $z = 1$. If cold-stream condensations are embedded in the hot filaments, a follow up search for OI or OII lines and HI absorption with SKA will provide estimates of the metallicity of the "cool" part of the systems. The hotter part of the WHIM system, spatially co-existing with its embedded cold part, will likely have the same chemical abundance, and so the extrapolation of such measurements enables estimates of the local cosmic gas density.

- In the soft X-rays, *Athena* will reach line centroid precisions $\geq$100-200 km s$^{-1}$. Therefore, *Athena* studies of the WHIM in the local Universe will greatly benefit from the identification of associated cold-gas interfaces detectable with the SKA at a resolution of 10-20 km s$^{-1}$. The kinematics of cooler interfaces, present also in outflows, will help to shed light on the role of shocks in the circumgalactic medium and diffuse IGM.

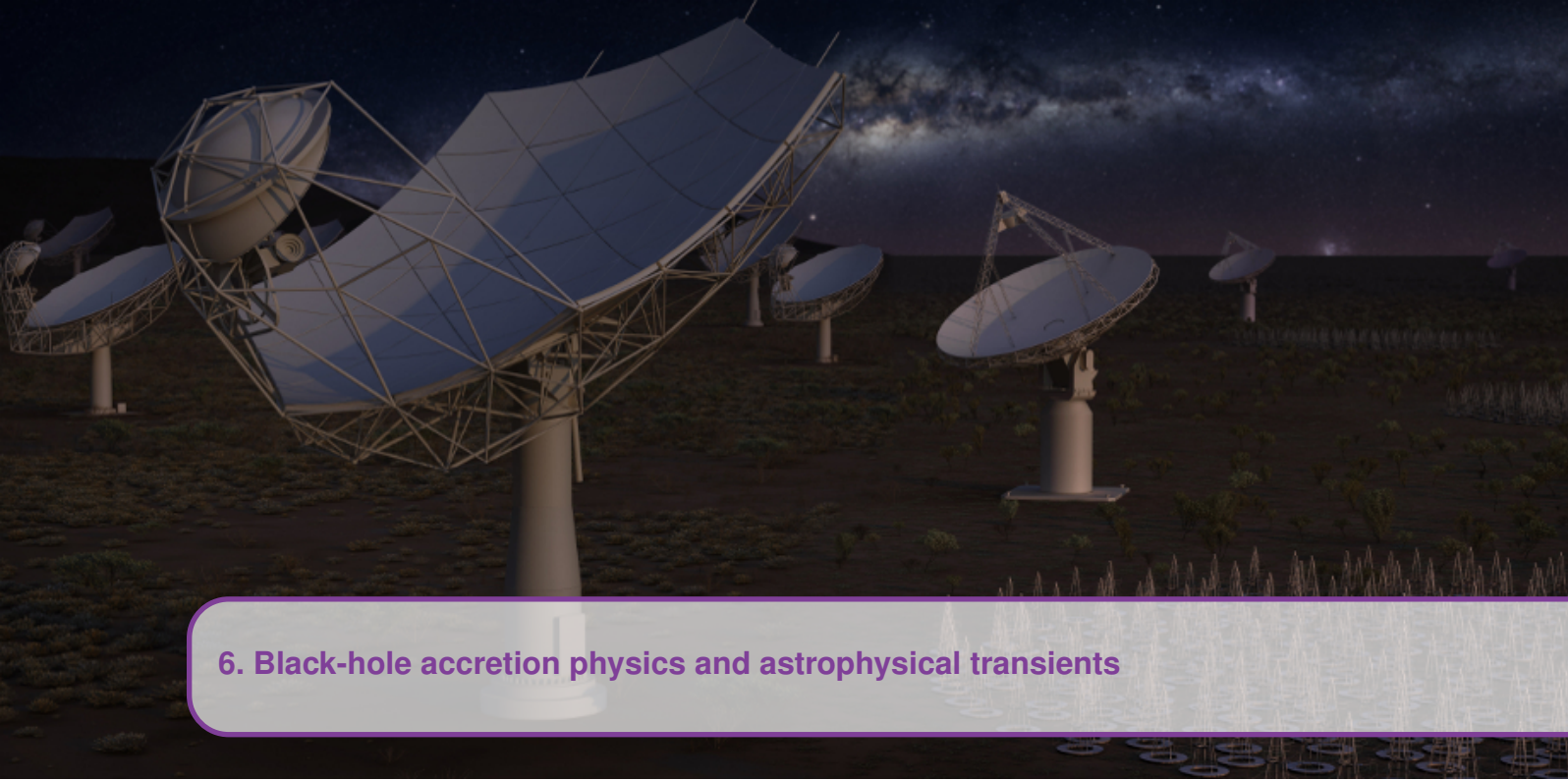

## 6.1   Introduction

The process of accretion onto black holes and neutron stars, the most efficient engine for the liberation of gravitational potential energy in the universe, reveals itself via strong X-ray emission. It is furthermore very often coupled to a relativistic outflow, which can be the dominant channel for the feedback of liberated energy to the local environment, and which emits strongly in the radio regime. In this way coupled radio and X-ray observations are extremely well suited to studying the relationship between accretion 'modes' (rate, geometry, optical depth, temperature) and the power and frequency of this relativistic feedback. Future combined *Athena* and SKA observations will be able to probe this coupling in unprecedented detail.

## 6.2   X-ray binary winds and jets

Tight correlations, spanning many orders of magnitude, are observed between the properties of BHs and their galactic bulges (Ferrarese and Merritt 2000; Gebhardt et al. 2000; Häring and Rix 2004; Kormendy and Ho 2013), indicating the presence of a mechanism, so-called feedback, which couples accretion power (Shakura and Sunyaev 1973) to the host galaxy and its bulge. The feedback process is still poorly understood, remaining one of the outstanding problems in modern astrophysics. However, observational evidence for feedback is clearly seen in clusters and groups of galaxies (Boehringer et al. 1993; Churazov et al. 2000; Fabian 2012). There is also a clear correlation between AGN activity and other properties of nearby galaxies (Kauffmann and Haehnelt 2000). It is thought that feedback occurs through outflows of two different kinds: i) collimated and relativistic ejections, in the form of jets; ii) un-collimated winds, which are non-relativistic. Despite the mounting observational evidence for winds and likely wind-driven feedback (Arav, Li, and Begelman 1994; Crenshaw et al. 1999; McKernan, Yaqoob, and Reynolds 2007; Tombesi et al. 2010; Feruglio et al. 2010; Sturm et al. 2011; Maiolino et al. 2012), we still lack a complete understanding of their key physical properties. What is the wind launching mechanism? How are winds and jets connected? How are winds and jets evolving during the intermediate states? How are they related to the accretion disc physics? What is the outflow evolution? Simultaneous SKA-*Athena* observations can answer these questions by looking at winds in XRBs.

It is widely recognised that, in many respects, XRBs are scaled down versions of AGN (Merloni, Heinz, and di Matteo 2003; McHardy et al. 2006). The link is so profound that, in the past, XRBs have been used as a template to: i) develop the models of accretion that are currently applied to AGN (Shakura and Sunyaev 1973); ii) demonstrate the identical nature of jets (Heinz and Sunyaev 2003; Merloni, Heinz, and di Matteo 2003) in XRBs and AGN and; iii) demonstrate the identical nature of their central engine (Merloni, Heinz, and di Matteo 2003; McHardy et al. 2006). One of the major steps forward in our understanding of XRBs has been the recognition of a picture where XRBs transit through a hysteresis loop in their X-ray properties during their outbursts, and that their relative position in this cycle determines the mode and power of their feedback (Fig. 6.2). Indeed, XRBs are observed in the hard state at the beginning and at the end of an outburst. Moreover, they can transit to the soft state at high luminosities and back to the hard state, when the luminosity decreases to a few per cent of the Eddington luminosity. It was initially recognised that hard states are associated with an optically thin and geometrically thick accretion flow, a compact-steady jet, large amplitude variability and the presence of quasi-periodic oscillations. The soft states are, instead, dominated by a standard (Shakura-Sunyaev) accretion disc, characterised by quenched jet emission and by low levels of variability. A series of intermediate states are also observed at the transit from the canonical hard to the soft state. During such intermediate states, XRBs show ballistic jet ejections, peculiar (type B) quasi-periodic oscillations and (sometimes) super-Eddington phases. Currently, the evolution in time and the connections between this complex phenomenology is not completely understood. Recent work has highlighted the importance of winds as an additional key component of the accretion system. Indeed, the past decade saw the discovery of the importance of winds in XRBs. Winds in XRBs originate from the accretion disc, they have an equatorial geometry, they are ubiquitous and they carry away more matter than is accreted (Ponti et al. 2012; Ponti, Muñoz-Darias, and Fender 2014; Ponti et al. 2015b; Ponti et al. 2015a; Ponti et al. 2016). Thanks to these discoveries, it is now well established that winds must have a close connection with the other components of the accretion-ejection system. Intriguingly, winds are observed only in the soft state, when a standard alpha-disc is present and the jet is quenched. Simultaneous SKA-*Athena* monitoring campaigns during the outbursts of XRBs (and, in particular during the intermediate states) will allow us to pin down the underlying physical processes and to determine the link between the wind, jets and accretion disc. Indeed, *Athena*/X-IFU will provide us with key measurements of the wind properties as well as of the accretion flow and its variability behaviour. At the same time, the unprecedented sensitivity in the radio band provided by SKA, will allow us to pin down the simultaneous behaviour of the jet, the third key ingredient of the accretion-outflow system. Deep and frequent monitoring of XRBs during the crucial intermediate states, while XRBs undergo major variations of the accreting systems, such as: i) the disappearance of the jet and ballistic mass ejections; ii) variations in the hot accretion flow; iii) quasi periodic oscillations; iv) variations in the appearance the wind; v) variations of the standard Shakura-Sunyaev accretion disc; etc., will be enabled by the fast triggering capabilities of *Athena* as well as the large field of view of SKA. This will be crucial to solve the connections between outflows (jet and winds) and the accretion disc. By measuring the energetics of each inflow-outflow component, we will trace the balance of power in accreting XRBs and by extension in AGN. In particular, this will allow us to trace the evolution of the wind and compare it to the evolution of the jet and accretion disc, to investigate the connection with super-Eddington phases, as experienced in some ultra luminous X-ray sources and TDEs, as well as to test transient events (such as baryonic loading of the jet; Kotani et al. 1994; Díaz Trigo et al. 2013). *Athena* observations will allow us to pin down the wind launching mechanism (Ponti et al. 2012; Ponti et al. 2016), to measure the radius at which the wind is launched from the accretion disc, and to determine the wind mass outflow rate, kinetic luminosity and the amount of angular momentum removed by each component of the wind. The SKA observations will allow us to trace the simultaneous evolution of the jet. By tracing the evolution of all components, the underlying physical link between accretion and outflows should emerge. Ultimately, this will allow us to measure the quantity of mass, momentum and energy transferred into the environment.

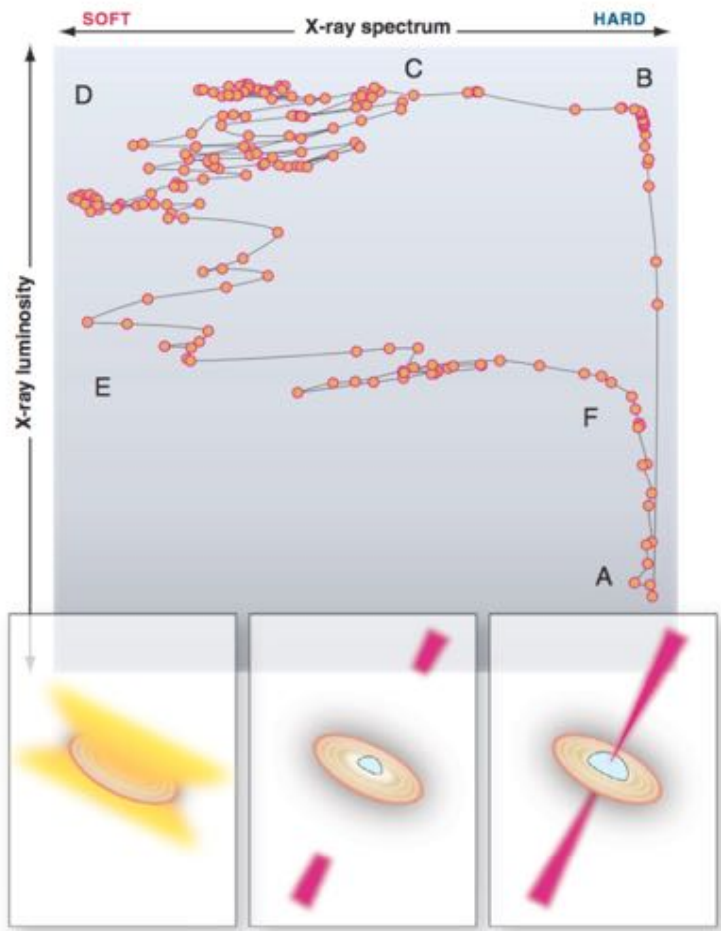

Figure 6.1: A simplified picture of the connection between accretion states, jets and winds in accreting stellar mass black holes, from Fender and Belloni 2012. The upper panel shows the track followed in a typical black hole binary outburst, in an anticlockwise direction from A to F, displaying clear hysteresis. The x-axis (abscissa) of this panel is the X-ray colour, in the sense that to the right are hard, Comptonised spectra, and to the left are softer, blackbody-like spectra. The y-axis (ordinate) is the X-ray luminosity. Modes of feedback show a very clear connection to position in the hysteresis cycle, most strongly with X-ray spectrum: hard states are associated with steady jets and weak or non-existent winds, and vice-versa in soft states. Intermediate states *may* show a mix of both, at times, and the high luminosity transition from hard to soft states (B → C → D) is usually associated with a bright radio flare and resolved, relativistic, ejections.

By building a deep understanding of the accretion-ejection process in the simpler case of XRBs, it will be possible to lay down the foundations for classifying and understanding the zoo of wind phenomena present in AGN. Indeed, in AGN the phenomenology is expected to be more complex, because more wind-launching mechanisms are at work (such as line driven winds or torus evaporation; Krolik and Kriss 2001; Proga 2005). The acquired knowledge on XRBs will provide us with a template to understand AGN winds and feedback. This will be instrumental to solve a serious uncertainty at the core of our current picture of how structure in the Universe evolves. Indeed, despite the fact that AGN outflows can control and shape the formation and evolution of galaxies in the Universe, the current implementations of AGN feedback in numerical models are typically simple prescriptions that do not yet incorporate the complex behaviour of real accreting systems.

## 6.3 X-ray binaries at the highest accretion rates

In stellar-mass black holes (BHs) and neutron stars (NSs) accreting from a donor star, the gravitational power of the infalling matter is partly released directly via radiation (mostly X-ray photons from hot gas), and partly converted to bulk kinetic energy of relativistic jets, which are detectable via their radio synchrotron emission. The relative fraction of power released in the two channels depends on the nature and properties of the compact object (BH: mass and possibly spin; NS: magnetic field), and on the rate of mass accretion. Contemporaneous (ideally, simultaneous) X-ray and radio measurements are needed to model the variable output power ratio.

The current generation of X-ray and radio facilities has greatly advanced our understanding of inflows and outflows at low or moderate (sub-Eddington) accretion rates, with spectral and timing studies of galactic XRBs and their jet activity (Gallo, Fender, and Pooley 2003; Fender, Belloni, and Gallo 2004; Falcke, Körding, and Markoff 2004). In contrast, the physics and phenomenology of super-Eddington accretion remain largely untested. This hampers our understanding of many astrophysical problems: for example, the contribution of accreting BHs to cosmic reionization; the growth timescale and feedback power of seed nuclear BHs; the observable rate and properties of tidal disruption events; the mass distribution of non-nuclear BHs. Stellar-mass BHs and NSs in the super-Eddington regime provide the best testing ground for the physics of super-critical accretion. However, such sources are rare in the local Universe: at most a few per galaxy. We need to identify and spatially resolve sources in a large region of sky up to a few 10s of Mpc, in order to have a sufficiently large sample, covering different sub-populations and different environments, and including enough sources with variable/transient properties. At those distances, *XMM-Newton*, *Chandra* and the VLA do not have enough sensitivity for reasonable exposure times.

Incidentally, although some members of the empirical class of Ultraluminous X-ray Sources (ULXs) are now solidly associated with stellar-mass super-Eddington accretors (see Sect. 6.4), a compact object can be in the super-critical regime without appearing as a ULX. This may happen for example for systems seen almost edge on, or if they are shrouded by optically thick winds (or both). Other sources may be surrounded by ionized bubbles, evidence of recent super-Eddington power, but may currently be in a faint state.

There are at least five mechanisms for radio emission from super-critical stellar-mass sources:

a) Magnetohydrodynamic (MHD) simulations (Kawashima et al. 2012; Jiang, Stone, and Davis 2014; Narayan, Sadowski, and Soria 2017) predict that a continuous, relativistic jet is launched and collimated inside the low-density polar funnel; in some models, part of the jet power is extracted from the spin of the compact object. The predicted 5-GHz radio luminosity of the flat-spectrum compact core is $10^{32}$ erg s$^{-1}$ $\lesssim L_{5\text{GHz}} \lesssim 10^{33}$ erg s$^{-1}$, consistent with the core luminosity in the Galactic super-Eddington source SS 433 (Vermeulen et al. 1993). No self-absorbed, compact radio jet has ever been detected in any super-Eddington source in external galaxies: the expected flux density is $\lesssim$ a few $\mu$Jy, achievable with SKA1-MID for distances $\lesssim$ a few Mpc (Fig. 6.2);

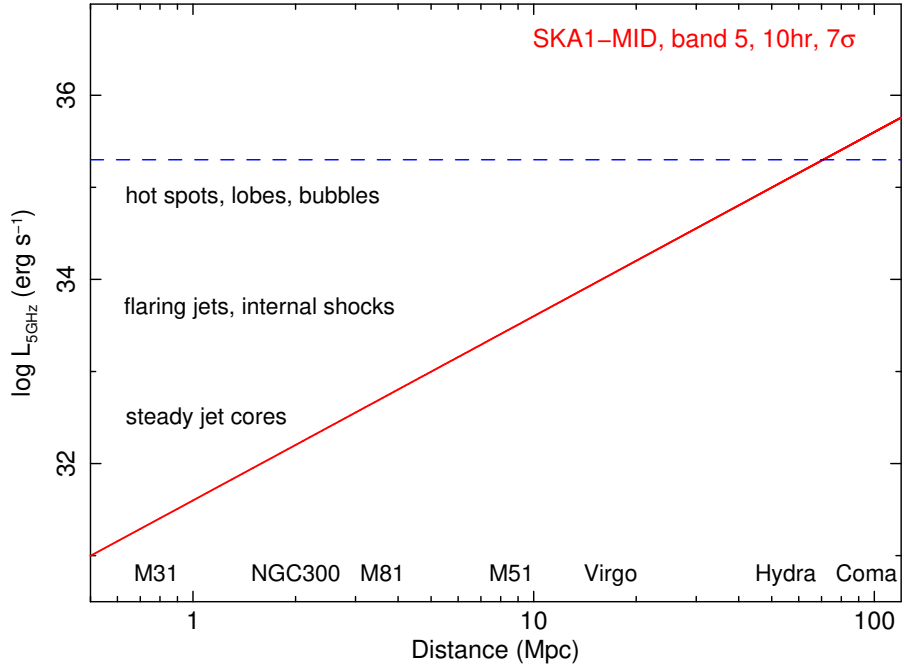

Figure 6.2: Expected SKA1-MID 7$\sigma$ detection limit for a 10-hr integration, in the frequency band 5 (corresponding to $\approx 3\mu$Jy rms in 1 hr, solid red line), plotted as the minimum detectable radio luminosity as a function of distance. Distances of characteristic galaxies and clusters are indicated along the X axis. We labelled the characteristic 5-GHz synchrotron luminosity of various types of physical structures associated with jetted super-Eddington stellar-mass accretors. The horizontal dashed blue line marks the approximate maximum luminosity ($L_{5\text{GHz}} \approx 2 \times 10^{35}$ erg s$^{-1}$) observed from super-Eddington jet bubbles in the local universe; we speculate that it corresponds to a maximum achievable jet power $\approx 10^{40}$ erg s$^{-1}$ in stellar-mass sources.

b) flaring jets, consisting of a series of adiabatically-cooling, discrete ejecta. The source is still spatially unresolved, but its radio spectrum is optically thin and rapidly varying. Expected luminosities are $\sim 10^{33}$–$10^{34}$ erg s$^{-1}$. A good example is the jetted ULX in the Holmberg II galaxy ($d \approx 3.2$ Mpc), peaking at $L_{5\text{GHz}} \approx 10^{34}$ erg s$^{-1}$ (Cseh et al. 2014; Cseh et al. 2015b). With *Athena* and the SKA we will discover similar sources as far as the Virgo cluster, and (for the nearest sources) we will correlate the radio and X-ray variability, to explain what changes in the accretion flow cause the ejections;

c) internal shocks, caused by variations in the velocity and density of the ejecta; shocks and consequent electron re-acceleration occur when faster material catches up with a slower part of the flow (Rees 1978; Kaiser, Sunyaev, and Spruit 2000). Bright knots consistent with this scenario are seen in the candidate super-Eddington microquasar NGC 300 S9, with $L_{5\text{GHz}} \approx 10^{34}$ erg s$^{-1}$ (Urquhart et al., in prep.);

d) hot spots and lobes, analogous to the structures of FRII radio galaxies (Begelman, Blandford, and Rees 1984; Begelman and Cioffi 1989). Large-scale ($\approx 100$–$300$ pc in diameter), jet-powered, shock-ionized bubbles are a tell-tale signature of the most powerful stellar-mass BHs, such as NGC 7793 S26 (Soria et al.

2010), M 83 MQ1 (Soria et al. 2014), IC 342 X-1 (Cseh et al. 2012). The interaction of the jets with the interstellar medium produces optically thin synchrotron emission from the hot spots and their back-flow, with a luminosity up to $L_{5GHz} \sim 10^{35}$ erg s$^{-1}$. Multi-band studies of the shock-ionized gas (Fig. 6.3) provide the most reliable method to measure the jet power in such sources (Pakull, Soria, and Motch 2010). With SKA1-MID we will be able to discover dozens of similar structures as far as ~50 Mpc;

e) free-free emission from the warm component ($T \approx 20,000$ K) of the shock-ionized plasma inside a wind- or jet-inflated bubble. The free-free radio luminosity is related to the optical Balmer line luminosity (Caplan and Deharveng 1986); the latter is roughly proportional to the input kinetic power (Dopita and Sutherland 1995). Combining the two relations, we expect a diffuse thermal bremsstrahlung luminosity $L_{5GHz} \approx$ a few $10^{33}$ erg s$^{-1}$ for the most powerful jet-inflated bubbles.

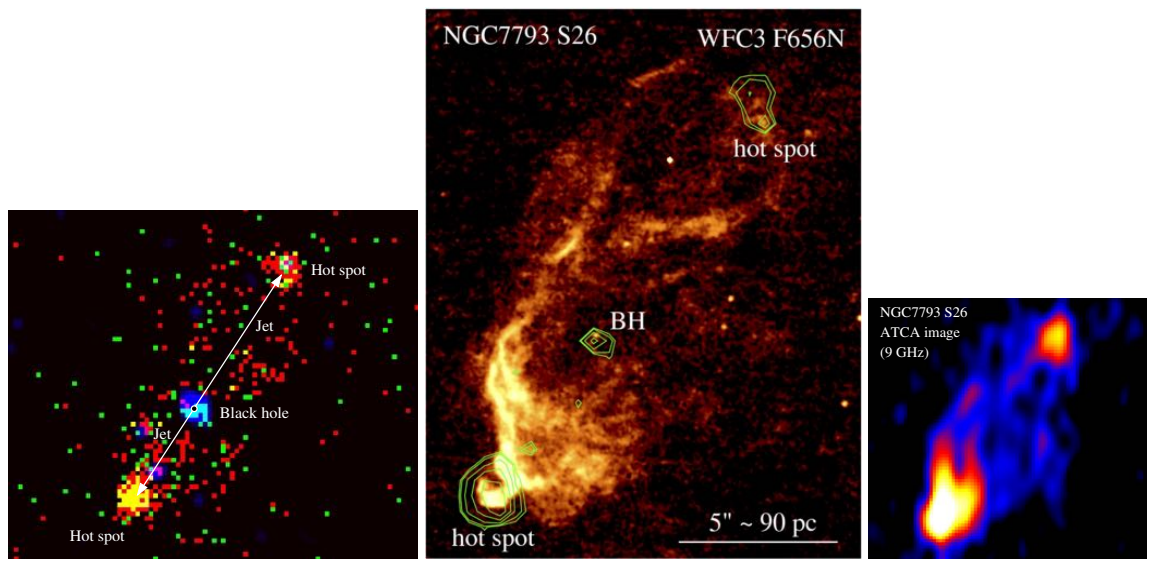

Figure 6.3: The bubble of ionized gas around the (likely) super-Eddington BH S26 in the galaxy NGC 7793 shows how such sources heat, ionize and sweep up the surrounding interstellar medium; combined X-ray, optical, and radio studies are required to model the interaction. *Left: Chandra* true-colour image (red = 0.3–1.0 keV; green = 1.0–2.0 keV; blue = 2.0–7.0 keV). *Middle*: optical *HST*/WFC3 image in the narrow-band Hα filter; green contours mark the location of the X-ray core (hard power-law spectrum) and of the X-ray hot spots (soft thermal spectrum). *Right*: ATCA map at 9 GHz; at a radio luminosity of $\approx 2 \times 10^{35}$ erg s$^{-1}$, NGC 7793 S26 is the most luminous jet-powered bubble known to-date in the local universe. (Adapted from Pakull, Soria, and Motch 2010 and Soria et al. 2010.)

In summary, understanding the physics of inflows and outflows (jets and winds) in the super-critical regime requires synergies between *Athena* and the SKA, in particular, high-resolution X-ray spectroscopy and simultaneous radio and X-ray monitoring. The increased sensitivity will enable us to probe a deeper volume of space, will provide a more representative sample of super-Eddington sources, and permit studies of flux variability over a larger luminosity range.

## 6.4 Ultraluminous X-ray sources

ULXs are defined very simply as off-nuclear X-ray sources that have X-ray luminosities exceeding the Eddington luminosity (L~$10^{39}$ erg s$^{-1}$) for a stellar mass black hole of reasonable (~10 $M_\odot$) proportions. For a long time, they have been assumed to contain accreting black holes, which were thought perhaps to be of intermediate mass ($10^{2-5}$ $M_\odot$), so as to explain the very high luminosities without invoking complicated scenarios such as super-Eddington accretion. However, the first results with *NuSTAR* (e.g. Bachetti et al. 2013; Walton et al. 2013), favoured a scenario in which these systems contain stellar mass black holes accreting above the Eddington limit. More recently, three ULXs have been shown to contain accreting neutron stars (Bachetti et al. 2014; Fürst et al. 2016; Israel et al. 2017), via the detection of X-ray pulsations; this confirms that at least some ULXs do undergo super-Eddington accretion. A number of authors have proposed that many ULXs may, therefore, contain neutron stars accreting at highly super-Eddington rates (e.g. Koliopanos et al. 2017; King, Lasota, and Kluźniak 2017; Mushtukov et al. 2017). However, in contrast, the extreme ULX ESO 243-49 HLX-1 (e.g. Farrell et al. 2009; Webb et al. 2010) has been shown to contain an Intermediate Mass Black Hole (IMBH) of $\sim 10^4$ $M_\odot$ (e.g. Webb et al. 2012; Godet et al. 2012). This means that ULXs are a very heterogeneous collection of objects. Further observations are required to elucidate their nature.

Whatever the nature of ULXs, their study is extremely useful for understanding the growth of SMBHs. For those ULXs accreting above the Eddington limit, studying them should help us to uncover the physical processes behind these extreme accretion rates that are thought to play an essential role in the formation of SMBHs that appear early on in the Universe, e.g. Mortlock et al. 2011. The SMBHs now found in galaxy centres could grow from IMBHs, or from super-Eddington accretion onto intermediate mass black holes (e.g. Greene 2012; Volonteri 2012). Observing the size and the distribution of the population of IMBHs will help us to probe open questions on IMBHs such as: how are IMBH formed and how do IMBH evolve (e.g. Miller and Colbert 2004)? There are many other open questions concerning super-Eddington accretion (Narayan, Sadowski, and Soria 2017), which could also be studied through observations of ULXs. These questions include:

- How viable is super-Eddington accretion?
- What is the geometry of the accretion flow? How does it impact observations as a function of inclination angle?
- How luminous are super-Eddington systems? Are they radiatively efficient?
- How much mechanical energy do super-Eddington disks produce in outflows? What role do the outflows play in feedback?
- How often do super-Eddington disks produce relativistic jets? How do these jets compare with blazar jets?

Observations with the SKA and *Athena* will be essential to elucidate the nature of ULXs. Firstly, the intrinsic power of the ULX can be estimated from the optical emission-line nebulae, created from shock-ionised driven jets, outflows or disc winds and/or because of photo-ionisation from the X-ray and UV emission around the black hole (Pakull and Mirioni 2002; Pakull and Mirioni 2003). Some of these nebulae show a radio counterpart. These radio nebulae can also be used to understand how outflows and photo-ionisation can play a role in the behaviour of the ULX, and how they affect their surrounding environment (Pakull and Mirioni 2002; Pakull and Mirioni 2003; Roberts et al. 2003; Kaaret, Ward, and Zezas 2004). The sensitivity (and resolution) of the SKA will allow us to detect these nebulae in the radio in galaxies out to 10s of Mpc. Using *Athena*/X-IFU observations, we will be able to search for emission and absorption lines, providing evidence for a ULX wind (e.g. Pinto, Middleton, and Fabian 2016) that may be at the origin of the nebula (Cseh et al. 2015b). We will also be able to search for the presence of any compact radio source on the (sub-)arcsecond scale to determine if these nebulae are blown by compact jets emanating from the compact object. With quasi-contemporaneous *Athena* X-ray observations, necessary as stellar mass black holes are known to vary rapidly and HLX-1 also can vary in both X-rays and radio by a factor 20 in a single day (Webb et al. 2012), we can also estimate the mass of the black hole. This can be done using the *fundamental plane of black hole activity*, a correlation relating the X-ray and the radio flux and the mass of the black hole (e.g. Merloni, Heinz, and di Matteo 2003; Körding, Falcke, and Corbel 2006; Plotkin et al. 2012).

HLX-1 was the first ULX to show jets in the same way as X-ray binaries, although they are too faint to detect in the low/hard state and can only be detected when discrete jet ejection events occur during the transition from the low/hard state to the high/soft state (Webb et al. 2012; Cseh et al. 2015a). HLX-1 may be detectable in the low/hard state with SKA and definitely with *Athena*, but closer ULXs (<95 Mpc away) will be much easier to detect than HLX-1 in the low/hard state. Other nearby ULXs have now also been seen to show radio jets close to the high/soft state, like XMMU J004243.6+412519 (Middleton et al. 2013) which shows radio variability on a timescale of tens of minutes, arguing that the source is highly compact and any X-ray and radio observations should be contemporaneous in order to constrain the mass of the compact object and understand the emission mechanisms. In addition, low state/quiescent IMBHs are expected to have steady jets. Madau and Rees 2001 propose that there may be many IMBHs in a Milky Way type galaxy. The SKA will be able to detect almost any quiescent IMBH in our Galaxy ($\sim \mu$Jy). Plotting the radio fluxes against quasi-contemporary X-ray fluxes (the fundamental plane of black hole accretion) will demonstrate their IMBH nature (Maccarone 2004). Identifying where IMBH reside will give us clues as to how they were formed (Miller and Colbert 2004).

In order to identify neutron stars as the compact objects in ULXs, the most clear-cut way is to detect pulsations. *Athena* will be sensitive enough and have a good enough time resolution to detect pulsations, if present, from almost all known ULXs within 100 Mpc. However, we know that ULXs can vary by factors of 100-1000 or more in X-ray luminosity (e.g. the accreting neutron star ULX, M 82 X-2, Brightman et al. 2016). Once the neutron star is no longer accreting at high (super-Eddington) rates, it may be possible to detect the radio pulsations, in the same way as was done for the accreting pulsars e.g. PSR 1023+0038 (Archibald et al. 2009) or PSR J1824-2452I (Papitto et al. 2013). The wide field of view, great time resolution and sensitivity of the SKA will be essential tools to identify (or put limits on) pulsations from known ULXs and help find new pulsating neutron stars that may become ULXs later in their lives.

Combined *Athena* and SKA observations will therefore enable us to distinguish the different types of ULXs that exist, helping us to understand this highly heterogeneous population. This will in turn allow us to study super-Eddington accretion in those systems where it is occurring, and therefore understand the physical mechanism behind it, in order to understand how SMBHs grow. New identifications will enable us to understand the populations and where they reside, which will give us clues to their origin, especially important for the IMBH population for which very few good examples exist today.

## 6.5 Tidal disruption events

TDEs involving stars torn apart by quiescent SMBHs are recognised as potentially powerful tools to study and constrain: i) the SMBH mass function; ii) quiescent SMBH environments; and iii) accretion and outflow physical processes. The first two of these tools result from the ability of disrupted stars to power flares that signpost otherwise undetectable SMBHs, providing information on their masses and spins and on the surrounding medium, where light may be reprocessed before reaching us (e.g., Rees 1988; Phinney 1989; Ulmer 1999; Leloudas et al. 2016; Metzger and Stone 2016). The last point comes about because observed flares and accompanying accretion and outflowing phenomena have lifetimes of weeks to (a few) years, contrary to steady sources (by human standards) like AGN.

In particular, insight into the formation of jets and the role of accretion and/or spin in powering them can be gained by observing and modelling TDEs that show a significant non-thermal component in the X-ray and radio bands. These *jetted* TDEs are a relatively newly discovered class, observed for the first time by *Swift*/BAT in 2011 (e.g. Burrows et al. 2011; Zauderer et al. 2011; Cenko et al. 2012). In at least 3 objects the non-thermal emission has been associated with the presence of a relativistic jet, and for two of them (Swift J1644+57 and Swift J2058+05) a precise localization around the centre of a quiescent galaxy supports the TDE interpretation. Because of their luminosity, such events can allow us to detect and study SMBHs around and beyond the peak of the star formation in the Universe. Moreover, the assumed dependence of the X-ray luminosity on the inverse of the SMBH mass makes them a unique tool to discover and constrain intermediate mass black holes in galaxy centres, which may still retain imprints of their (as yet unknown) formation mechanism. However, in stark contrast with the (few tens of) thermally emitting TDEs, the paucity of these events is preventing theoretical progress. Increasing the sample of jetted TDEs is therefore of paramount importance, and we propose a synergetic use of *Athena* with radio telescopes to reach this goal.

### 6.5.1 The role of X-rays and future radio surveys

The discovery of jetted TDEs in X-rays and the persistence of light in this band for a few years (e.g., Levan et al. 2016) may suggest that upcoming X-ray surveys would be the ideal discovery tool. In addition, (thermal and non-thermal) X-ray observations may have the advantage to quickly identify the TDEs from the expected and observed temporal index $t^{-5/3}$, which is a unique signature of TDE emission. The temporal behaviour of TDE emission at lower optical/UV frequencies is instead more diverse, and its interpretation controversial (e.g., Metzger and Stone 2016; Piran et al. 2015). These facts make lower frequencies a more uncertain tracer of TDEs.

Unfortunately, the predicted detection rates in X-ray surveys are somewhat disappointing. Our attempt to estimate how many J1644-like objects would be discovered by future X-ray surveys found that only a few per year will be detected by eROSITA, while several tens per year may possibly be detected by *Einstein Probe* up to redshift $\sim 1$. While several uncertainties in the modelling of X-ray emission (e.g. the value of the jet bulk Lorentz factor $\Gamma$, the value of the jet efficiency, the TDE luminosity function) affect the estimates of jetted TDE rate, much higher rates still appear unlikely.

The prospect improves when considering future radio surveys that will reach unprecedented sensitivity in the continuous monitoring of a large fraction of the sky. The SKA in survey mode at 1.4 GHz may serendipitously discover several *hundreds*, even *thousands*, of events per year up to $z \sim 2.5 - 3$ (see Donnarumma et al. 2015). The precise rates will depend on the SKA observing strategy, the cadence of observations of the same region of the sky, and the corresponding flux limit, but the expectations still remain much higher than those for future X-ray surveys, and so present a unique chance to extend the accessible cosmic volume.

However, the radio emission is quite featureless. Therefore it will be of primary importance to give a first rough identification of the transient as a TDE through localization. The localisation of an event within the nucleus of a quiescent host galaxy is a strong indicator of a TDE, although a low level of AGN activity may mimic such an event, and rarely TDEs may also happen outside the galactic nucleus (Komossa and Merritt 2008). The sub-arcsec angular resolution of SKA1-MID (Dewdney et al. 2016) will achieve this requirement, providing that the host galaxy is promptly identified by quick optical follow-up. Moreover, synergy with VLBI will be of primary importance for monitoring the transient (Paragi et al. 2015), providing us with a broader energy coverage and higher resolution imaging, with a better estimate of the bulk Lorentz factor ($\Gamma$) and how it changes as the TDE evolves.

Despite the limitations discussed above, the role of X-rays for the identification and detecting the X-ray counterpart of radio selected object is crucial.

### 6.5.2   The role of *Athena* in studying jetted TDE

*Athena* will offer a unique chance to follow up and characterize SKA-triggered TDEs. Donnarumma and Rossi 2015 find that the *Athena* sensitivity lies in the saturation branch of Figure 6.4, which implies that the observed rate of X-ray jetted TDEs will be crucially linked to its follow-up efficiency. This is mainly influenced by the *Athena* sky accessibility, which is of the order of $\sim 50\%$ (Nandra et al. 2013), resulting in a rate of TDEs which can be of the order of *hundreds per year reaching* $z \sim 2$, by assuming $\Gamma \sim 2$ as estimated from radio observations (Berger et al. 2012). In fact, it is not clear if $\Gamma \sim 2$ applies to the X-ray emission as well, because the high energy SED indicates $\Gamma \lesssim 20$ (Burrows et al. 2011). If the blazar-like nature of the X-ray emission is confirmed then $\Gamma \sim 10$ will be a more reasonable choice, reducing the expected rate by a factor of $(2/10)^2$. This would imply a more compact region responsible for the X-ray emission, in agreement with the fast variability observed in Sw J1644 ($\sim 200$ s), which was not detected in radio.

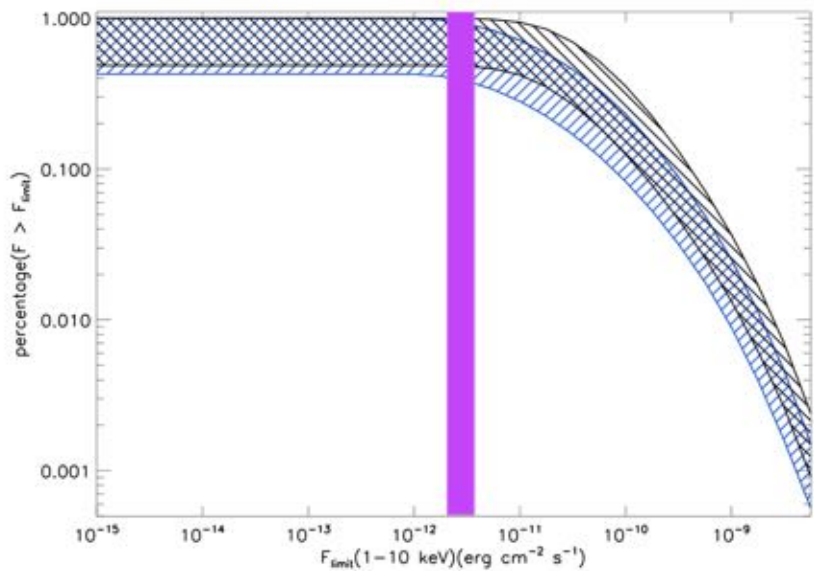

Figure 6.4: Fraction of radio selected TDEs identified in X-rays as a function of the (unabsorbed) flux limit in 1-10 keV. The flux limit is an average over 4 days of X-ray observations started within a 1-day delay since the radio trigger.The blue and black shaded areas are obtained by different assumptions in the modelling of radio light curve. See Donnarumma and Rossi 2015 for more details. The figure shows that an unabsorbed flux limit of $\sim 5 \times 10^{-12}$ erg cm$^{-2}$ s$^{-1}$ in 1-10 keV (purple box) should allow identifying any radio detected TDE. This figure has been adapted from Donnarumma and Rossi 2015.

Given the small sample detected so far, several uncertainties in the modelling of the X-ray emission remain, preventing us from predicting confidently the expected X-ray emission whenever a relativistic radio jet is observed (see e.g., Kara et al. 2016). Moreover, the recent discovery of a mildly relativistic outflow (van Velzen et al. 2016) associated with a *thermal* TDE (ASASSN-14li) makes the picture more complex, and highlights the relevant role of radio and X-ray synergy in these studies. The question is therefore what are the emission regions/mechanisms of thermal and non-thermal X-ray and of the radio light, and what is their relationship?

*Athena* follow-up of radio triggered events will help to clarify these open questions, thanks to its sensitivity and spectroscopic capability (the latter being more relevant for thermal TDEs). Moreover, its soft X-ray band extending up to 10 keV will play a crucial role in distinguishing between thermal and non-thermal X-ray emission. These observations will be crucial to constrain the nature of radio emitting TDEs and will provide a unique opportunity to study the underlying engine.

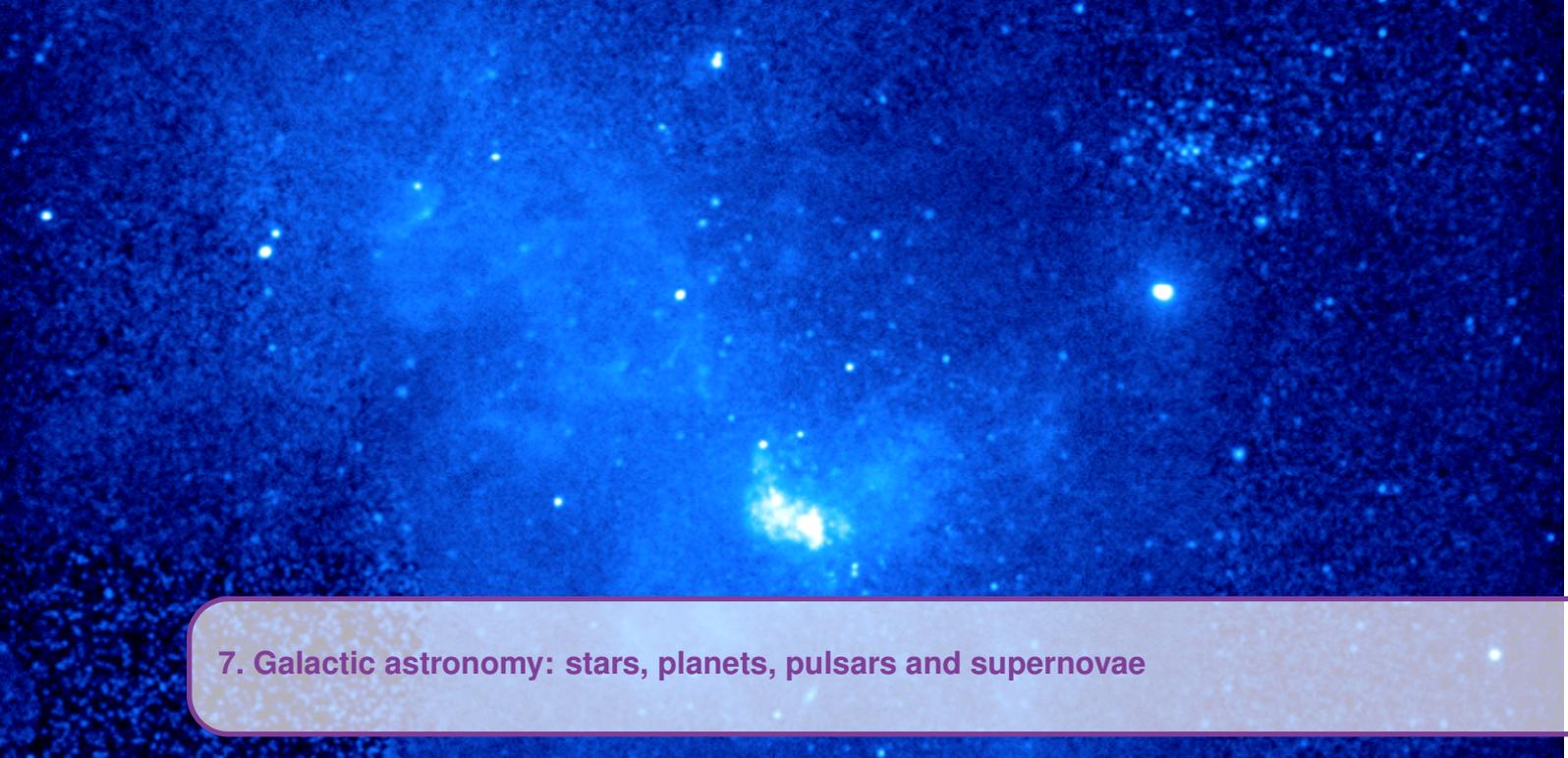

## 7. Galactic astronomy: stars, planets, pulsars and supernovae

### 7.1 Introduction

In this chapter, we consider the astrophysics of individual stars, of all types from low-mass protostars to the most massive stars, and finally the relativistic remnants, neutron stars. The combined use of SKA and *Athena* has the potential to unlock a great deal of new science in these areas. In particular, radio and X-ray emission identify the sites of the highest energy processes and of interactions between the different components.

### 7.2 Young stellar objects and dwarf stars

#### 7.2.1 Low-mass protostars

Even the earliest evolutionary stages of YSOs show high-energy emission in both the radio and the X-ray wavelength ranges (e.g., Feigelson and Montmerle 1999). The fact that their time-averaged radio and X-ray luminosities follow the empirical Güdel-Benz (GB) relation (Güdel and Benz 1993) for active stars points to an underlying connection, but the details remain unclear, particularly for the least evolved protostars (e.g., Forbrich, Osten, and Wolk 2011). The non-thermal radio emission is thought to trace the energy injection in magnetic energy release, with the heating then resulting in thermal X-ray emission. This scenario is sometimes observed through correlated light curves: the so-called Neupert effect (Neupert 1968). Additionally, thermal radio emission from the ionized material is produced at the base of jets. In the time domain, spectacular X-ray flares from YSOs are a frequently studied phenomenon (e.g., Getman et al. 2008). These flares are an important source of high-energy irradiation of both protoplanetary disks and young exoplanets. Contrary to X-ray flares, only a few large YSO radio flares have been seen (Bower et al. 2003; Forbrich, Menten, and Reid 2008). This phenomenon and its relation to X-ray flares are finally becoming more observationally accessible with the advent of the upgraded VLA. Forbrich et al. 2016b obtained deep simultaneous observations of the Orion Nebula Cluster (ONC), targeting hundreds of YSOs with both the VLA and *Chandra*. First results from the time-domain analysis of this dataset indicate the presence of a few strongly correlated radio and X-ray flares, as well as significant uncorrelated strong X-ray and radio variability. The results demonstrate that both wavelength ranges indeed provide us with complementary views of high-energy processes in YSOs (Forbrich et al. 2017).

#### 7.2.2 Ultracool dwarfs

Since the first radio detection almost twenty years ago, it is known that very low mass stars and brown dwarfs (hereafter, Ultracool Dwarfs, UCDs) produce unexpectedly strong radio emission, demonstrating that these objects are generating and dissipating kG-strength magnetic fields (e.g., Berger et al. 2001; Osten et al. 2006; Hallinan et al. 2008). The radio luminosity remains nearly uniform from early-M to mid-L dwarfs, even though other magnetic activity indicators (Hα and X-rays) decline by about two orders of magnitude over the same spectral type range (e.g., West et al. 2004; Berger et al. 2010). As a result, these objects significantly exceed the empirical GB relation (Güdel and Benz 1993) for active stars, indicating that the underlying physical processes are different. Williams, Cook, and Berger 2014 established that stars with spectral types <M6.5 are still compatible with the GB relation, while there is an increasing departure for cooler and generally lower mass stars. Additionally, highly polarized, non-thermal radio bursts have been detected toward UCDs (e.g., Berger et al. 2005; Hallinan et al. 2006), and it is possible that some show both quiescent gyrosynchrotron emission and flaring due to Electron-Cyclotron Maser Instability (ECMI), providing a powerful probe of the magnetic field properties. UCDs are a stepping stone between stellar and planetary radio emission, and intriguingly, ECMI is also thought to explain radio emission of planets in our own solar system, Jupiter and Saturn (e.g., Zarka 1998). The presence of non-thermal radio emission also makes UCDs interesting targets to look for sub-stellar companions through VLBI sub-milliarcsecond astrometry and/or direct imaging (e.g., Forbrich et al. 2016a).

### 7.2.3 SKA–*Athena* synergies

The main synergies between the SKA and *Athena* will be in the systematic exploration of high-energy processes in the time domain, utilizing simultaneous observations (e.g., Forbrich et al. 2011). While separate studies at radio and X-ray wavelengths already expand the currently accessible parameter space, this combination has only recently become accessible with the VLA and *Chandra*/*XMM-Newton*, and first results are intriguing. The SKA and *Athena* will make such observations accessible for a larger sample of objects and at higher S/N than before. One constraint to keep in mind, however, is that the SKA will typically have better angular resolution than *Athena*. Targets will therefore have to be carefully selected taking this into account. Therefore, the main benefit of SKA-*Athena* synergies will be for nearby UCDs that are known not to be in multiple systems. It will be possible not only to study time-integrated radio and X-ray properties into the brown dwarf regime, where emission is often faint, but it will also be possible to obtain time-resolved spectral information from both radio and X-ray observations, enabling more detailed study of the unusual high-energy processes of UCDs. While the SKA will allow us to better disentangle various non-thermal radio emission processes, *Athena* will provide X-ray spectral fit parameters like plasma temperatures at unprecedented sensitivity, including in the time domain.

Studies of YSOs will be more limited in sample size than those with *Chandra*, owing to the larger PSF and the fact that most nearby YSOs are in clusters; however, the radio–X-ray properties of a range of sufficiently isolated YSOs will be accessible at high sensitivity. While smaller, this sample will still cover a wide range of stellar masses and evolutionary stages. Simultaneous X-ray and radio observations for these sources will in particular allow us to disentangle thermal and non-thermal radio emission, while obtaining time-resolved X-ray spectral fit parameters such as plasma temperatures, and also measuring the variability of spectral lines that constrain the accretion process and/or disk heating.

As a result, using the SKA and *Athena*, we will be able to compare an unprecedented number of sufficiently isolated YSOs and UCDs with the GB relation to trace their overall high-energy characteristics. A crucial requirement for this work is the ability to disentangle non-thermal and thermal emission in YSOs. Additionally, the SKA and *Athena* will enable radio/X-ray time-domain studies of these sources with spectral information in both bands at unprecedented sensitivity.

## 7.3 Stars and star-planet interactions

The study of stellar and planetary magnetospheres is an important field of research in modern astrophysics. However, current research in this area is severely limited by a lack of instrument sensitivity, which will be solved by next-generation facilities like *Athena* and the SKA. A new and promising route to obtaining magnetospheric constraints is via star-planet magnetic interaction (SPMI), which is predicted to give rise to observable phenomena directly connected to magnetospheres in several bands of the electromagnetic spectrum, and in particular at X-ray (flaring-like emission) and radio wavelengths (auroral radio emission). These observations will, therefore, provide a unique route to a deeper comprehension of stars and planets, supplying different and complementary information with respect to other currently available methods.

In solar-type stars, the large-scale magnetic field, generated by a dynamo process in their convection zones, has a great impact in the immediate proximity of the star, where close-in planets may orbit. SPMI causes multiple effects, which are usually time-dependent, because close-in planets encounter an inhomogeneous environment along their orbit (Strugarek 2017), and/or because of geometrical effects due to the directionality of the resulting emission. As previously stated, the SPMI produces observable effects at both X-ray and radio wavelengths. X-ray emission is predicted where the presence of a hot Jupiter triggers flaring-like enhanced activity in the stellar corona. This effect has been tentatively observed in a couple of stars, but we still lack high quality data, in terms of both sensitivity and orbital phase coverage, to study this phenomenon in detail. Two examples are provided by HD 17156 and by HD 189733. In HD 17156, which is orbited by a hot Jupiter in an eccentric orbit, an X-ray flare was observed by *XMM-Newton* when the planet was at periastron (Maggio et al. 2015). In HD 189733, a periodic enhancement of the X-ray emission was explained by the photo-evaporation of the atmosphere of a close-in planet, colliding with the magnetized stellar wind, spiraling down toward the star and finally impacting its surface ahead of the sub-planetary point (Pillitteri et al. 2015).

Radio observations are another method to probe the magnetic environment of stars and planets. Radio emission, at very low frequency and variable in time, has been detected from all the magnetized bodies of the solar system (Zarka, Lazio, and Hallinan 2015). The main mechanism is the cyclotron maser instability, and the corresponding emission is referred to as the Auroral Radio Emission (ARE). ARE is observed not only in planets but also in stars, where it is caused by the interaction of the stellar magnetosphere with the stellar wind or orbiting planets that act as inductors (Trigilio et al. 2011; Leto et al. 2016). The ARE is therefore a powerful probe for studying extrasolar planetary systems, both searching directly for planetary ARE or looking for the planetary signature in stellar ARE (Zarka, Lazio, and Hallinan 2015).

Although first attempts to detect clear evidence of SPMI have been made at both X-ray and radio wavelengths, very few cases have resulted in unambiguous detections. The main reason is that current facilities do not allow this kind of study for a significantly large sample of objects. These phenomena are expected to be intrinsically faint and variable in time. Therefore, a successful observational strategy is to systematically monitor a single object for at least an entire orbital period, rather than collecting sparse "snapshots" of many different systems. In addition, the instruments need to be sufficiently sensitive to achieve robust detections in a short and well-defined amount of time. Only in this situation is it possible to detect and sample such transient emission. For example, it is expected that the typical ARE from nearby exoplanets (like for example HD 189733b discussed above) will be characterized by flux densities between a few microJansky and tens of milliJansky at frequencies around 100 MHz, depending on the star-planet distance. This sensitivity limit is beyond the capabilities of current facilities. Therefore while SPMI is potentially a powerful probe to study both stellar and planetary magnetosphere, our models are still poorly constrained because of observational difficulties.

Joint observations of stars hosting a hot Jupiter and systems with a massive planet in high eccentric orbits with *Athena* and SKA will be extremely valuable, since emission is also predicted in the radio band. Further coordinated simultaneous SKA and *Athena* observations of the best known systems hosting a hot Jupiter will provide a tremendous insight into the physical process at work, enabling us to refine the physical parameters characterizing the SPMI, and shedding light on the planetary system.

## 7.4 Massive stars

The stellar winds of massive stars are a defining component, driving their evolution and determining their endpoint. Massive stars are also key to our understanding of galactic "ecology", and it is vital to understand the details of how they provide radiative, mechanical and chemical feedback to the galactic ISM. Here we will focus on O/early B-type and Wolf-Rayet (WR) stars. Observations at X-ray and radio wavelengths are key to understanding different components of the stellar wind. At both wavelengths, the observed emission comes solely from the wind, and while it is free-free emission in both cases, the locations and processes traced by the emission are quite different. It is worth noting that the mass-loss rates of massive stars remain uncertain, with methods of estimating mass loss being subject to a range of uncertainties, including the level of asymmetry and clumping (Oskinova, Hamann, and Feldmeier 2007).

Characteristic properties of the winds depend on stellar class. O-stars have mass-loss rates $\dot{M} \sim 10^{-8} - 10^{-6} M_\odot$ yr$^{-1}$ and wind terminal velocities, $V = 1000 - 3000$ km s$^{-1}$, For WR stars, values of $\dot{M} = 10^{-5} M_\odot$ yr$^{-1}$ or higher are typical – that is WR stars have mass-loss rates that are an order of magnitude greater than O-stars, and the wind abundances are dominated by He (or C+O) rather than H. A large fraction of massive stars are in binary systems where both components have strong winds. These winds collide and give rise to copious (and harder) X-ray emission (as compared to single stars) and often non-thermal radio emission (Stevens, Blondin, and Pollock 1992).

We now know that a fraction (10%) of O-stars have magnetic fields with $B > 100$ G. Fields of this magnitude are capable of modifying the wind dynamics. An interesting new result concerns the B-type binary system $\varepsilon$ Lupi, where both components have magnetic fields ($B_* \sim 100 - 200$G), and the B-fields are roughly parallel to the rotation axes of the star, but are anti-aligned to each other (Shultz et al. 2015).

### 7.4.1 X-ray/radio emission processes and characteristics

**X-ray:** Massive stars are X-ray sources with a characteristic thermal temperature of $kT \sim 0.5$ keV. Radiatively driven winds are highly unstable, and this leads to shocks within the wind. A post-shock gas temperature of 0.5 keV corresponds to a strong shock with $\Delta V \sim 650$ km s$^{-1}$. Massive stars have a complex X-ray line spectrum (Cassinelli et al. 2001). The presence of a strong magnetic field may alter the wind dynamics, to give rise to a wind-compressed disk, where X-rays comes from a region in the magnetic equatorial plane (Güdel and Nazé 2009).

In terms of variability, to date this has been poorly sampled. For single stars, the variability is typically quite low-level and stochastic. One important recent exception is the discovery of X-ray pulsations (with the same period as the optical pulsations) from the massive B-type star $\zeta^1$ CMa, a $\beta$ Cepheid type star, which also has a strong magnetic field (Oskinova et al. 2014). For binary systems, orbital X-ray variability may be very marked. An example is WR140, a very eccentric WR+O-star system with a period of 7.9 years, where significant X-ray variability is seen around periastron passage (Sugawara et al. 2015).

**Radio:** Massive stars have ionised winds and these will be thermal radio sources. The flux from such winds scale as $S_\nu \propto \dot{M}^{4/3} V^{-4/3} d^{-2} \nu^{0.6}$, where $\dot{M}$ is the wind mass-loss rate, $V$ the terminal velocity, $d$ the distance and $\nu$ the frequency (Wright and Barlow 1975). In addition, some stars (usually binary) also show non-thermal radio emission believed to be associated with shock acceleration from the wind collision region.

Most single O-stars are not very variable at radio wavelengths (though this is poorly studied, with sparse observations). Some binary systems do show orbital variability (Dougherty et al. 2005). Radio Recombination Lines (RRLs) have been seen from a small number of massive stars with slow winds (Fenech et al. 2017).

### 7.4.2 SKA-*Athena* synergies

With their enhanced sensitivity, both the SKA and *Athena* will greatly advance our knowledge of massive stars, extending our coverage to a larger number of objects. It is worth highlighting a few specific points regarding SKA-*Athena*.

1. Single stars: Both *XMM-Newton* and *Chandra* grating instruments have observed a rich X-ray line spectrum in several massive stars. *Athena*/X-IFU has a much larger collecting area than the existing instruments, and substantially better spectral resolution and energy coverage (including the Fe lines at $\sim 6.5$ keV). The improved sensitivity of SKA will enable line profiles of more stars to be observed.
   We know that the winds of massive stars are clumpy, and these clumps will propagate on characteristic timescales of a few hours to days (as has been seen in the UV). The X-ray emission region is typically a few stellar radii, whereas the radio emission region is much larger (and frequency dependent – larger at low frequencies). Monitoring of bright single O-stars, such as Zeta Puppis, in both bands, should reveal sequential variability (both continuum and line), and this will be a great boost to modelling.

2. Single stars – rotational modulation: We have hints of this at radio and X-ray wavelengths (Leto et al. 2017), but for magnetic winds (where the magnetic axis is different from the rotational axis) we expect to see phase-related variability, due to the non-symmetric wind geometry. We will also see pulsationally driven modulation of the wind in X-rays; whether it will persist to radio wavelengths is unclear.

3. Magnetic Peculiar Stars: In a small number of cases we see Electron Cyclotron Emission (ECM) at radio wavelengths – this has been seen in M-type low mass stars, and in one massive star (CU Vir/A0Vp) (Ravi et al. 2010). The observational signature is short duration, highly polarised bursts (with the beamed emission coming from polar regions). The SKA, having a much greater focus on short timescale transients, will detect a larger number of these objects. CU Vir is a weak X-ray source (Robrade 2016). This will become a much larger area of research with SKA-*Athena*.

4. Binary systems: Orbital X-ray spectral line variability has been predicted for binary stars – for example, we expect very substantial variations in the shape of the Fe XXV triplet lines ($\sim 6.7$ keV), which will diagnose the hot gas from the wind collision (Rauw, Mossoux, and Nazé 2016). This line is not accessible to current grating instruments, but it will be to *Athena*. We may be able to see similar radio RRL variations, as well as changes in continuum flux and shape (as the free-free optical depth changes). Consequently, coordinated radio/X-ray observations of colliding wind binary systems should provide extremely powerful diagnostics via continuum and line variations.

5. Surveys: The rapid surveying capability of SKA and *Athena*/WFI will generate results for a substantial population of massive stars, enabling identification of unusual objects, as well as systematic studies of individual objects. It is worth noting that much can be done in the meantime with MeerKAT and *XMM-Newton*, such as the MeerGAL survey of the Galactic Plane. Arranging simultaneous observations is hard at the best of times. Having a mechanism whereby time can be awarded on both facilities, with flexible scheduling to allow simultaneous observation, would be extremely desirable for optimal exploitation of SKA-*Athena* synergies in this area.

In summary, there are a wide range of massive star phenomena that are accessible to SKA and *Athena*, and both missions promise exciting new results, especially when combined.

## 7.5 Pulsars

Pulsars are the NS remnants that result from the collapse of massive stars. They produce beams of radio emission from their magnetic poles, which we detect once per rotation, every time the beam sweeps past our line of sight. Pulsars are among the most strongly magnetised objects in the Universe, and often display outstanding rotational stability, approaching that of atomic clocks over timescales of several years. The high-precision timing of pulsars – and especially of Millisecond Pulsars (MSPs), whose rapid spin results from matter and angular momentum transfer from a binary companion donor – allows us to address many burning questions in fundamental physics and astrophysics.

Pulsar research will reach exciting new heights once SKA is operational (see, e.g., Kramer and Stappers 2015). For example, nanohertz-frequency gravitational wave detection will certainly be accomplished through pulsar timing observations with the SKA. Its high sensitivity will help to usher in a new era of Gravitational Wave (GW) astrophysics, in particular probing SMBH formation and galaxy evolution in the early, distant Universe. In addition, we expect the first discovery of a pulsar-black hole binary, which will provide unprecedentedly precise strong-field tests of gravity. The SKA will also provide a census of the Galactic pulsar population, from which a comprehensive understanding of the formation and evolutionary histories of neutron stars, both isolated and within binary systems, can be developed. Indeed, the revolutionary potential of pulsar observations to probe strong-field relativity and provide a wide variety of ultra-precise science is a key driver for Phase 1 of the SKA.

Research with NSs will additionally be boosted by exploiting high-profile areas of X-ray and radio observational overlap, where many strong synergies between SKA and *Athena* are anticipated. In what follows, we itemise and briefly discuss several of these key areas of study:

- **Emission from rotation-powered neutron stars:** At present, less than 10% of known rotation-powered radio pulsars are also detected in the X-ray band (Prinz and Becker 2015), via magnetospheric and/or thermal emission. SKA will increase the number of known pulsars by a factor of 10, while *Athena* will strongly enhance the chances of detecting many X-ray counterparts. This much larger available sample will lead the way to significantly constraining the long-debated processes and geometry of non-thermal emission from rotation-powered NSs. Focused X-ray timing campaigns and spectroscopic observations of young pulsars, many of which are known to frequently glitch, will allow for more comprehensive studies of the physics of the NS interior and magnetospheric evolution.

- **Core-collapse supernovae:** The standard picture of core-collapse supernovae is that of an explosion producing both a NS, whose rotational energy powers a Pulsar Wind Nebula (PWN) and a SNR. One therefore expects most young neutron stars to be associated with both a PWN and a SNR. However, there are very few such systems seen to have a pulsar, PWN, and SNR. This situation will change significantly thanks to the orders-of-magnitude upgrade in sensitivity offered by *Athena* and SKA (Gelfand et al. 2015), which will greatly improve our understanding of how NSs are formed in these explosions.

- **Evolution of high-mass X-ray binaries and their progenitors:** At present, only a handful of pulsars orbiting massive stars are known. This number will increase with the SKA, while *Athena* will provide new source samples and further observations and characterisation of high-mass X-ray binaries (HMXBs). Combined, this will enable a population flow study and a better understanding of the link between pulsars with massive companions (including the double NS systems) and the HMXBs. This will be crucial to significantly improve constraints on NS evolution and NS kicks (Noutsos et al. 2013; Tauris et al. 2015).

- **X-ray dim isolated neutron stars (XDINSs) and magnetars:** There are 8 cataloged young NSs that are apparently radio quiet, but detected in X-rays, thanks to the thermal emission produced by their hot surfaces; they are dubbed XDINSs, and they will be a prime target for sensitive radio observations with the SKA, while *Athena* is expected to discover a number of additional nearby XDINSs. A combined radio/X-ray investigation will enable the comparison of the XDINSs' ages derived from both the surface temperature (via the NS cooling curves) and, where detected, radio timing. This combination of observing strategies will clarify how they fit into the full picture of the so-called NS "zoo" (Tauris et al. 2015). One possibility is that they are descendants of the magnetars, i.e. NSs powered by the decay of their ultra-strong magnetic fields (Kaspi and Beloborodov 2017). The study of magnetars will also greatly benefit from coordinated SKA-*Athena* observations. In fact, the onset of pulsed radio emission in a sub-group of these objects (the so-called transient magnetars) is linked with the occurrence of X-ray outbursts (e.g., Camilo et al. 2006; Rea et al. 2012) by some mechanism that is not yet understood.

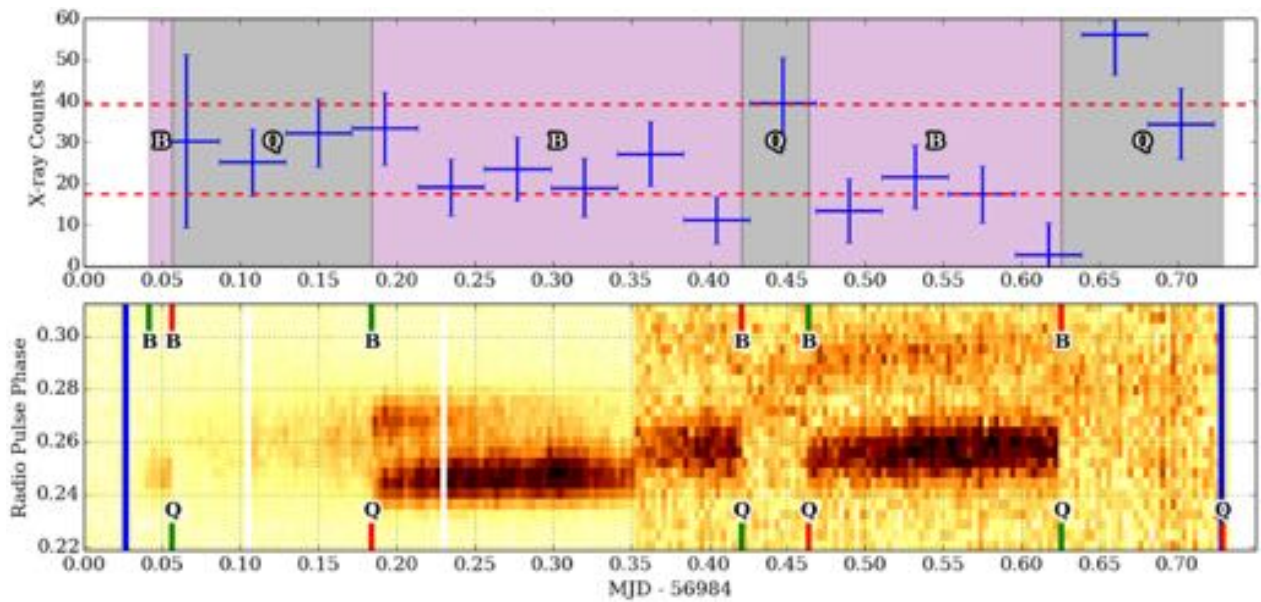

Figure 7.1: X-ray (top; *XMM-Newton* 0.2-10 keV) and radio (bottom; LOFAR for $MJD < 56984.35$/LWA for $MJD > 56984.35$) light curves for an observing session of PSR B0943+10. One can clearly observe the radio profile changes coincident with abrupt changes in X-ray counts (labelled as green/red ticks at start/end of so-called B/Q modes). Blue lines represent *XMM-Newton* observation start/stop times (From Mereghetti et al. 2016).

- **Moding pulsars:** A (so far small) number of "moding" pulsars (e.g., Hermsen et al. 2013; Mereghetti et al. 2016) show dramatic changes in their radio emission profiles that coincide with strong variations in their X-ray emission properties. These are particularly favourable targets for understanding the conditions under which various non-thermal emission processes can occur in the strong NS magnetosphere. In order to study these targets in detail one needs high-quality simultaneous radio and X-ray data, as well as a much larger sample of moding pulsars. SKA and *Athena* will together provide both of these requirements.

- **Transitional millisecond pulsars:** Radio millisecond pulsars (MSPs) are old NSs that have been spun up via accretion of mass from a donor star in an X-ray binary. This "recycling" scenario, for so long theoretical, has now been strengthened by the discovery of the transitional MSPs (tMSPs; e.g., Archibald et al. 2009; Papitto et al. 2013), which are detected alternately as radio and X-ray emitters. They are seen to switch between modes that exhibit the properties of rotation-powered MSPs and low-mass X-ray binaries. These tMSPs are key probes of various accretion phenomena and of rapid variations in the pulsar magnetosphere. The combination of SKA and *Athena* is eagerly anticipated, as it will reveal a large number of additional tMSPs to investigate with unprecedented accuracy. This will enable a deep understanding of the evolution of NSs into full-fledged, radio MSPs.

- **X-ray point sources in globular clusters:** MSPs in globular clusters (GCs) are uniquely valuable objects. Timing in the radio band allows the investigation of binary evolution in conditions of very high stellar density, as well as constraints on many otherwise difficult-to-measure properties of the host GC. Unfortunately, a large fraction of the MSPs in GCs are still undetected in the radio band (e.g., Turk and Lorimer 2013). Observations with the excellent spatial resolution of *Athena* will be critical for precisely highlighting the positions of many additional MSPs in the targeted GCs. The high sensitivity of the SKA will then be fully exploited to easily provide radio detections of these pulsars, and their follow-up timing and characterisation (e.g., Hessels et al. 2015).

## 7.6 Supernova remnants

SNRs are objects that are formed by the strong shock waves of a supernova explosion. The shock heats and ionises the ambient gas and the ejected material, which makes SNRs bright X-ray sources. In addition, particles are accelerated in the shocks. The emission from non-thermal electrons with GeV energies makes them bright at radio wavelengths. In the last two decades, new observations from X-rays up to the TeV band have shown that electrons can gain TeV energies in SNR shocks, while heavier particles can be also accelerated and become highly relativistic. Non-thermal X-ray and TeV emission has been detected in particular from young nearby SNRs (e.g., Tycho's SNR, see Fig. 7.2, left).

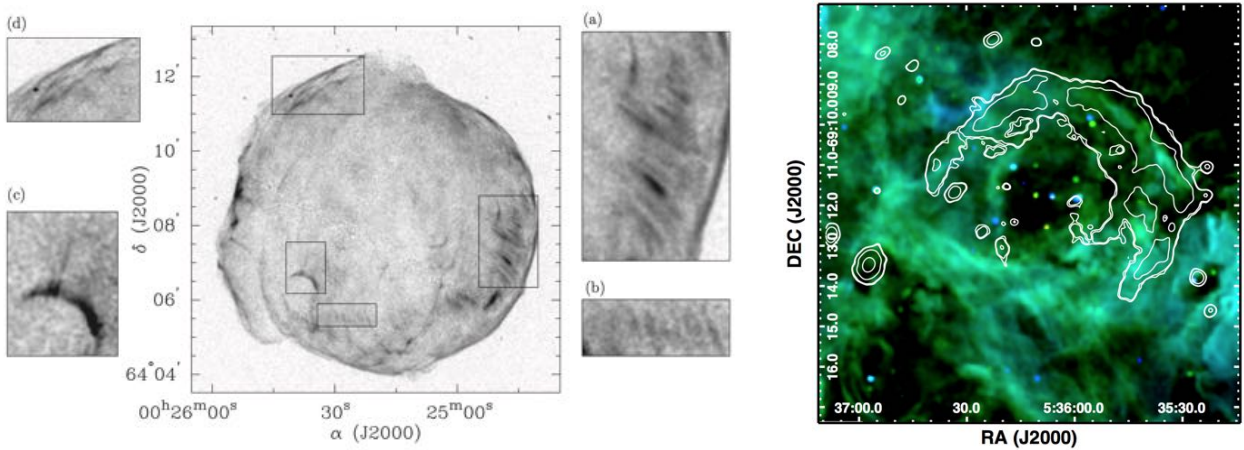

Figure 7.2: *Left:* X-ray image of Tycho's SNR taken with the *Chandra* X-ray Observatory ($4 - 6$ keV). There are non-thermal filaments and stripes indicative of amplified magnetic fields. Taken from Eriksen et al. 2011. *Right:* Three-color image of 30 Dor C using images of the Magellanic Cloud Emission Line Survey (MCELS) with red: [S II], green = H$\alpha$, blue = [O III] and contours of the X-ray emission taken with *XMM-Newton* ($2 - 7$ keV).Taken from Kavanagh et al. 2015.

By observing supernova remnants, we can study the processes responsible for the acceleration of particles to ultra-relativistic energies. In particular, the following questions can be addressed: What is the origin of the TeV emission? How does the injection of particles work, and how do they escape the shocks? What are the magnetic field configurations, and how are magnetic fields amplified?

In addition, interstellar structures called superbubbles are also powered by strong shock waves, and can thus be expected to act as cosmic particle accelerators as well. Superbubbles are created by the stellar winds of massive stars. Typically, superbubbles are also thermal X-ray sources, surrounded by HII regions. However, there is a superbubble in the Large Magellanic Cloud (LMC), 30 Doradus C (30 Dor C), which is a hard and non-thermal X-ray source. It shows a nice ring-like structure in radio, X-ray, and in H$\alpha$ (see Fig. 7.2, right). Based on the analysis of SEDs from radio to X-rays, Kavanagh et al. 2015 have shown that the non-thermal emission of 30 Dor C is mainly synchrotron emission.

In addition, 30 Dor C is one of only three TeV sources confirmed so far in the LMC (H.E.S.S. Collaboration et al. 2015). The possible scenarios for their TeV emission are the same as those discussed for SNRs: in the leptonic scenario, inverse Compton scattering of photons from the existing radiation fields by high-energy electrons can give rise to TeV emission, or protons and nuclei that have been accelerated in the shock fronts interact with matter in the surroundings producing neutral pions that decay by emitting GeV to TeV photons. Both scenarios require extreme conditions in the ambient interstellar medium (high magnetic fields or high densities) and high shock velocities. In summary, the TeV emission from 30 Dor C indicates that there must be a young SNR with a shock velocity of $> 3000$ km s$^{-1}$ expanding inside.

### 7.6.1 Studying interstellar shocks with SKA and *Athena*

Combining SKA and *Athena* will enable a complete census of SNRs and superbubbles in the Milky Way and the Magellanic Clouds to be achieved, so that the population of high-energy particles and their emission cam be studied. In particular, the *Athena*/X-IFU will allow us to study the thermal plasma properties in detail and to constrain the emission of the non-thermal processes. The radio observations will improve our understanding of the interstellar electron population and the structure and the properties of the magnetic fields. In particular, we will be able to study the lower energy population of non-thermal electrons, which are responsible for the low-energy end of the SED.

To summarize, SKA and *Athena* will allow us to get a better understanding of the microphysics of interstellar shocks and their interaction with the ambient interstellar medium, study the population of high-energy particles and their acceleration mechanisms, and the interstellar magnetic fields. These investigations will also be crucial for the study of the same objects and physics in the TeV energy band with the CTA.

## Acknowledgements

- Akahori thanks Dongsu Ryu and Bryan Gaensler for their significant contributions to the presented work. This work is supported in part by JSPS KAKENHI Grants: 15H03639, 15K17614, and 17H01110.
- Brunetti and Markevitch thank T.W. Jones, A. Lazarian, F. Vazza, S. Peng Oh, S. Ettori, C. Pfrommer, F. Gastaldello, J. Donnert for insightful discussions.
- Ferdman acknowledges J. Gelfand, J. W. T. Hessels, T. J. W. Lazio, M. Kramer, A. Possenti, I. H. Stairs and T. M. Tauris.
- Feretti and Govoni acknowledge extensive discussions with colleagues who contributed to the realization of this project: A. Bonafede (INAF Istituto di Radioastronomia), S. Ettori (INAF Osservatorio Astronomico), F. Gastaldello (INAF Istituto di Astrofisica Spaziale), G. Giovannini (University of Bologna), M. Gitti (University of Bologna), M. Murgia (INAF Osservatorio Astronomico), V. Vacca (INAF Osservatorio Astronomico), and F. Vazza ((Università di Bologna, Italy/Universität Hamburg, Germany).
- Ghirlanda acknowledges supporters of this science case: A. Ferrara (SNS - Pisa), M. Giroletti (INAF-IRA Bologna), L. Piro (INAF-IAPS Roma), R. Salvaterra (INAF-IASF Milano), T. Venturi (INAF-IRA Bologna), S. Vergani (Observatoire de Paris).
- Ingallinera acknowledges the contributions of Salvatore Sciortino, Antonio Maggio and Ignazio Pillitteri from INAF - Osservatorio Astronomico di Palermo and of Corrado Trigilio, Grazia Umana and Carla Buemi from INAF - Osservatorio Astrofisico di Catania.
- Panessa acknowledges her collaborators: Ranieri Baldi (University of Southampton); Loredana Bassani (IASF/INAF); Ehud Behar (Israel institute of Technology); Rob Beswick (Jodrell Bank Centre for Astrophysics and Jodrell Bank Observatory); Piergiorgio Casella (Osservatorio Astronomico di Roma); Poshak Gandhi (University of Southampton); Marcello Giroletti (IRA/INAF); Lorena Hernandez-Garcia (IAPS/INAF); Ari Laor (Israel institute of Technology); Anna Lia Longinotti (CONACYT, INAOE Puebla); Ian Mc Hardy (University of Southampton); Raffaella Morganti (Netherlands Institute for Radio Astronomy); Rodrigo Nemmen (University of São Paulo); Miguel Perez-Torres (IAA-CSIC); Luigi Piro (IAPS/INAF); Hayden Rampadarath (Jodrell Bank Centre for Astrophysics); Francesco Tombesi (NASA/University of Maryland); Phil Uttley (University of Amsterdam)
- Reinout van Weeren wishes to thank Timothy W. Shimwell, Ralph P. Kraft, and Annalisa Bonafede for their input.
- Sasaki thanks her collaborators Miroslav Filipović (Penrith Observatory, Western Sydney University, Australia), Patrick Kavanagh (Dublin Institute for Advanced Studies, Ireland), Stefan Ohm (Deutsches Elektronen-Synchrotron DESY, Germany), and Jacco Vink (Astronomical Institute Anton Pannekoek, University of Amsterdam, Netherlands).
- Soria thanks James Miller-Jones, Gemma Anderson, Richard Plotkin, Vlad Tudor, Ryan Urquhart.
- Vazza et al. acknowledge precious scientific support by Marcus Brüggen (Universität Hamburg), Annalisa Bonafede (IRA/INAF) and Chiara Ferrari (Nice Observatory) for the preparatory work for this contribution. These computations were produced on Piz Daint (ETHZ-CSCS, Lugano) under allocation

s701. Vazza acknowledges financial support from the grant VA 876-3/1 by DFG, and from the European Union's Horizon 2020 research and innovation programme under the Marie-Sklodowska-Curie grant agreement no.664931.

- Webb acknowledges the help of F. Koliopanos, M. Coriat, O. Godet and M. Bachetti.

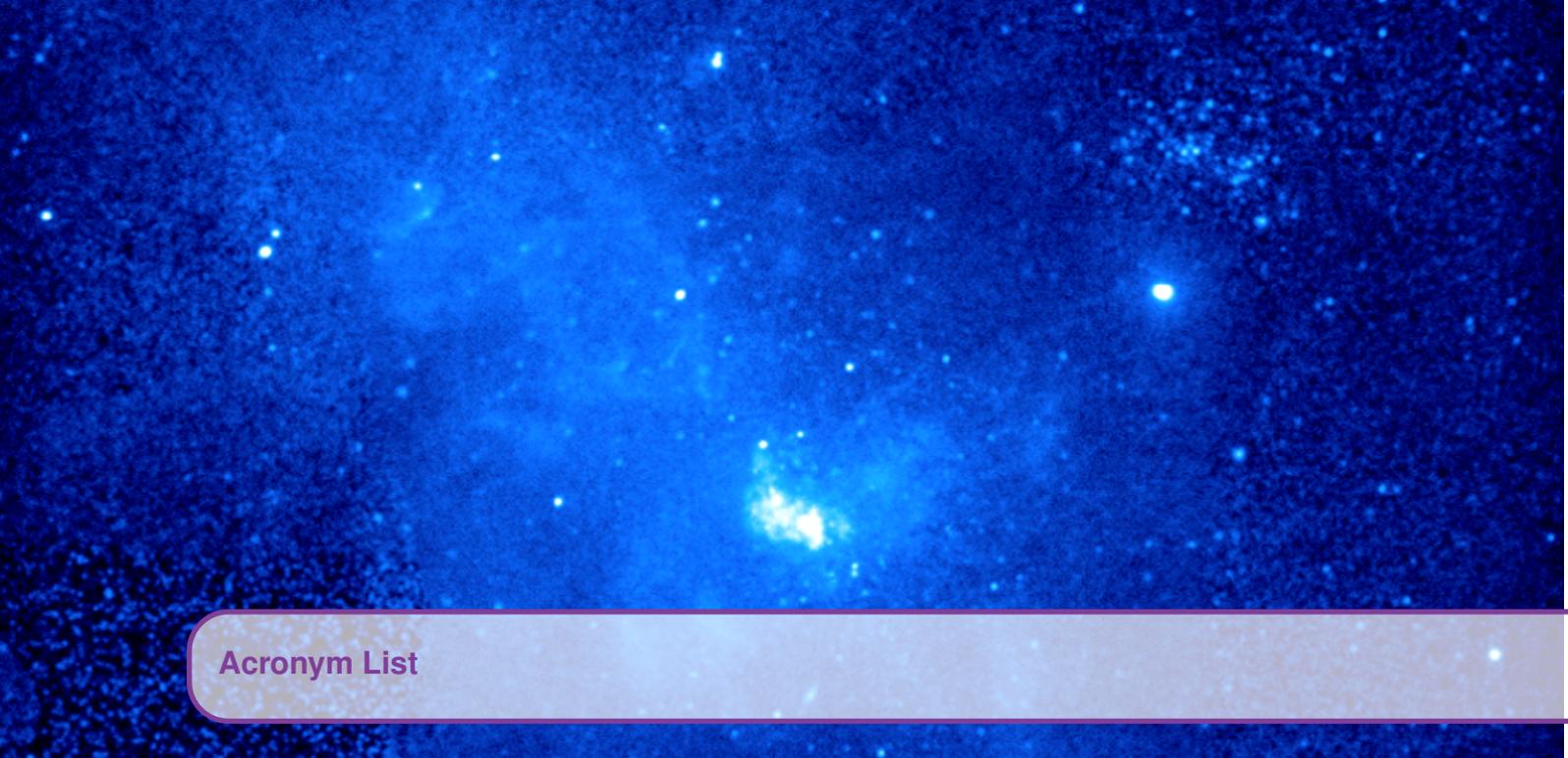

**Acronym List**

| | |
|---|---|
| AGN .. | Active Galactic Nucleus/Nuclei |
| ARE ... | Auroral Radio Emission |
| ASKAP | Australian Square Kilometre Array Pathfinder |
| ASST . | *Athena* Science Study Team |
| ATCA . | Australia Telescope Compact Array |
| *Athena* | Advanced Telescope for High ENergy Astrophysics |
| BH .... | Black Hole |
| CD .... | Cosmic Dawn |
| cD .... | central Dominant |
| CDFS . | *Chandra* Deep Field South |
| CMB .. | Cosmic Microwave Background |
| CR .... | Cosmic Ray |
| CTA ... | Cherenkov Telescope Array |
| DSA ... | Diffusive Shock Acceleration |
| ECM .. | Electro Cyclotron Emission |
| ECMI .. | Electron-Cyclotron Maser Instability |
| ELT ... | Extremely Large Telescope |
| EoH ... | Epoch of Heating |
| EoR ... | Epoch of Reionization |
| ESA ... | European Space Agency |
| FIR .... | Far-Infrared |
| FoV ... | Field of View |
| FRII ... | Fanaroff and Riley Class II |
| FRB ... | Fast Radio Burst |
| GB .... | Güdel-Benz |
| GC .... | Globular Cluster |
| GBH .. | Galactic Black Hole |
| GMT .. | Giant Magellan Telescope |
| GRB .. | Gamma Ray Burst |
| GW ... | Gravitational Wave |
| *HST* .. | Hubble Space Telescope |
| ICM ... | Intracluster Medium |
| IGM ... | Intergalactic Medium |
| IMBH .. | Intermediate Mass Black Hole |
| ISM ... | Interstellar Medium |
| HERG . | High Excitation Radio Galaxy |
| HEW .. | Half Energy Width |
| HMXBs | High-Mass X-ray Binaries |
| *JWST* | James Webb Space Telescope |
| ΛCDM | Lambda cold dark matter |
| LERG . | Low Excitation Radio Galaxy |
| LF .... | Luminosity Function |
| LMC .. | Large Magellanic Cloud |
| LOFAR | LOw Frequency ARray |
| MHD .. | Magnetohydrodynamic |
| MIR ... | Mid-Infrared |
| MSP .. | Millisecond Pulsar |
| NIR ... | Near-Infrared |
| NS .... | Neutron Star |
| ONC .. | Orion Nebula Cluster |
| Pop-I .. | Population I |
| Pop-II . | Population II |
| Pop-III | Population III |
| PSF .. | Point Spread Function |
| PWN .. | Pulsar Wind Nebula |
| QSO .. | Quasi-Stellar Object |
| RH .... | Radio Halo |
| RM ... | Rotation Measure |
| RRLs . | Radio Recombinations Lines |
| SAST . | SKA-*Athena* Synergy Team |
| SED .. | Spectral Energy Distribution |
| SFG .. | Star-Forming Galaxy |
| SFR .. | Star Formation Rate |
| SKA .. | Square Kilometer Array |
| SKAO . | SKA Organization |
| SMBH | Super-massive Black Hole |
| S/N ... | Signal-to-Noise ratio |
| SNe .. | Supernovae |
| SNR .. | Supernova Remnant |
| SPMI . | Star-Planet Magnetic Interaction |
| SZ .... | Sunyaev-Zel'dovich |
| TDE .. | Tidal Disruption Event |
| TMT .. | Thirty Meter Telescope |
| ToO ... | Target of Opportunity |
| UCD .. | Ultracool Dwarf |
| ULX .. | Ultraluminous X-ray source |
| UVB .. | Ultraviolet Background |
| VLA ... | Very Large Array |
| VLBI .. | Very Long Baseline Interferometry |
| WFI ... | Wide Field Imager |
| WHIM . | Warm-Hot Intergalactic Medium |
| WR ... | Wolf-Rayet |
| X-IFU . | X-ray Integral Field Unit |
| XDINS | X-ray Dim Isolated Neutron Star |
| XRBs . | X-ray Binary |
| YSO .. | Young Stellar Object |